\documentclass[11pt,leqno]{article}

\usepackage[margin=1.2in]{geometry}
\usepackage[T1]{fontenc}
\usepackage{amsmath,amssymb,amsthm}
\numberwithin{equation}{section}
\usepackage{graphicx}
\usepackage{float}
\usepackage{placeins}
\usepackage{array}
\usepackage{xcolor}
\usepackage{mathbbol}   % blackboard bold for lower-case and Greek letters, e.g. \mathbb{p} and \mathbb{\Lambda}
\usepackage{prodint}    % product integral
\usepackage{csquotes}   % \enquote
\usepackage{booktabs}
\usepackage{subcaption}
\usepackage{algorithm}
\usepackage{algpseudocode}
\usepackage{tikz}
\usetikzlibrary{arrows.meta, positioning, calc, fit, backgrounds, shapes.geometric}
\usepackage[authoryear,round]{natbib}
\usepackage[hypertexnames=false]{hyperref} % Unique page anchors when supplementary numbering restarts.
\hypersetup{
  colorlinks=true,
  citecolor=black,     % citations
  linkcolor=black,      % internal cross-references: equations, sections, results
  urlcolor=black,   % external hyperlinks
  filecolor=black,
}
\newcommand{\doi}[1]{\href{https://doi.org/#1}{doi:\nolinkurl{#1}}}

\newtheorem{theorem}{Theorem}[section]
\newtheorem{corollary}[theorem]{Corollary}
\newtheorem{lemma}[theorem]{Lemma}
\newtheorem{proposition}[theorem]{Proposition}
\theoremstyle{definition}
\newtheorem{remark}[theorem]{Remark}
\newtheorem{definition}[theorem]{Definition}
\newtheorem{assumption}{Assumption}
\newtheorem{example}[theorem]{Example}

\makeatletter
\renewenvironment{proof}[1][]{\par\pushQED{\qed}\normalfont
  \topsep6\p@\@plus6\p@\relax
  \trivlist\item[\hskip\labelsep\itshape Proof\if\relax\detokenize{#1}\relax\else\ #1\fi\@addpunct{.}]\ignorespaces
}{\popQED\endtrivlist\@endpefalse}
\makeatother

\newcommand{\startsupplement}{%
  \clearpage
  \pagenumbering{arabic}%
  \setcounter{section}{0}%
  \renewcommand{\thesection}{S\arabic{section}}%
  \renewcommand{\theHsection}{supplement.\arabic{section}}%
  \setcounter{assumption}{0}%
  \renewcommand{\theassumption}{S\arabic{assumption}}%
  \renewcommand{\theHassumption}{supplement.\arabic{assumption}}%
  \counterwithin{figure}{section}%
  \counterwithin{table}{section}%
  \addtocontents{toc}{\protect\setcounter{tocdepth}{2}}%
  \begin{center}
    {\large\bfseries Supplementary material for ``Random Jump Forests''}\\[8pt]
    {\normalsize Martin Bladt\qquad Rasmus Frigaard Lemvig}\\[2pt]
    {\normalsize Department of Mathematical Sciences, University of Copenhagen}
  \end{center}
  \begingroup
    \hypersetup{linkcolor=black}%
    \tableofcontents
  \endgroup
  \clearpage
}

\def\CC{\mathcal{C}}
\def\CD{\mathcal{D}}
\def\CW{\mathcal{W}}
\def\CF{\mathcal{F}}
\def\CG{\mathcal{G}}
\def\CH{\mathcal{H}}
\def\CI{\mathcal{I}}

\def\CN{\mathcal{N}}
\def\CX{\mathcal{X}}
\def\CZ{\mathcal{Z}}

\def\BC{\mathbb{C}}
\def\BD{\mathbb{D}}
\def\BE{\mathbb{E}}
\def\BF{\mathbb{F}}
\def\BI{\mathbb{I}}
\def\BN{\mathbb{N}}
\def\BR{\mathbb{R}}
\def\BP{\mathbb{P}}

\def\dd{\mathrm{d}}
\def\Var{\operatorname{Var}}
\def\Cov{\operatorname{Cov}}

\newcommand{\indep}{\perp\!\!\!\!\!\!\perp}
\makeatletter
\newcommand\pint{\DOTSI\if@display\Prodi\else\prodi\fi\ilimits@}
\makeatother

\begin{document}

% Hide main-paper entries in the supplement table of contents.
\addtocontents{toc}{\protect\setcounter{tocdepth}{-5}}

\title{Random Jump Forests}
\author{Martin Bladt\qquad Rasmus Frigaard Lemvig\\[2pt]
\normalsize Department of Mathematical Sciences, University of Copenhagen\\[2pt]
\normalsize\texttt{martinbladt@math.ku.dk}\quad\texttt{rfl@math.ku.dk}}
\date{}
\maketitle

\begin{abstract}
Random jump forests are introduced as a nonparametric tool for estimating conditional cumulative transition rates and occupation probabilities in multi-state models. The main idea is that a single tree partitions the covariate space for all transitions simultaneously, using partially observed trajectories, and with Nelson--Aalen estimators at the terminal nodes. Occupation probabilities are recovered through the product integral without requiring the Markov assumption. Uniform consistency and a functional central limit theorem for the estimators of cumulative rates can be established under the assumption that the covariate vector takes finitely many values, and under suitable splitting conditions. In particular, pointwise confidence intervals and simultaneous bands are available in the Markov case. Simulation studies show a predictive accuracy close to the correctly specified benchmark models and marked improvements when benchmarks are misspecified, with further illustration on four biomedical data sets. In the coverage study, confidence bands attain near-nominal coverage in several settings, with pooling of heterogeneous rates being the main source of possible undercoverage. The method, including prediction assessment and variable importance, is implemented in the R package \texttt{JumpForests}.
\end{abstract}



\section{Introduction}

\subsection{Motivation}

Multi-state models describe how an individual moves between a finite number of states over time. A patient who has received a bone marrow transplant may experience platelet recovery, relapse or death; an insured person may become disabled, recover or die; a firm may migrate between several credit ratings before defaulting. The quantities of interest are the transition rates between the states and the resulting occupation probabilities, both as functions of covariates given at baseline levels. Such models are used throughout biostatistics \citep[see e.g.][]{ABGK, AndersenRavn, Aalenetal, PutterFioccoGeskus}, and are widely used in life insurance mathematics \citep[see][]{SteffensenAsmussen, Christiansen, StatIns}. They also appear in credit risk \citep{JarrowLandoTurnbull, Lando}, reliability \citep{LimniosOprisan}, epidemiology \citep{Commenges} and queueing theory \citep{Fabens}.

Our aim is to estimate all transition rates using a common covariate partition, allowing the splitting criterion to use information from every transition. Under the classical Markov assumption, transition probabilities follow from the transition rates through the Kolmogorov equations; see \citet{Amsler} and \citet{HoemMarkov, HoemCongress} for early insurance applications and \citet{Koller} for a modern account. Without covariates, the rates and the transition probabilities are estimated nonparametrically by the Nelson--Aalen and Aalen--Johansen estimators \citep{Nelson, Aalen, AalenJohansen}. With covariates, the standard practice is to fit a Cox proportional hazards regression to each transition separately \citep{AndersenRavn}. This transition-wise approach requires the analyst to specify the functional form of every covariate effect and every interaction and it imposes a proportional hazards restriction on each transition. It also relies on the Markov assumption to combine the fitted transitions into occupation probabilities. However, the Markov assumption itself is often questionable, since the time already spent in a state frequently affects the rate of leaving it; the duration of a disability period, for instance, affects the probability of returning to work. Semi-Markov models \citep{JanssenDominicis, Nollau, Helwich, JanssenLimnios, Korolyuketal} address duration dependence and have a long history in insurance \citep{Janssen, HoemSemiMarkov}, in credit risk \citep{VasileiouVassiliou} and in nonparametric estimation \citep{LagakosSommerZelen, VoelkelCrowley}, but retain assumptions on the dependence of transition intensities on the process history. Machine learning offers an automated alternative to hand-specified regression models. The random survival forest of \citet{RSF} can be fitted in an ad-hoc fashion to each transition separately, resetting the clock whenever a new state is entered. This transition-wise forest serves as a benchmark in our numerical studies.

Two lines of work estimate multi-state quantities nonparametrically without a Markov assumption. Landmarking \citep{vanHouwelingen, PutterSpitoni} fits the Aalen--Johansen estimator to the individuals occupying a given state at a landmark time. Its consistency was studied by \citet{DattaSatten} and, with covariates, by \citet{Hoffetal}, and its weak convergence by \citet{Glidden} and \citet{Niessletal}; see also \citet{Overgaard} and \citet{Munchetal} on state occupation probabilities in non-Markov models and \citet{Sandqvist} for a combination of landmarking with boosted trees. Landmarking targets transition probabilities conditional on the state at a fixed landmark time; here we target occupation probabilities conditional on baseline covariates. Kernel smoothing targets the rates directly. It goes back to \citet{Beran} and \citet{Dabrowska} for survival data and to \citet{RamlauHansen}, \citet{McKeagueUtikal}, \citet{NielsenLinton} and \citet{DabrowskaCox} for counting processes and multi-state models; see \citet{Ghosh} for background. The conditional Aalen--Johansen estimator of \citet{BladtFurrer} targets conditional occupation probabilities without a Markov assumption. The present paper retains this setup and replaces kernel smoothing by adaptive tree partitions shared across transitions, so that covariate interactions can be represented through successive splits.

\subsection{Contributions}

We introduce the \emph{random jump forest}, called the jump forest for short, an extension of random forests \citep{BreimanBagging, BreimanRF, CART} to general multi-state models. Random forests have previously been adapted to quantile regression \citep{Meinshausen}, survival analysis \citep{RSF} and competing risks \citep{RFCompetingRisks}; the jump forest grows a single forest for all transitions of a multi-state process at once. Our contributions are the following.
\begin{enumerate}
    \item[(i)] \emph{Methodology.} Each tree partitions the covariate space using splitting rules that merge information across all transitions, and each terminal node stores Nelson--Aalen estimators of every transition rate together with the empirical initial distribution. Occupation probabilities are recovered from the forest-averaged rates through the product integral. No Markov assumption is needed, out-of-bag Brier and Kullback--Leibler scores provide built-in error assessment, and a permutation variable importance applies to general multi-state models.
    \item[(ii)] \emph{Theory and inference.} Assuming that the covariate vector takes finitely many values and that suitable splitting conditions hold, we establish uniform consistency of the estimators of cumulative rates and occupation probabilities obtained from the forest, together with a functional Gaussian limit for the estimator of cumulative rates. In the Markov case, a consistent variance estimator yields pointwise confidence intervals and simultaneous confidence bands. The results account for both sampling variation and the additional randomness from subsampling and aggregation.
    \item[(iii)] \emph{Empirical results.} The numerical studies assess prediction under a threshold interaction, irrelevant covariates and non-proportional hazards, together with four biomedical applications. They demonstrate adaptation to covariate structure and time-varying effects that the fitted Cox models cannot represent. Complementary studies examine correctly specified proportional hazards, variable importance, and the benefits and limitations of pooling for interval coverage.
    \item[(iv)] \emph{Software.} All methods are implemented in the R package \texttt{JumpForests}.
\end{enumerate}
Theoretical guarantees for adaptive random forests remain challenging even in regression. Consistency and asymptotic normality have been established for non-adaptive or simplified splitting schemes \citep{Biau2012, ScornetAsymp, Klusowski}, for honest forests that separate the observations used for splitting from those used for estimation \citep{WagerAthey}, through the theory of generalised U-statistics \citep{PengColemanMentch}, or, as in the consistency proof for random survival forests of \citet{RSFcon}, under the assumption that the covariates take finitely many values. We assume that the covariate vector takes finitely many values and adapt the assumptions of \citet{RSFcon} to general multi-state models. With finitely many covariate values and a splitting scheme satisfying Assumption~\ref{asm:splitting}, terminal nodes eventually separate those values, reducing the analysis to stratified Nelson--Aalen estimation and its aggregation over subsamples. The theory does not describe how forests borrow strength across neighbouring covariate values, which is central to their use with continuous covariates. The numerical studies also consider continuous covariates. The supplementary coverage study examines the intervals under the practical fitting procedure with discrete covariates.

The paper is organised as follows. Section~\ref{sec:preliminaries} introduces the multi-state setup and the target quantities. Section~\ref{sec:rf} presents the algorithm, its estimators and the default splitting rule. Section~\ref{sec:theory} develops the asymptotic theory and inference, Section~\ref{sec:simulations} the simulation studies and biomedical applications, and Section~\ref{sec:discussion} a discussion of limitations and recommendations. The supplement contains additional methodological components, supporting asymptotic results, all proofs and further numerical studies, including the coverage assessment.

%%%%%%%%%%%%%%%%%%%%%%%%%%%%%

\section{Preliminaries}\label{sec:preliminaries}
\emph{Notation.} Throughout, $s \land t = \min(s, t)$, $1_{A}$ denotes the indicator of an event $A$, and $f(t-)$ the left limit at $t$ of a càdlàg function $f$. Estimators are written in blackboard bold, so that $\mathbb{\Lambda}$ estimates $\Lambda$ and $\mathbb{p}$ estimates $p$. The data consist of independent and identically distributed (iid) copies of a generic observation, indexed by a superscript $i = 1, \dots, n$, and all limits are taken as $n \to \infty$ unless otherwise stated.

Let $Z = (Z_t)_{t \geq 0}$ be a pure-jump process on the finite state space $\CZ$, defined on a probability space $(\Omega, \CF, \BP)$. We assume that $\CZ \subseteq \BR$ and let $\tau_Z$ denote the (possibly infinite) absorption time of $Z$. To $Z$ is associated a multivariate counting process $N$ with components $N_{jk} = (N_{jk}(t))_{t \geq 0}$ given by
\begin{equation*}
    N_{jk}(t) = \#\{s \in (0, t]: Z_{s-} = j, Z_s = k\}, \quad t \geq 0
\end{equation*}
for $j, k \in \CZ$ and $j \neq k$. We assume that
\begin{equation*}
    \BE[N_{jk}(t)] < \infty
\end{equation*}
for all $t \geq 0$ and $j \neq k$ in $\CZ$ which also guarantees that $Z$ is non-explosive. Let $X$ be a covariate vector with values in $\BR^d$. Define \emph{conditional occupation probabilities}
\begin{equation*}
    p_j(t \mid x) = \BP(Z_t = j \mid X = x)
\end{equation*}
and let $p(t \mid x)$ denote the row vector of these probabilities. We define auxiliary quantities
\begin{equation*}
    p_{jk}(t \mid x) = \BE[N_{jk}(t) \mid X = x]
\end{equation*}
along with \emph{conditional transition rates}
\begin{equation*}
    \Lambda_{jk}(t \mid x) = \int_0^t \frac{1}{p_j(s- \mid x)}p_{jk}(\dd s \mid x), \quad \Lambda_{jj}(t \mid x) = -\sum_{k: k \neq j}\Lambda_{jk}(t \mid x).
\end{equation*}
Whenever we write $\Lambda_{jk}(t \mid x)$, we implicitly assume that $\Lambda_{jk}(t \mid x) < \infty$. The identity
\begin{equation*}
    p(t \mid x) = p(0 \mid x) + \int_0^t p(s- \mid x)\Lambda(\dd s \mid x)
\end{equation*}
can be verified for any dependence structure of $Z$. From this relation, it follows from the \emph{inhomogeneous equation} for the product integral, see Theorem 10 in~\citet{GillJohansen}, that
\begin{equation}\label{eq:occprob}
    p(t \mid x) = p(0 \mid x)\pint\limits_0^t(\text{Id} + \Lambda(\dd s \mid x)).
\end{equation}
This recovers the occupation probabilities from the initial distribution and the cumulative transition rates. If $Z$ is Markov conditionally on $X$, the relation \eqref{eq:occprob} also holds for transition probabilities. To be precise, if we have $t \geq s$ and let
\begin{equation*}
    p(s, t \mid x) = \{p_{jk}(s, t \mid x)\}_{j, k \in \CZ}, \quad p_{jk}(s, t \mid x) = \BP(Z_t = k \mid Z_s = j, X = x)
\end{equation*}
denote the matrix of transition probabilities, we have the relation
\begin{equation}\label{eq:Kolmogorov}
    p(s, t \mid x) = \pint\limits_s^t (\text{Id} + \Lambda(\dd v \mid x)).
\end{equation}
The case of \emph{smooth} Markov models where $Z$ allows for transition intensities $\mu_{jk}(\cdot \mid x)$ is of particular practical interest. In that case, 
\begin{equation*}
    \Lambda_{jk}(t \mid x) = \int_0^t \mu_{jk}(s \mid x) \dd s,
\end{equation*}
and \eqref{eq:Kolmogorov} becomes the classical \emph{Kolmogorov differential equations}. Whenever $Z$ is Markov conditionally on $X$, smooth or not, $t \mapsto 1_{(Z_{t-} = j)}\Lambda_{jk}(\dd t \mid X)$ is the compensator of $N_{jk}$ with respect to the natural filtration of $(X, Z)$. When $Z$ is not Markov, \eqref{eq:Kolmogorov} no longer applies, and transition probabilities need to be estimated via conditioning on internal covariates. In this case, the transition rates $\Lambda_{jk}$ have no interpretation as compensators for the counting processes $N_{jk}$. They are population rates instead: $\Lambda_{jk}(\dd t \mid x)$ is the expected number of transitions from $j$ to $k$ during $[t, t + \dd t)$ among the individuals with covariate value $x$, divided by the fraction of these individuals currently occupying state $j$. These population rates are the targets of Nelson--Aalen estimation even when $Z$ is not Markov, and \eqref{eq:occprob} shows that they determine the occupation probabilities in either case; see \citet{DattaSatten} and \citet{Overgaard} for the corresponding results on the Aalen--Johansen estimator in non-Markov models.

It remains to consider right-censoring, which is modelled by a random variable $R$ with $\BP(R > 0) = 1$. We observe the covariates $X$ and the path of $Z$ on $[0, R]$, hence also the time $\tau_Z \land R$ and the indicator $1_{(\tau_Z \leq R)}$ of absorption before censoring. We write the observed data as
\begin{equation*}
    (X, (Z_t)_{0 \leq t \leq R}, \tau_Z \land R).
\end{equation*}
To handle censoring, introduce the quantities
\begin{align*}
    p_j^c(t \mid x) &= \BP(Z_t = j, t < R  \mid X = x), \\
    p_{jk}^c(t \mid x) &= \BE[N_{jk}(t \land R) \mid X = x].
\end{align*}
We impose the following censoring assumption to identify the transition rates from censored observations in non-Markov models.

\begin{assumption}\label{asm:rrc}
Right-censoring is conditionally entirely random,
\begin{equation*}
    Z \indep R \mid X.
\end{equation*}
\end{assumption}

The assumption allows censoring to depend on the covariates but not to carry further information about the trajectory. In a Markov model an unconditional independent censoring condition of \citet{ABGK} suffices. Instead, the occupation probabilities above can only be recovered from censored data if censored and uncensored individuals with the same covariate information have the same distribution. Under this assumption,
\begin{equation*}
    \Lambda_{jk}(t \mid x) = \int_0^t \frac{1}{p_j^c(s- \mid x)}p_{jk}^c(\dd s \mid x).
\end{equation*}
Since $p_j^c(t \mid x)$ and $p_{jk}^c(t \mid x)$ can be estimated from right-censored data, so can $\Lambda_{jk}(t \mid x)$ under Assumption~\ref{asm:rrc}.
 
\section{The jump forest}\label{sec:rf}

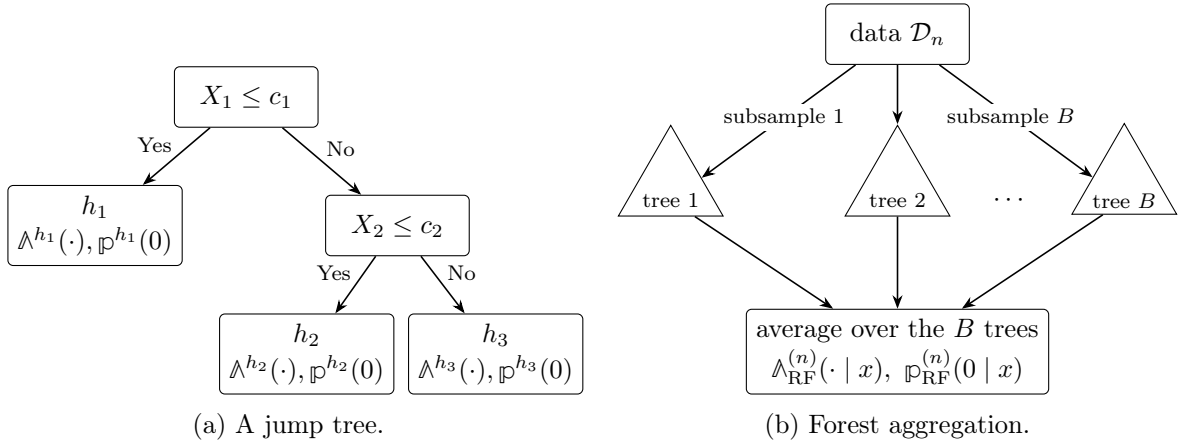
\begin{figure}[tbp]
    \centering
    \begin{subfigure}[b]{0.48\textwidth}
        \centering
        \begin{tikzpicture}[
          split/.style={rectangle, draw, rounded corners=2pt, align=center,
            minimum width=1.9cm, minimum height=0.8cm, font=\small},
          leaf/.style={rectangle, draw, rounded corners=2pt, align=center,
            minimum width=2.3cm, minimum height=0.95cm, font=\small},
          arr/.style={-{Stealth[length=1.8mm]}, semithick},
          lab/.style={font=\scriptsize, inner sep=1.5pt}
        ]
          \node[split] (root) at (0,0) {$X_1 \le c_1$};
          \node[leaf]  (r1) at (-2.0,-1.7) {$h_1$\\[0.5mm]$\mathbb{\Lambda}^{h_1}(\cdot), \mathbb{p}^{h_1}(0)$};
          \node[split] (s2) at (2.0,-1.7) {$X_2 \le c_2$};
          \node[leaf]  (r2) at (0.8,-3.4) {$h_2$\\[0.5mm]$\mathbb{\Lambda}^{h_2}(\cdot), \mathbb{p}^{h_2}(0)$};
          \node[leaf]  (r3) at (3.3,-3.4) {$h_3$\\[0.5mm]$\mathbb{\Lambda}^{h_3}(\cdot), \mathbb{p}^{h_3}(0)$};
          \draw[arr] (root) -- node[lab, midway, above left] {Yes} (r1);
          \draw[arr] (root) -- node[lab, midway, above right] {No} (s2);
          \draw[arr] (s2) -- node[lab, midway, above left] {Yes} (r2);
          \draw[arr] (s2) -- node[lab, midway, above right] {No} (r3);
        \end{tikzpicture}
        \caption{A jump tree.}
        \label{fig:jumptree-a}
    \end{subfigure}
    \hfill
    \begin{subfigure}[b]{0.48\textwidth}
        \centering
        \begin{tikzpicture}[
          box/.style={rectangle, draw, rounded corners=2pt, align=center,
            minimum width=1.9cm, minimum height=0.8cm, font=\small},
          tree/.style={isosceles triangle, isosceles triangle apex angle=60,
            shape border rotate=90, draw, minimum height=1.2cm, inner sep=1pt,
            font=\scriptsize},
          arr/.style={-{Stealth[length=1.8mm]}, semithick},
          lab/.style={font=\scriptsize, inner sep=1.5pt, fill=white}
        ]
          \node[box] (data) at (0,0) {data $\CD_n$};
          \node[tree] (t1) at (-3.0,-2.2) {tree 1};
          \node[tree] (t2) at (0,-2.2) {tree 2};
          \node[font=\small] at (1.5,-2.2) {$\cdots$};
          \node[tree] (tB) at (3.0,-2.2) {tree $B$};
          \node[box] (avg) at (0,-4.2) {average over the $B$ trees\\[0.5mm]$\mathbb{\Lambda}^{(n)}_{\operatorname{RF}}(\cdot \mid x),\ \mathbb{p}^{(n)}_{\operatorname{RF}}(0 \mid x)$};
          \draw[arr] (data) -- (t1);
          \draw[arr] (data) -- (t2);
          \draw[arr] (data) -- (tB);
          \node[lab] at (-1.5,-1.1) {subsample 1};
          \node[lab] at (1.5,-1.1) {subsample $B$};
          \draw[arr] (t1) -- (avg);
          \draw[arr] (t2) -- (avg);
          \draw[arr] (tB) -- (avg);
        \end{tikzpicture}
        \caption{Forest aggregation.}
        \label{fig:jumptree-b}
    \end{subfigure}
    \caption{Jump-tree construction and forest aggregation. (a) Splits on $X_1$ and $X_2$ produce three terminal nodes. Each node $h$ provides Nelson--Aalen estimates $\mathbb{\Lambda}^h$ of the cumulative transition rates and an empirical initial distribution $\mathbb{p}^h(0)$. (b) For a covariate value $x$, these quantities are averaged over the corresponding terminal nodes of $B$ trees grown on independent subsamples.}
    \label{fig:jumptree}
\end{figure}

The jump forest extends the random survival forest of \citet{RSF} to multi-state data. A single forest is grown for all transitions. In each terminal node, cumulative transition rates are estimated nonparametrically from the observed event histories. Algorithm~\ref{alg:jf} summarises the procedure and Figure~\ref{fig:jumptree} illustrates it.

\begin{algorithm}[tbp]
\caption{The random jump forest}\label{alg:jf}
\begin{algorithmic}[1]
\Require Data $\CD_n$ of $n$ right-censored trajectories with covariates; number of trees $B$; subsample size $m$; number $q$ of candidate covariates per split; minimum numbers $d_j \geq 1$ of observed jumps out of each non-absorbing state $j$; a splitting statistic $L$ (see subsection~\ref{subsection:splittingrules}).
\For{$b = 1, \dots, B$}
    \State Draw a subsample of size $m$ from $\CD_n$, with or without replacement.
    \State Grow a tree on the subsample by treating each node $h$, starting from the root, as follows:
    \State \hspace{\algorithmicindent} draw $q$ candidate covariates at random;
    \State \hspace{\algorithmicindent} among all splits $(f, c)$ of $h$ on a candidate covariate $f$ specified by a threshold or subset of levels $c$ that leave at least $d_j$ observed jumps out of every non-absorbing state $j$ in both children, choose the one maximising $|L(f, c)|$;
    \State \hspace{\algorithmicindent} if no such split exists, declare $h$ terminal and compute the Nelson--Aalen estimators $\mathbb{\Lambda}^h$ of all transition rates and the empirical initial distribution $\mathbb{p}^h(0)$ computed from the observations in $h$ (as per subsection~\ref{subsection:estimators}).
\EndFor
\Ensure For a covariate value $x$, the forest predictions $\mathbb{\Lambda}_{\operatorname{RF}}^{(n)}(\cdot \mid x)$ and $\mathbb{p}^{(n)}_{\operatorname{RF}}(0 \mid x)$ are the averages over the $B$ trees of the estimators stored in the terminal node containing $x$; occupation probabilities follow from formula \eqref{eq:occprob}.
\end{algorithmic}
\end{algorithm}

The tuning parameters are the number of trees $B$, the subsample size $m$, the number $q$ of candidate covariates per split and the minimum numbers $d_j$ of observed jumps in a terminal node. Trees are grown as deep as the constraint on observed jumps allows; the implementation instead uses a minimum number of observations per node, as specified in Subsection~\ref{subsection:design}. These are distinct stopping constraints. We require from the splitting rule that a node can split whenever an admissible split exists, as is formalised in Assumption~\ref{asm:splitting}. A split on a continuous covariate $f$ is described by a threshold $c$, where observations with $X_f^i \leq c$ go to the left child and the remaining ones to the right child. For a categorical covariate, a split is a partition of its levels into two subsets, one of which defines the left child. We use the term bootstrap sample for a subsample drawn with or without replacement. We next describe the node estimators, predictions and default splitting rule.

\subsection{Estimators and prediction}\label{subsection:estimators}
Suppose that a terminal node $h$ contains the $n_h$ observations
\begin{equation*}
    (X^i, (Z_t^i)_{0 \leq t \leq R^i}, \tau_{Z^i}\land R^i)_{i \in h}.
\end{equation*}

The estimator of $p_j^c(s \mid x)$ in $h$ is
\begin{equation*}
    \BI_j^h(s) = \frac{1}{n_h}\sum_{i \in h}1_{(Z_s^i = j)} 1_{(s < R^i)},
\end{equation*}
and analogously, the estimator of $p_{jk}^c(s \mid x)$ in $h$ is
\begin{equation*}
    \BN_{jk}^h(s) = \frac{1}{n_h}\sum_{i \in h} N^i_{jk}(s \land R^i).
\end{equation*}
With the convention $0/0 = 0$ at times with no observations at risk, these yield the \emph{Nelson--Aalen} estimator of $\Lambda_{jk}$ in terminal node $h$,
\begin{equation*}
    \mathbb{\Lambda}_{jk}^{h}(t) = \int_0^t \frac{1}{\BI_j^h(s-)}\BN_{jk}^h(\dd s), \quad j \neq k,
\end{equation*}
and the initial probability vector $p(0 \mid x)$ is estimated in a terminal node $h$ by
\begin{equation*}
    \mathbb{p}^h(0) = (\mathbb{p}_j^h(0))_{j \in \CZ}, \quad \mathbb{p}_j^h(0) = \frac{1}{n_h}\sum_{i \in h} 1_{(Z_0^i = j)}.
\end{equation*}
Note that $x$ does not appear in these expressions since the estimator is the same for all observations in a node; conditioning has nonetheless happened by growing the tree. For the case $j = k$, we define $\mathbb{\Lambda}_{jj}^h(t) = -\sum_{k:k \neq j}\mathbb{\Lambda}_{jk}^h(t)$. All these quantities are represented together in the matrix $\mathbb{\Lambda}^h(t)$. For a covariate value $x$, the predicted rate matrix $\mathbb{\Lambda}^{(n)}(t \mid x)$ of a single tree is obtained by dropping $x$ down the tree, namely $\mathbb{\Lambda}^{(n)}(t \mid x) = \mathbb{\Lambda}^h(t)$ when $x$ belongs in $h$ (and similarly, $\mathbb{p}^{(n)}(0 \mid x) = \mathbb{p}^{h}(0)$). For a forest of $B$ trees, let $\mathbb{\Lambda}_b^{(n),*}$ and $\mathbb{p}_b^{(n),*}$ denote the predictions from the tree grown on the $b$th bootstrap sample. The forest predictions are the averages
\begin{equation*}
    \mathbb{\Lambda}_{\operatorname{RF}}^{(n)}(t \mid x) = \frac{1}{B}\sum_{b = 1}^B \mathbb{\Lambda}_b^{(n),*}(t \mid x), \quad \mathbb{p}_{\operatorname{RF}}^{(n)}(0 \mid x) = \frac{1}{B}\sum_{b = 1}^B \mathbb{p}_b^{(n),*}(0 \mid x).
\end{equation*}
These predictions are defined for any covariate value $x$ and are used for new data. For error estimation on the training data, one instead uses the \emph{out-of-bag} (OOB) prediction of an observation $i$, which averages only over the trees whose bootstrap sample does not contain $i$; we say that $i$ is out-of-bag for such a tree. Letting $I_{i, b}$ be equal to one precisely when $i$ is OOB for bootstrap data set $b$, the OOB estimate of $\Lambda(t \mid X^i)$ for observation $i$ is given by
\begin{equation*}
    \mathbb{\Lambda}^{(n), \text{OOB}}_{\operatorname{RF}}(t \mid X^i) = \frac{\sum_{b = 1}^B I_{i,b} \mathbb{\Lambda}_b^{(n),*}(t \mid X^i)}{\sum_{b = 1}^B I_{i,b}},
\end{equation*}
and OOB estimates of the initial probabilities are computed analogously. The OOB error estimation is a built-in analogue of cross-validation, since the OOB errors can be computed while the forest is fitted, without requiring refitting on validation subsets \citep{BreimanOOB}. 

\subsection{Log-rank splitting}\label{subsection:splittingrules}

The default splitting rule sums transition-specific log-rank statistics comparing the child nodes induced by a split. We refer to Chapter V.2 of \citet{ABGK} for further background on these tests. Fix a node $h$ and a split $(f, c)$ on a covariate; we exemplify with notation for continuous covariates, with discrete, categorical or mixed cases being analogous. Let $0 < t_1 < \cdots < t_N$ denote the distinct times at which uncensored transitions occur in $h$. For a transition from $j$ to $k$ and $m = 1, \dots, N$, the numbers at risk in state $j$ just before $t_m$ in the left and right child are
\begin{equation*}
    Y_{m, l}^j := \sum_{i \in h} 1_{(Z_{t_m-}^i = j)}1_{(t_m \leq R^i)}1_{(X^i_f \leq c)}, \quad Y_{m, r}^j := \sum_{i \in h} 1_{(Z_{t_m-}^i = j)}1_{(t_m \leq R^i)}1_{(X^i_f > c)},
\end{equation*}
with total $Y_m^j = Y_{m, l}^j + Y_{m, r}^j = n_h \BI^h_j(t_m-)$, and the numbers of observed transitions from $j$ to $k$ at $t_m$ are
\begin{equation*}
    d_{m, l}^{jk} := \sum_{i \in h} 1_{(t_m \leq R^i)}\Delta N_{jk}^i(t_m) 1_{(X^i_f \leq c)}, \quad d_{m, r}^{jk} := \sum_{i \in h} 1_{(t_m \leq R^i)}\Delta N^i_{jk}(t_m) 1_{(X^i_f > c)},
\end{equation*}
with total $d_m^{jk} = d_{m, l}^{jk} + d_{m, r}^{jk} = n_h \Delta \BN_{jk}^h(t_m)$. The log-rank statistic of the split is
\begin{equation}\label{eq:logrank}
    L(f, c) = \sum_{\substack{j, k \in \CZ \\ j \neq k}} \frac{\sum_{m = 1}^N \Big(d_{m,l}^{jk} - Y_{m,l}^j \frac{d_m^{jk}}{Y_m^j} \Big)}{\sqrt{\sum_{m = 1}^N \frac{Y_{m,l}^j}{Y_m^j} \Big(1 - \frac{Y_{m,l}^j}{Y_m^j} \Big)\frac{Y_m^j - d_m^{jk}}{Y_m^j - 1}\, d_m^{jk}}},
\end{equation}
and the split maximising $|L(f, c)|$ over the candidate covariates $f$ and thresholds $c$ is selected. Terms with no variance contribution are assigned zero. Each remaining summand is the standardised log-rank statistic for the transition $j$ to $k$, with the associated hypergeometric variance in the denominator. For a survival model with a single transition, \eqref{eq:logrank} reduces to the log-rank splitting rule of \citet{RSFR} and \citet{RFCompetingRisks}.

Summing the signed standardised statistics over transitions rewards splits that separate the two child nodes consistently across transitions. When a covariate changes different transition rates in opposite directions, the transition-specific statistics may partially cancel; a sum of squared statistics would be a natural alternative in such situations; we do not pursue the idea further.

\subsection{Prediction assessment, variable importance and extensions}\label{subsection:assessmentoverview}
Prediction assessment and covariate importance complement the rate and occupation-probability estimators developed above. Supplementary Section~\ref{sec:methodcomponents} provides their full construction. Censoring-adjusted Brier and logarithmic scores assess occupation-probability predictions, and their integrated versions summarise performance over time. Out-of-bag evaluation provides a criterion for tuning without reserving a separate validation sample. Permutation importance measures the contribution of each covariate to occupation-probability prediction and is available both on held-out data and internally through out-of-bag observations. Weighted and approximate alternatives to the log-rank splitting criterion are also developed. These components form part of the implemented jump forest procedure.

\section{Asymptotic theory and inference}\label{sec:theory}

We establish uniform consistency and a Gaussian limit for the forest, then derive confidence intervals and simultaneous bands in the Markov case. Supplementary Section~\ref{sec:additionalasymptotics} develops the supporting single-tree and conditional bootstrap results, the general Markov variance formula and additional transition-probability results. Proofs of the three main results are collected in Supplementary Section~\ref{sec:mainproofs}.

\subsection{Assumptions and scope}\label{subsection:assumptions}

Together with Assumption~\ref{asm:rrc} on censoring, the assumptions below are the standing assumptions for the asymptotic results. We assume that the covariate vector takes finitely many values, as in the survival setting of \citet{RSFcon}. Under the splitting condition below, terminal nodes eventually separate these values, reducing the asymptotic analysis to stratified estimation. The argument applies to the log-rank splitting statistic and its weighted and approximate variants, subject to Assumption~\ref{asm:splitting}.

\begin{assumption}\label{asm:featurespace}
The image of the covariate vector $X$ is contained in a subset $\CX \subseteq \BR^d$ such that $\CX$ is finite and discrete. Furthermore, $\BP(X = x) > 0$ for every $x \in \CX$.
\end{assumption}

Under Assumptions~\ref{asm:featurespace}--\ref{asm:splitting}, every terminal node of a tree eventually contains a single covariate value (see the proof of Theorem~\ref{thm:constree}), so that the tree prediction for $x$ coincides, for all large $n$, with the Nelson--Aalen estimator computed from the observations with $X^i = x$. The results below therefore establish that trees and forests converge to the fully stratified Nelson--Aalen estimator, and that the aggregation over bootstrap samples preserves its Gaussian limit, and also its $\sqrt{n}$ rate when the subsample size is proportional to $n$.

In order to state our results without unnecessary restrictions on the transition rates, we introduce the following notation. Let
\begin{equation*}
    T(x) = \sup\{t \geq 0: \Lambda_{jk}(t \mid x) < \infty \text{ for all } j \neq k\}
\end{equation*}
denote the largest time horizon on which all transition rates for the covariate value $x$ are finite, and let $T_{R,x} = \sup\{t \geq 0: F_R(t \mid x) < 1\}$ denote the right endpoint of the conditional distribution of the censoring variable $R$ given $X = x$, that is, the largest horizon over which individuals with covariate value $x$ remain under observation with positive probability. We let $T = \min_{x \in \CX}T(x)$ and $T_R = \min_{x \in \CX} T_{R,x}$; Assumption~\ref{asm:featurespace} ensures that these minima make sense. Some regularity of $Z$ is also needed. For a fixed state $j$, consider the total number of observed jumps out of $j$,
\begin{equation*}
    N_j^\bullet = \sum_{k:k \neq j} N_{jk}(R).
\end{equation*}
We impose the following condition.

\begin{assumption}\label{asm:jumpsout}
It holds that $\BP(N_j^\bullet > 0 \mid X = x) > 0$ for every $x \in \CX$ and every non-absorbing state $j$.
\end{assumption}

A sufficient condition for Assumption~\ref{asm:jumpsout} is given in Proposition~\ref{prop:jumpsout}.

The asymptotic results also require that the tree-growing algorithm keeps splitting a node as long as it contains several distinct covariate values. With $q < d$ randomly drawn candidate covariates, or with randomly drawn candidate thresholds, an implementation may declare a node terminal although a covariate separating its covariate values was not examined. We therefore impose the following condition on the splitting scheme.

\begin{assumption}\label{asm:splitting}
A node containing at least two distinct covariate values is split whenever an admissible split exists.
\end{assumption}

It is sufficient that every threshold is a candidate split point and that candidate covariates are drawn until a covariate that varies within the node has been examined. This holds in particular when all $d$ covariates are candidates at every split.
The condition is vacuous at a node admitting no such split, so it does not conflict with a minimum node size in finite samples; it specifies a behaviour as $n$ grows, when every value in $\CX$ eventually has enough observed jumps to be split.

Finally, asymptotic results for the Nelson--Aalen estimator require the denominator $p_j^c(s- \mid x)$ not to vanish on the time interval under consideration.

\begin{assumption}[Non-vanishing occupation probabilities]\label{asm:nonvanishing}
For the covariate value $x$ and the time horizon $t$ under consideration, $\inf_{s \in [0, t]}p_j(s \mid x) > 0$ for all non-absorbing states $j$.
\end{assumption}
The above assumption requires every non-absorbing state to be occupied with positive probability throughout $[0, t]$, including at time zero. It excludes, for instance, an illness--death model in which all individuals start in the healthy state. The condition is needed for the continuity of the Nelson--Aalen functional at the limit; see the proof of Theorem~\ref{thm:constree}. It can be relaxed by fixing $t_0 \in (0, t)$ with $\inf_{s \in [t_0, t]}p_j(s \mid x) > 0$ for all non-absorbing $j$. All limit theorems then hold verbatim for the increments $\Lambda_{jk}(\cdot \mid x) - \Lambda_{jk}(t_0 \mid x)$ on $[t_0, t]$ and their estimators, with the same proofs, and occupation probabilities on $[t_0, t]$ may be recovered from \eqref{eq:occprob} started at $t_0$, with $p(t_0 \mid x)$ estimated again by the empirical frequencies.

\emph{Standing assumptions.} Throughout Section~\ref{sec:theory}, Assumptions~\ref{asm:rrc} to~\ref{asm:nonvanishing} are taken as valid, $x \in \CX$ is a fixed covariate value and $t \in (0, T_{R,x} \land T(x))$ is a fixed time horizon; in particular, Assumption~\ref{asm:nonvanishing} is related to this $x$ and $t$. Beyond this fixed $x$, the presented results may extend to all $x \in \CX$ simultaneously, whenever $t \in (0, T_R \land T)$ and Assumption~\ref{asm:nonvanishing} holds for every $x \in \CX$; routine restatements are omitted. For a matrix $A = (a_{ml})_{ml}$, we let $|A| = \max_m \sum_l |a_{ml}|$ denote the maximum row-sum norm.

\subsection{Uniform consistency}\label{subsection:consistency}
The forest estimates both cumulative rates and occupation probabilities consistently. The proof proceeds through consistency of single trees and bootstrap trees; the occupation probabilities are obtained from the forest-averaged rates and initial distribution through \eqref{eq:occprob}.

\begin{theorem}[Consistency of forests]\label{cor:forestconsistency}
Suppose that $\BE[N_{jk}(t)^2] < \infty$ for all $j \neq k$. Consider a forest of a fixed number $B$ of trees, each grown on a bootstrap sample obtained by multinomial subsampling or subsampling without replacement. Let the subsample size satisfy $m_n\to\infty$. Then
\begin{equation*}
    \sup_{s \in [0, t]}|\mathbb{\Lambda}_{\operatorname{RF}}^{(n)}(s \mid x) - \Lambda(s \mid x)| \overset{\BP}{\longrightarrow} 0
    \quad \text{and} \quad
    \sup_{s \in [0, t]}|\mathbb{p}^{(n)}_{\operatorname{RF}}(s \mid x) - p(s \mid x)| \overset{\BP}{\longrightarrow} 0.
\end{equation*}
If, in addition, $\BE[N_{jk}(t)^4] < \infty$ for all $j \neq k$ and the subsample size satisfies $\sum_{n = 1}^\infty m_n^{-2} < \infty$, both convergence statements hold almost surely.
\end{theorem}

\subsection{Gaussian limits for the forest}\label{subsection:normality}

All process limits below are in the uniform norm on $[0,t]$. Write $p_x=\BP(X=x)$ and let $\zeta_{jk}^x$ denote the tight mean-zero Gaussian limit of the single-tree estimation error scaled by the square root of its terminal-node sample size. Its representation and the supporting conditional bootstrap limits are given in Supplementary Section~\ref{subsection:additionallimits}.

The limit is centred at the rate for a fixed covariate value $x$. In finite samples, a terminal node may pool values with different rates, introducing bias not represented by this Gaussian approximation.

\begin{theorem}[Unconditional weak convergence of forests]\label{thm:weakconvergencebootstrapunconditional}
Suppose that $\BE[N_{jk}(t)^2] < \infty$ for all $j \neq k$. Consider a forest of a fixed number $B$ of trees, each grown on an independent bootstrap sample of size $m_n\to\infty$. Assume either multinomial subsampling with $m_n/n\to\alpha\in[0,1]$, or subsampling without replacement with $m_n/n\to\alpha\in[0,1)$. Define
\begin{equation}\label{eq:kappa}
    \kappa_n := \begin{cases} \dfrac{1}{n} + \dfrac{1}{B m_n} & \text{for multinomial subsampling}, \\[2ex] \dfrac{1}{n} + \dfrac{1-m_n/n}{B m_n} & \text{for subsampling without replacement}. \end{cases}
\end{equation}
Then, for each $j \neq k$,
\begin{equation*}
    \kappa_n^{-1/2}\big(\mathbb{\Lambda}_{\operatorname{RF}, jk}^{(n)}(\cdot \mid x) - \Lambda_{jk}(\cdot \mid x) \big) \overset{d}{\longrightarrow} \frac{1}{\sqrt{p_x}}\zeta_{jk}^x.
\end{equation*}
\end{theorem}

The factor $\kappa_n$ separates the common full-data contribution $1/n$ from the Monte Carlo contribution, which is proportional to $1/B$. Conditional on the data, averaging independent trees reduces Monte Carlo fluctuation, but does not remove the full-data estimation error. Sampling without replacement reduces the Monte Carlo contribution by the finite-population factor $1-m_n/n$. This is a first-order covariance decomposition for fixed $B$. For a single tree sampled without replacement, $\kappa_n=1/m_n$.

\subsection{Confidence intervals and simultaneous bands}\label{subsection:inference}

When $Z$ is Markov conditionally on $X$ and $\Lambda_{jk}(\cdot\mid x)$ is continuous, the Gaussian limit has covariance $\Cov(\zeta_{jk}^x(s_1),\zeta_{jk}^x(s_2))=V_{jk}(s_1\land s_2\mid x)$, where
\begin{equation*}
    V_{jk}(s\mid x)=\int_0^s\frac{1}{p_j^c(u-\mid x)}\Lambda_{jk}(\dd u\mid x).
\end{equation*}
For the $b$th tree, let $h_b(x)$ be the terminal node containing $x$, let $n_{h_b(x)}^*$ be its sample size, counting multiplicities, and let $\BI_j^{h_b(x),*}$ and $\BN_{jk}^{h_b(x),*}$ be its empirical at-risk proportion and mean observed transition count, respectively. Define
\begin{align*}
    \mathbb{V}_{jk, b}^{(n),*}(s \mid x) &:= \int_0^s \frac{1}{\BI_j^{h_b(x),*}(u-)^2}\BN_{jk}^{h_b(x),*}(\dd u), \\
    \tilde{\mathbb{V}}_{\operatorname{RF}, jk}^{(n)}(s \mid x) &:= \frac{1}{B}\sum_{b = 1}^B \frac{m_n}{n_{h_b(x)}^*} \mathbb{V}_{jk, b}^{(n),*}(s \mid x),
\end{align*}
with zero contribution at times when no observations are at risk. Let $z_{1-\delta/2}$ be the standard normal quantile and $c_{1-\delta}$ the $(1-\delta)$-quantile of $\sup_{0\leq u\leq1}|W(u)|$ for a standard Brownian motion $W$.

\begin{corollary}[Confidence intervals and simultaneous bands]\label{cor:forestCI}
Suppose that the assumptions of Theorem~\ref{thm:weakconvergencebootstrapunconditional} hold, that $Z$ is Markov conditionally on $X$, and that $\Lambda_{jk}(\cdot\mid x)$ is continuous on $[0,t]$. Then
\begin{equation*}
    \sup_{s\in[0,t]}\big|\tilde{\mathbb{V}}_{\operatorname{RF},jk}^{(n)}(s\mid x)-V_{jk}(s\mid x)/p_x\big|\overset{\BP}{\longrightarrow}0.
\end{equation*}
For every $\delta\in(0,1)$, the following coverage limits hold:
\begin{enumerate}
    \item[(i)] For every $s \in (0, t]$ with $V_{jk}(s \mid x) > 0$,
    \begin{equation*}
        \BP\Big(\big|\mathbb{\Lambda}_{\operatorname{RF}, jk}^{(n)}(s \mid x) - \Lambda_{jk}(s \mid x)\big| \leq z_{1 - \delta/2}\big(\kappa_n \tilde{\mathbb{V}}_{\operatorname{RF}, jk}^{(n)}(s \mid x)\big)^{1/2}\Big) \longrightarrow 1 - \delta.
    \end{equation*}
    \item[(ii)] If $V_{jk}(t \mid x) > 0$, then
    \begin{equation*}
        \BP\Big(\sup_{s \in [0, t]}\big|\mathbb{\Lambda}_{\operatorname{RF}, jk}^{(n)}(s \mid x) - \Lambda_{jk}(s \mid x)\big| \leq c_{1 - \delta}\big(\kappa_n \tilde{\mathbb{V}}_{\operatorname{RF}, jk}^{(n)}(t \mid x)\big)^{1/2}\Big) \longrightarrow 1 - \delta.
    \end{equation*}
\end{enumerate}
\end{corollary}

Part (i) yields pointwise intervals and part (ii) a constant-width band simultaneous over time for the fixed transition $(j,k)$ and covariate value $x$. Since $\Lambda_{jk}(\cdot \mid x) \geq 0$, replacing a negative lower limit by zero does not reduce the coverage. The band in part (ii) rests on the fact that, in the Markov case with continuous rates, $\zeta_{jk}^x$ is a Gaussian process with independent increments and variance function $V_{jk}(\cdot \mid x)$, that is, a time-changed Brownian motion. All quantities entering $\tilde{\mathbb{V}}_{\operatorname{RF}, jk}^{(n)}$ are computed in the terminal nodes when the Nelson--Aalen estimators are formed, so the intervals and bands require no additional model fitting or resampling.

The coverage study in Supplementary Section~\ref{subsection:app-coverage} examines these intervals and bands under the practical split search and prediction-oriented tuning. Forest bands attain coverage close to nominal in several settings, and pooling substantially improves coverage over fully stratified estimation for sparse transitions, although it does not always restore the nominal level. Conversely, pooling heterogeneous rates can introduce bias and undercoverage. Thus, improved precision from pooling does not by itself ensure nominal coverage at a fixed covariate value.

\section{Numerical studies}\label{sec:simulations}
We examine prediction under a threshold interaction, its robustness to irrelevant covariates, and prediction under non-proportional hazards, followed by four biomedical applications. Two complementary studies in the supplement assess the cost of flexibility and variable importance under correctly specified proportional hazards, and the finite-sample coverage of confidence intervals and simultaneous bands under practical fitting. Table~\ref{tab:simdesign} summarises all four simulation designs. All simulated processes are Markov conditionally on the covariates.

\subsection{Design and evaluation}\label{subsection:design}
All analyses use the \texttt{JumpForests} package described in Section~\ref{sec:software}. In the prediction studies, each jump forest (JF) has $B=500$ trees, $q=\lceil\sqrt d\,\rceil$ candidate covariates and $10$ randomly drawn thresholds per split, and uses subsamples without replacement of size $m=0.7n$. The log-rank rule is used with minimum node size $\lceil\sqrt n\,\rceil$, where $n$ is the number of training observations. Because candidate covariates and thresholds are sampled, the implementation need not satisfy Assumption~\ref{asm:splitting}.

Cox denotes transition-wise proportional hazards models with main effects, whose occupation probabilities are obtained by product integration of the Breslow cumulative-hazard estimators. Cox $X_1{:}X_2$ adds $X_1X_2$ to the linear predictor of every transition, and Cox interactions add all two- and three-way products of $X_1$, $X_2$ and $X_3$. Cox lasso uses an $L^1$ penalty chosen by five-fold cross-validation within each training set.

RSF is an ad-hoc method derived from the random survival forest fitted separately to each transition with the clock reset on entry into a new state (in terms of software, we use the survival implementation in \texttt{JumpForests}, with the same number of trees and minimum node size as JF; occupation probabilities are computed from the transition-specific cumulative hazards). AJ null is the Aalen--Johansen estimator without covariates.

Predictive accuracy is measured on held-out data by the normalised integrated Brier score (IBS), as formally defined in the Supplementary Section~\ref{subsection:error}, with uniform state weights $c_j=1/|\CZ|$ and horizon $\bar t$ equal to the largest observed event time in the training data. We report the paired relative difference $100(\mathrm{IBS}_{\mathrm{model}}-\mathrm{IBS}_{\mathrm{JF}})/\mathrm{IBS}_{\mathrm{JF}}$, so positive values favour JF. Each prediction simulation uses $100$ independently generated data sets split $80/20$ into training and test observations; the biomedical comparisons use $100$ such splits of each fixed data set. For inverse probability of censoring weighting (IPCW), the censoring distribution is estimated by a marginal Kaplan--Meier estimator across all models.

For the biomedical data and the study in Subsection~\ref{subsection:phinteraction}, all subjects start in state $1$, so the initialisation is the same across methods. In the Cox sample-size, noise and variable-importance studies, occupation-probability predictions for all methods, including JF, are initialised at the observed initial state of each subject to ensure comparability. The internal OOB importance of JF instead uses the estimated initial distribution.

\subsection{Proportional hazards with a threshold interaction}\label{subsection:phinteraction}
We first consider the three-state time-homogeneous Markov model with intensity matrix
\begin{equation*}
    Q(X_1, X_2, X_3) = c(X_1, X_2, X_3)\, Q_0, \qquad Q_0 = \begin{pmatrix} -0.40 & 0.25 & 0.15 \\ 0.20 & -0.45 & 0.25 \\ 0 & 0 & 0 \end{pmatrix},
\end{equation*}
with $c(x_1, x_2, x_3) := 1 + 3x_1 1_{(|x_2| \leq 1/3)}(1 + 2x_3)$,
where $X_1 \sim \text{Bern}(1/3)$, $X_2 \sim U(-1, 1)$ and $X_3 \sim U(0, 1)$ are independent. All individuals start in state $1$, and we impose no censoring. Every transition intensity is of the form $c(x)q_{jk}^0$, so that the ratio of the intensities for two covariate values does not depend on time and the hazards are proportional for every transition. The population consists of two groups: individuals with $X_1 = 1$ and $|X_2| \leq 1/3$, whose intensities are between four and ten times those of $Q_0$ depending on $X_3$, and the remaining individuals, whose intensities are those of $Q_0$. The covariates therefore act through a threshold interaction. A tree can represent the partition into the two groups exactly, through a split on $X_1$ and splits on $X_2$ at $\pm 1/3$, and approximates the linear dependence on $X_3$ within the high-intensity group by a step function, whereas none of the Cox models of Subsection~\ref{subsection:design} can represent the threshold.

\begin{figure}[tbp]
    \centering
    \includegraphics[width=\linewidth]{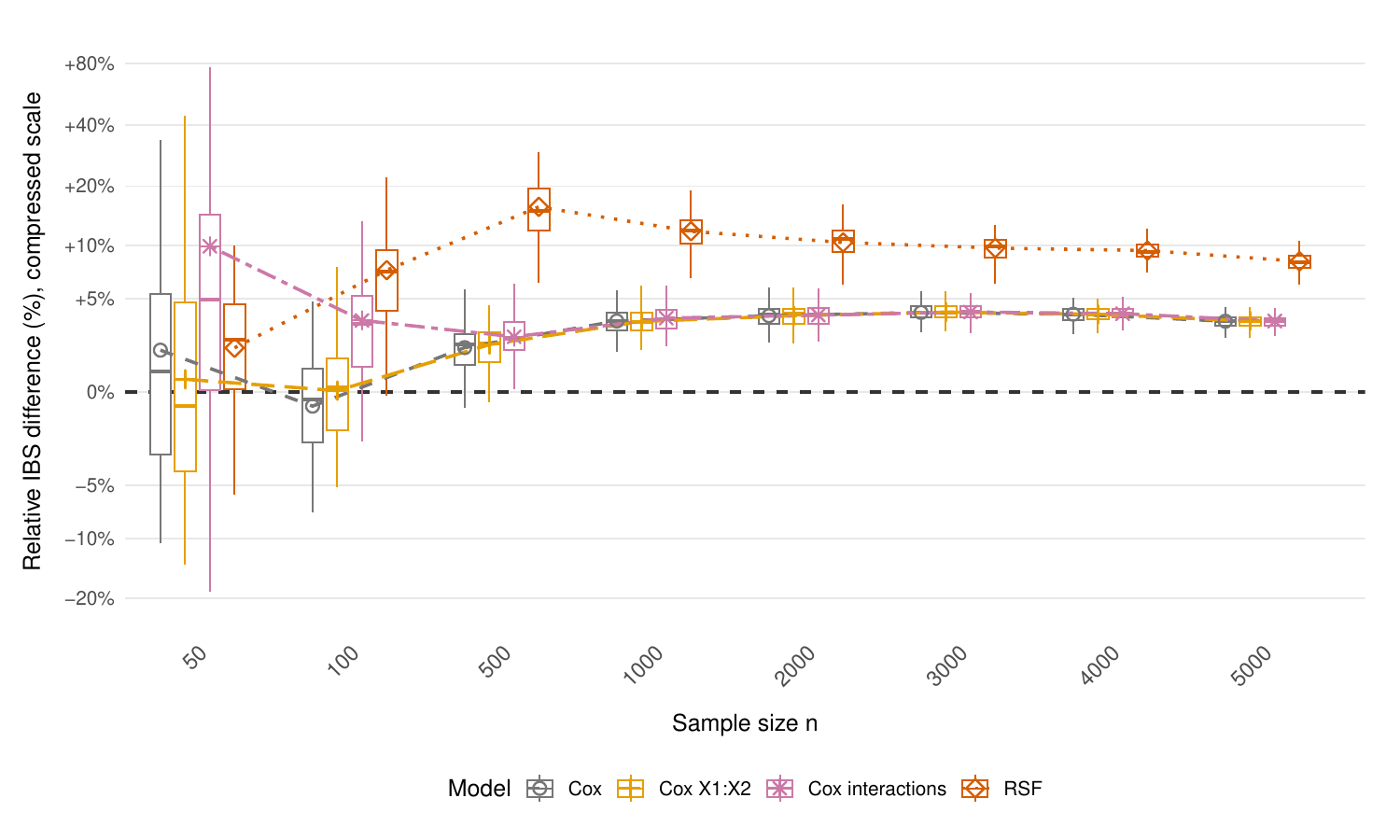}
    \caption{Threshold-interaction study, $100$ replications. Relative difference (percent) in test-set IBS of the transition-wise Cox models (Cox, Cox X1:X2 and Cox interactions) and random survival forest (RSF) with respect to the jump forest (JF), as a function of sample size $n$. The vertical axis uses the transformation $y\mapsto\operatorname{arsinh}(y/4)$. Positive values favour JF.}
    \label{fig:phint-relative}
\end{figure}

Figure~\ref{fig:phint-relative} compares the models for $n=50$, $100$, $500$, $1000$, $2000$, $3000$, $4000$ and $5000$. Adding products of the original covariates does not compensate for the missing threshold: at $n\geq1000$, the three Cox specifications have nearly identical median relative differences of $3.5$--$4.1$ percent in favour of JF. RSF has larger median relative differences, decreasing from $15.0$ percent at $n=500$ to $8.1$ percent at $n=5000$. This comparison concerns the complete prediction procedures; it does not isolate the contribution of sharing splits across transitions.

\begin{figure}[tbp]
    \centering
    \includegraphics[width=\linewidth]{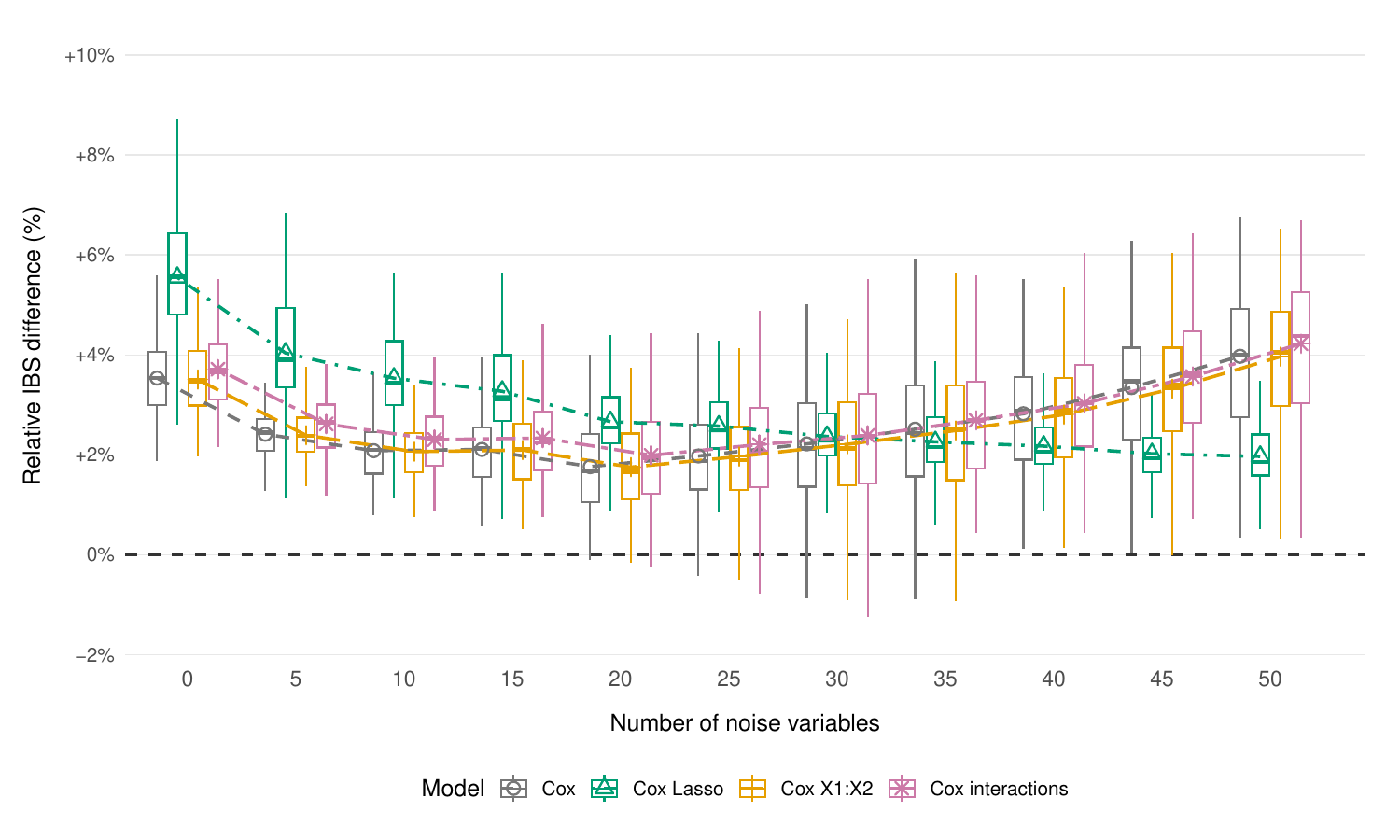}
    \caption{Threshold-interaction study, $n=1000$, $100$ replications. Relative difference (percent) in test-set IBS of the transition-wise Cox models (Cox, Cox X1:X2, Cox interactions and Cox Lasso) with respect to the jump forest (JF), as a function of the number of added noise covariates. Positive values favour JF.}
    \label{fig:phint-noise}
\end{figure}

To assess robustness to irrelevant predictors, we fix $n=1000$ and add $0,5,\dots,50$ noise covariates, drawn independently from the standard normal distribution. The advantage of JF under threshold misspecification persists across the noise levels considered, against every fitted Cox specification including lasso (Figure~\ref{fig:phint-noise}). Median relative differences range from $1.7$ to $5.6$ percent, and every lower quartile exceeds $1$ percent.

\subsection{Prediction under non-proportional hazards}\label{subsection:nonph}
We next examine time-varying covariate effects in a three-state Markov process with absorbing state $3$ and transition intensities
\begin{align*}
    \mu_{12}(t \mid X) &= \begin{cases} 0.52 + 0.50\, g(X_2) 1_{(t \geq 1.2)}, & X_1 = 1, \\ 0.02 + 3.00 \cdot 1_{(t \geq 3.3)}, & X_1 = 0, \end{cases} \\
    \mu_{23}(t \mid X) &= \begin{cases} 0.08 + 0.70 \cdot 1_{(t \geq 1.5)}, & X_3 = 1, \\ 0.45 - 0.40 \cdot 1_{(t \geq 2.3)}, & X_3 = 0, \end{cases} \\
    \mu_{13}(t \mid X) &= 0.02, \qquad \mu_{21}(t \mid X) = 0.06,
\end{align*}
where $g(1)=-1$, $g(2)=0$, $g(3)=1$ and $g(4)=4$. Independently, $X_1,X_3\sim\operatorname{Bern}(1/2)$ and $X_2$ is uniform on $\{1,2,3,4\}$. All individuals start in state $1$, and censoring is independent of $(X,Z)$ and uniform on $[5,25]$; JF and both Cox models use $X_1$, $X_2$ and $X_3$, without noise covariates. The design combines crossing hazards with a time-varying interaction: $X_2$ affects $1\to2$ only when $X_1=1$ and $t\geq1.2$, while the $2\to3$ hazards also cross. Hence no fitted Cox model with time-constant coefficients can represent the design, even with a coefficient for every covariate combination.

\begin{figure}[tbp]
    \centering
    \includegraphics[width=\linewidth]{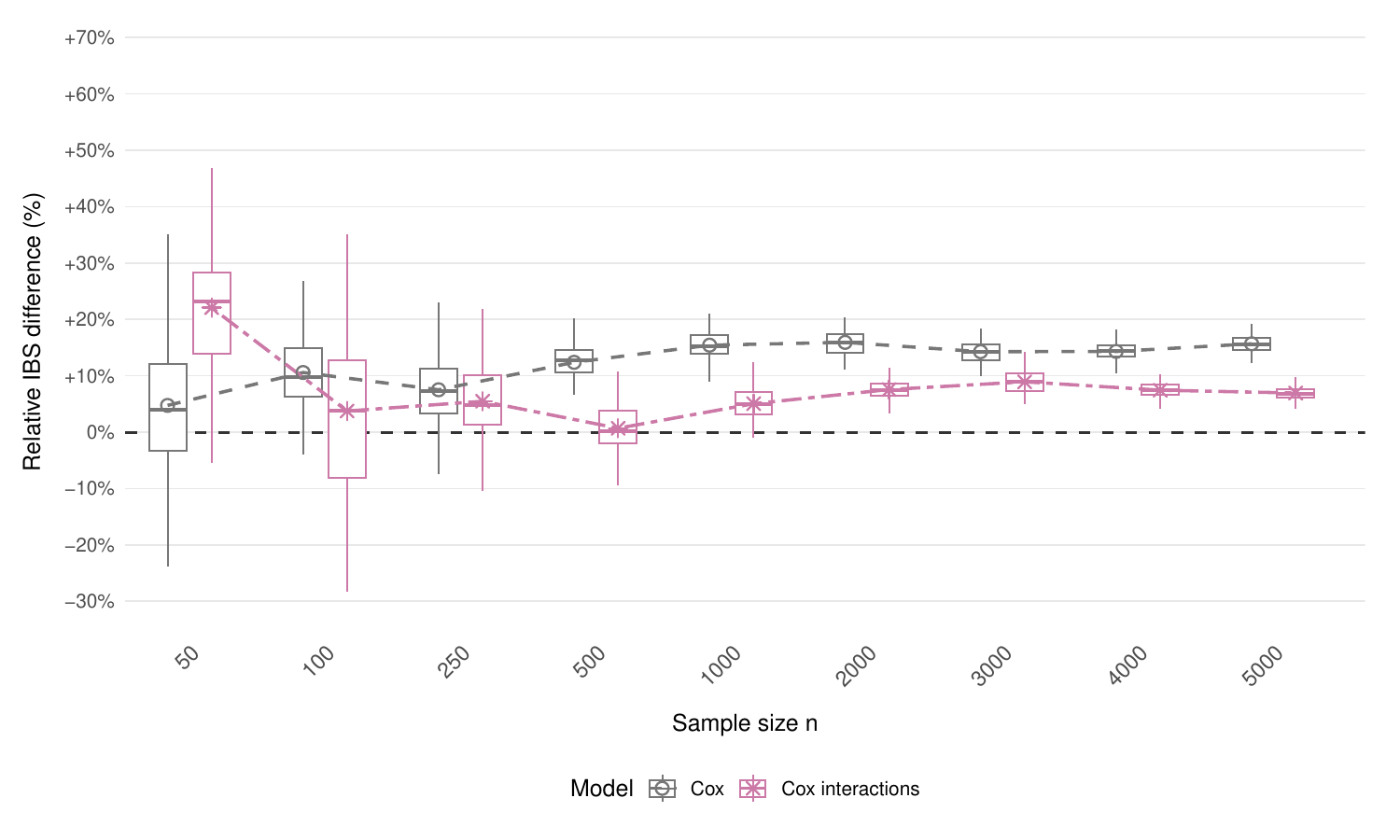}
    \caption{Non-proportional hazards study, $100$ replications. Relative difference (percent) in test-set IBS of the transition-wise Cox models (Cox and Cox interactions) with respect to the jump forest (JF), as a function of sample size $n$. Positive values favour JF.}
    \label{fig:nonph}
\end{figure}

Figure~\ref{fig:nonph} compares JF with Cox and Cox interactions for $n=50$, $100$, $250$, $500$, $1000$, $2000$, $3000$, $4000$ and $5000$. Because all individuals start in state $1$, Assumption~\ref{asm:nonvanishing} fails at time zero; the supplementary coverage study changes the initial distribution to satisfy it.

Adding interactions improves Cox prediction, but the median paired differences still favour JF at every sample size. For $n\geq2000$, the median paired relative differences favour JF by $14.2$--$15.8$ percent against Cox and $6.9$--$8.9$ percent against Cox interactions, consistent with the remaining misspecification of time-varying effects.

Under correctly specified proportional hazards, Cox instead has a small advantage without noise covariates (median relative difference $-1.6$ percent), while JF and cross-validated lasso remain comparable as noise is added (Supplementary Section~\ref{subsection:Cox}).

\subsection{Biomedical applications}\label{sec:data}\label{subsection:datasets}
We compare the methods on four data sets from the literature (Table~\ref{tab:datasets}). The \texttt{ebmt3} bone marrow transplantation data, distributed with the R package \texttt{mstate} \citep{mstate}, studies platelet recovery and relapse or death after transplantation. The remaining data and covariate descriptions accompany \citet{AndersenRavn}: \texttt{bmt} studies graft-versus-host disease, relapse and death after transplantation; \texttt{cphholter} follows atrial fibrillation, stroke and death in the Copenhagen Holter Study; and \texttt{prova} studies bleeding and death in a four-arm trial for liver cirrhosis.

\begin{table}[tbp]
    \centering
    \small
    \begin{tabular}{lcccc}
        \toprule
        & \texttt{ebmt3} & \texttt{bmt} & \texttt{cphholter} & \texttt{prova} \\
        \midrule
        States & 3 & 4 & 4 & 3 \\
        Individuals & 2204 & 2009 & 678 & 286 \\
        Covariates & 4 & 5 & 10 & 7 \\
        Not absorbed & 1363 (62\%) & 1272 (63\%) & 417 (62\%) & 211 (74\%) \\
        \addlinespace
        \multicolumn{5}{l}{\emph{Observed transitions}} \\
        \quad $1 \to 2$ & 1169 & 976 & 70 & 50 \\
        \quad $1 \to 3$ & 458 & 168 & 64 & 46 \\
        \quad $1 \to 4$ & --- & 198 & 190 & --- \\
        \quad $2 \to 3$ & 383 & 91 & 9 & 29 \\
        \quad $2 \to 4$ & --- & 307 & 27 & --- \\
        \quad $3 \to 2$ & --- & --- & 7 & --- \\
        \quad $3 \to 4$ & --- & 232 & 44 & --- \\
        \quad all & 2010 & 1972 & 411 & 125 \\
        \addlinespace
        \multicolumn{5}{l}{\emph{Jump forest}} \\
        \quad Minimum node size & 42 & 41 & 24 & 16 \\
        \quad Splitting rule & \multicolumn{4}{c}{log-rank} \\
        \bottomrule
    \end{tabular}
        \caption{The four data sets and the settings of the jump forest. Observed transitions are counted from the available data using the state definitions as depicted in the Supplementary Figure~\ref{fig:datasets}; a dash marks a transition that the corresponding model does not contain. \enquote{Not absorbed} is the number (in percentage) of individuals censored before reaching the absorbing state. All covariates counted in the table enter every model except for \texttt{bmt}, where the 255-level categorical variable \texttt{team} is excluded. The settings follow Subsection~\ref{subsection:design}; no tuning parameter is selected from the data.}
        \label{tab:datasets}
\end{table}

Using the protocol of Subsection~\ref{subsection:design} and the node sizes in Table~\ref{tab:datasets}, we compare JF with Cox and AJ null. We repeat the comparison with $10$, $25$ and $50$ added independent standard normal noise covariates, also including Cox lasso in these settings.

\begin{figure}[tbp]
    \centering
    \includegraphics[width=\linewidth]{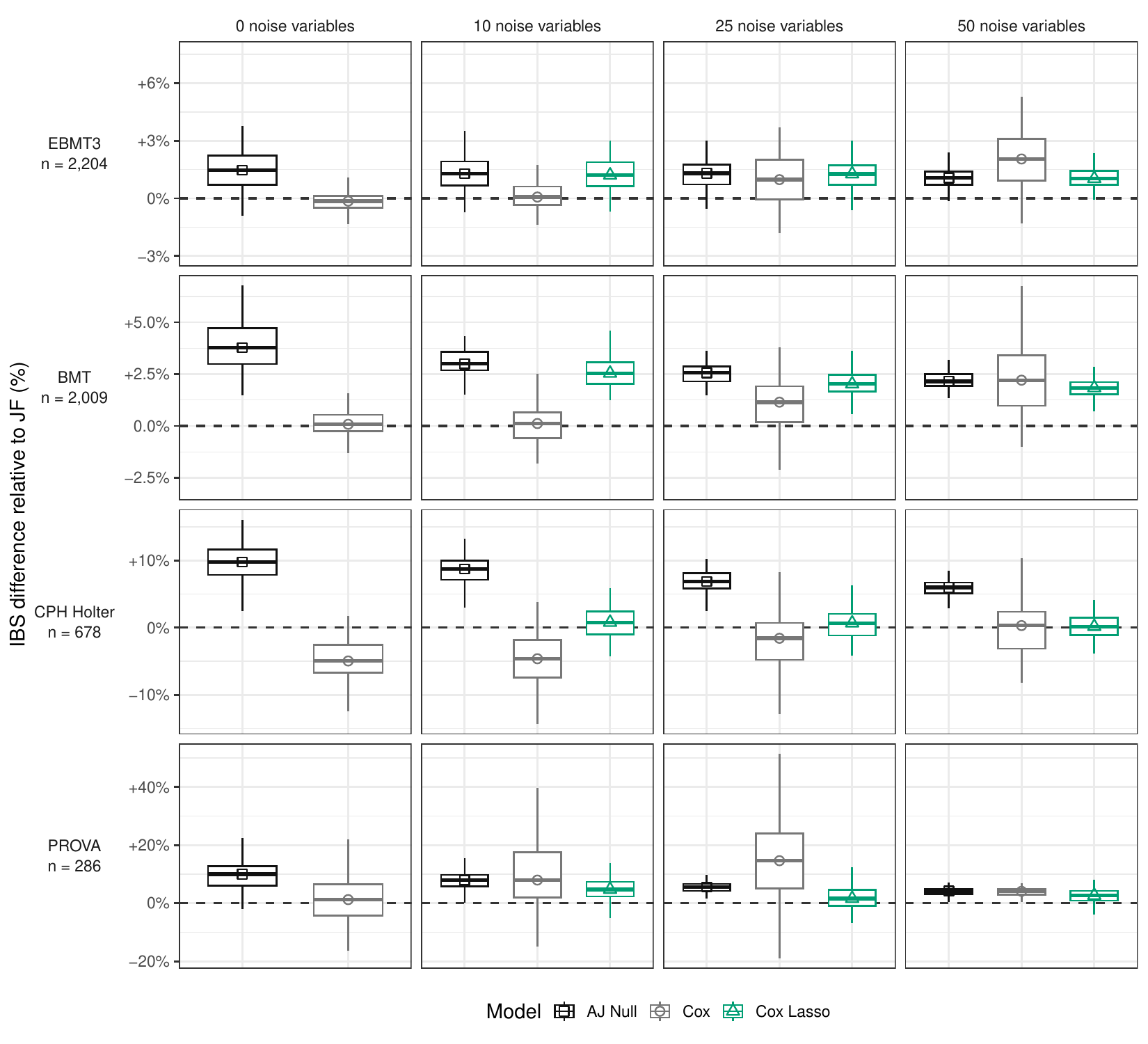}
    \caption{Biomedical data, $100$ random $80/20$ training--test splits. Relative difference (percent) in test-set IBS of the Aalen--Johansen estimator without covariates (AJ Null), the transition-wise Cox model (Cox) and its lasso-penalised version (Cox Lasso) with respect to the jump forest (JF). Rows correspond to the four data sets; columns to $0$, $10$, $25$ and $50$ added noise covariates. Positive values favour JF. Symbols inside the boxes mark the means.}
    \label{fig:ibsnoiserelative}
\end{figure}

Without added noise, JF and Cox have comparable IBS on the transplantation data: median relative differences are $-0.1$ percent on \texttt{ebmt3} and $0.1$ percent on \texttt{bmt} (Figure~\ref{fig:ibsnoiserelative}). Results are more variable for the small \texttt{prova} data, while the median paired relative IBS difference on \texttt{cphholter} is $-5.0$ percent, favouring Cox. This data set has fewer than ten observations for each of two transitions, a setting in which the regression structure may stabilise estimation.

With added noise, JF generally improves on unpenalised Cox. Its comparisons with lasso are closer: median differences favour JF in all twelve settings, by $0.2$--$4.7$ percent, although several interquartile ranges include zero. AJ null has higher median paired relative IBS in all sixteen settings, suggesting predictive information in the covariates. Supplementary Section~\ref{subsection:interactions} explores fitted covariate interactions.

\FloatBarrier
\section{Discussion}\label{sec:discussion}
The jump forest estimates the cumulative rates of a multi-state process within a common covariate partition and recovers occupation probabilities without a Markov assumption. Its clearest empirical advantage is prediction when the fitted Cox specifications cannot represent the covariate effects: under a threshold interaction, including with irrelevant covariates, and under non-proportional hazards. The complementary proportional-hazards study shows only a small loss relative to correctly specified Cox regression. Performance on the biomedical data varies across data sets.

The theory establishes consistency and Gaussian limits when covariates take finitely many values and the split search eventually separates them. Practical inference also depends on the bias from pooling different covariate values within terminal nodes. The coverage study finds near-nominal forest band coverage in several settings and gains over stratified estimation for sparse transitions, but substantial undercoverage when this bias dominates the reported uncertainty. The study uses practical fitting without censoring on the evaluation interval; it does not assess calibration under the theoretical splitting conditions.

For prediction, the studies use the log-rank rule and a minimum node size of $\lceil\sqrt{n}\,\rceil$. OOB IBS permits tuning without a separate validation sample, but optimising prediction error does not ensure valid confidence intervals at a given covariate value. Fitted rate curves can help explore departures from a Cox specification, but should not be interpreted as formal evidence of interaction.

Two extensions follow from these findings. Extending the theory to continuous covariates requires control of the error from pooling within data-dependent nodes; honest splitting, which separates the observations used for partitioning from those used for estimation, is one possible starting point. For inference, smaller nodes and a more exhaustive split search warrant study as ways to reduce pooling bias, with their effect on coverage assessed separately from prediction error.

\section{Software and reproducibility}\label{sec:software}
All methods are implemented in the R package \href{https://github.com/martinbladt/JumpForests}{\texttt{JumpForests}}, whose repository provides installation instructions and worked examples. The package is written in C++ through \texttt{Rcpp}, parallelises over trees, and provides fitting, prediction, OOB and test-set error estimation and permutation variable importance for regression, classification, survival and multi-state forests, including all five splitting rules for multi-state forests. The \texttt{ebmt3} data are obtained with the R package \texttt{mstate} \citep{mstate}, and the \texttt{bmt}, \texttt{cphholter} and \texttt{prova} data are available from the companion website of \citet{AndersenRavn}. Scripts for all numerical results in the paper and supplement are available in the \href{https://github.com/martinbladt/JumpForests}{\texttt{JumpForests/manuscript}} folder.

\section*{Acknowledgements}
The research of both authors was supported by the Carlsberg Foundation, grant CF23-1096, and was carried out within the project framework InterAct.

\clearpage

\startsupplement

\section{Splitting rules, prediction assessment and variable importance}\label{sec:methodcomponents}
This supplement section develops the additional methodological components introduced in Subsection~\ref{subsection:assessmentoverview}: alternative splitting rules, censoring-adjusted prediction scores and permutation importance.

\subsection{Alternative splitting rules}\label{subsection:alternativesplitting}
Using the risk-set and event-count notation of Subsection~\ref{subsection:splittingrules}, non-negative weights $w_m^j$ give the weighted log-rank statistic
\begin{equation}\label{eq:weightedlogrank}
    L(f, c) = \sum_{\substack{j, k \in \CZ \\ j \neq k}} \frac{\sum_{m = 1}^N w_m^j \Big(d_{m,l}^{jk} - Y_{m,l}^j \frac{d_m^{jk}}{Y_m^j} \Big)}{\sqrt{\sum_{m = 1}^N (w_m^j)^2 \frac{Y_{m,l}^j}{Y_m^j} \Big(1 - \frac{Y_{m,l}^j}{Y_m^j} \Big)\frac{Y_m^j - d_m^{jk}}{Y_m^j - 1}\, d_m^{jk}}},
\end{equation}
with the same maximisation and zero-variance convention as in Subsection~\ref{subsection:splittingrules}. Table~\ref{tab:weights} lists the weights corresponding to the classical two-sample tests. The notation follows \citet{RSFRS} and \citet{RFCompetingRisksS}; for a survival model with a single transition, \eqref{eq:weightedlogrank} reduces to their weighted log-rank splitting rules. The log-rank weight has optimal power against proportional hazards alternatives \citep{PetoPetoS, ABGKS}, whereas the Gehan, Tarone--Ware and Peto--Prentice weights emphasise differences at early times.

\begin{table}[h]
    \centering
    \caption{Weights $w_m^j$ in the splitting statistic \eqref{eq:weightedlogrank}. Here $\mathbb{p}_j^h(t_m-)$ denotes the occupation probability of state $j$ estimated within the node $h$ just before time $t_m$, that is, the $j$th entry of $\mathbb{p}^h(0)\pint_0^{t_m-}(\text{Id} + \mathbb{\Lambda}^h(\dd s))$, see \eqref{eq:occprob}.}
    \label{tab:weights}
    \begin{tabular}{ll}
        \toprule
        Splitting rule & Weight $w_m^j$ \\
        \midrule
        Log-rank \citep{MantelS, CoxS, PetoPetoS} & $1$ \\
        Gehan \citep{GehanS} & $Y_m^j$ \\
        Tarone--Ware \citep{TaroneWareS} & $\sqrt{Y_m^j}$ \\
        Peto--Prentice \citep{PetoPetoS, PrenticeS} & $\mathbb{p}_j^h(t_m-)\, Y_m^j/(1 + Y_m^j)$ \\
        \bottomrule
    \end{tabular}
\end{table}

Following \citet{RSFRS}, we also consider an \emph{approximate log-rank} rule, which keeps the numerator of the log-rank statistic ($w_m^j = 1$) but replaces its denominator by the approximation of Section 7.7 of \citet{CoxOakesS}. Concretely, with $N_{jk}^{\text{LR}} := \sum_{m = 1}^N (d_{m, l}^{jk} - Y_{m, l}^j d^{jk}_m/Y_m^j)$ the numerator and $D_l := \sum_{m = 1}^N d_{m, l}^{jk}$ and $D_r := \sum_{m = 1}^N d_{m,r}^{jk}$ the numbers of observed transitions from $j$ to $k$ in the two children, it is defined as
\begin{equation*}
    L^\text{ALR}(f, c) = \sum_{\substack{j, k \in \CZ \\ j \neq k}} \frac{\sqrt{D_l + D_r}\, N_{jk}^\text{LR}}{\sqrt{(D_l - N_{jk}^\text{LR})(D_r + N_{jk}^\text{LR})}},
\end{equation*}
and it avoids summing the exact variance contributions over event times. Zero denominators are assigned a contribution of zero.

\subsection{Error estimation}\label{subsection:error}
For the special cases of survival and competing risks models, a common error metric is Harrell's C-index \citep{HarrellCS, LongatoetalS, WolbersetalS, RFCompetingRisksS}, which measures the probability that the model correctly ranks the event times of a randomly chosen pair of individuals. Here we use Brier and Kullback--Leibler scores for occupation probabilities, following \citet{SpitoniLammensPutterS}. Let $\mathbb{p}_{\operatorname{RF}}^{(n),\text{OOB}}(t \mid x)$ denote the OOB prediction for the vector of occupation probabilities for covariate $x$, that is,
\begin{equation*}
    \mathbb{p}_{\operatorname{RF}}^{(n),\text{OOB}}(t \mid x) = \mathbb{p}_{\operatorname{RF}}^{(n),\text{OOB}}(0 \mid x) \pint\limits_0^t (\text{Id} + \mathbb{\Lambda}_{\operatorname{RF}}^{(n), \text{OOB}}(\dd v \mid x)).
\end{equation*}
The basic metrics of interest are generically given by
\begin{align*}
    \ell_\text{Brier}(t) &= \sum_{j \in \CZ} c_j (1_{(Z_t = j)} - \pi_j(t))^2, \\
    \ell_\text{KL}(t) &= -\sum_{j \in \CZ} c_j 1_{(Z_t = j)} \log \pi_j(t),
\end{align*}
where $\pi_j(t)$ denotes a prediction of the occupation probability in state $j$ at time $t$ and $(c_j)_{j \in \CZ}$ is a vector of non-negative weights summing to one chosen a priori. We remark that for the survival model with two states and a single transition, any choice of weights summing to one will yield the usual survival Brier score \citep[see][]{GrafetalS}, with equal predictions $\pi_j=1/2$ giving a score of $0.25$. Because of right-censoring, these scores cannot be estimated by plugging in the OOB predictions directly, and we use inverse probability of censoring weighting (IPCW) instead \citep[see e.g.][for the survival case]{GerdsSchumacherS, GrafetalS}. For observation $i$, let $\tilde{T}^i := \tau_{Z^i} \land R^i$ and denote by $G_R(t \mid X) = \BP(R > t \mid X)$ the conditional survival function of the censoring variable. The IPCW weight of observation $i$ at time $t$ is
\begin{equation*}
    w_i(t) = \begin{cases}
        1/\mathbb{G}_R^{(n),\text{OOB}}(\tilde{T}^i- \mid X^i) & \text{if } \tau_{Z^i} \leq R^i \text{ and } \tilde{T}^i \leq t \quad \text{(absorbed before $t$)}, \\
        1/\mathbb{G}_R^{(n),\text{OOB}}(t \mid X^i) & \text{if } \tilde{T}^i > t \quad \text{(under observation at $t$)}, \\
        0 & \text{otherwise} \quad \text{(censored before $t$ and before absorption)},
    \end{cases}
\end{equation*}
where $\mathbb{G}_R^{(n),\text{OOB}}$ is an OOB estimator of $G_R$, computed in each terminal node as the Kaplan--Meier estimator of the censoring distribution based on the data $(\tilde{T}, 1_{(R \leq \tau_Z)})$.  

An individual absorbed before $t$ is known to satisfy $R^i \geq \tau_{Z^i}$, an individual under observation at $t$ satisfies $R^i > t$, and an individual censored before $t$ and before absorption cannot contribute to the score at time $t$. With these weights, the OOB error estimators are
\begin{align*}
    \ell_\text{Brier}^{(n), \text{OOB}}(t) &= \frac{1}{n}\sum_{i = 1}^n w_i(t) \sum_{j \in \CZ} c_j (1_{(Z_t^i = j)} - \mathbb{p}_{\operatorname{RF},j}^{(n),\text{OOB}}(t \mid X^i))^2, \\
    \ell_\text{KL}^{(n), \text{OOB}}(t) &= -\frac{1}{n}\sum_{i = 1}^n w_i(t) \sum_{j \in \CZ} c_j 1_{(Z_t^i = j)} \log \mathbb{p}_{\operatorname{RF},j}^{(n),\text{OOB}}(t \mid X^i). 
\end{align*}
To summarise prediction error over time, we use the normalised integrated Brier and Kullback--Leibler scores, defined respectively by
\begin{equation*}
    \frac{1}{\bar{t}}\int_0^{\bar{t}} \ell_\text{Brier}(t) \dd t \quad \text{and} \quad \frac{1}{\bar{t}}\int_0^{\bar{t}} \ell_\text{KL}(t)\dd t,
\end{equation*}
which we compute with the respective plug-in estimators. Here $\bar{t}$ could for example be the largest observed event time in the data; the choice made in the numerical studies is stated in Subsection~\ref{subsection:design}, where the integrated Brier score is reported.

\subsection{Variable importance}\label{subsection:vimp}

Beyond prediction, a forest ranks the covariates by their contribution to predictive accuracy. We use the permutation importance of \citet{BreimanRFS}, adapted to the multi-state setting using the scores of Supplementary Section~\ref{subsection:error}.

Fix a covariate index $f \in \{1, \dots, d\}$, let $X_f'$ be an independent copy of $X_f$ which is independent of $(X, Z, R)$, and write
\begin{equation*}
    X^{(f)} := (X_1, \dots, X_{f - 1}, X_f', X_{f + 1}, \dots, X_d)
\end{equation*}
for the covariate vector whose $f$th entry has been resampled from its own marginal distribution. This preserves the marginal distribution of the $f$th covariate while destroying its dependence with the other elements of the vector. For a prediction rule $\pi$, which assigns to each covariate value $x$ a vector $\pi(\cdot \mid x)$ of occupation probability curves, let
\begin{equation*}
    \mathcal{R}(\pi) := \frac{1}{\bar{t}}\int_0^{\bar{t}} \BE\Big[\sum_{j \in \CZ} c_j \big(1_{(Z_t = j)} - \pi_j(t \mid X)\big)^2 \Big] \dd t
\end{equation*}
denote its integrated Brier risk, and let $\mathcal{R}_f(\pi)$ be given by the same expression with $\pi_j(t \mid X)$ replaced by $\pi_j(t \mid X^{(f)})$. The \emph{importance} of the $f$th covariate for $\pi$ is
\begin{equation}\label{eq:vimp}
    \operatorname{Imp}_f(\pi) := \mathcal{R}_f(\pi) - \mathcal{R}(\pi),
\end{equation}
which measures the increase in risk after replacing the $f$th covariate by an independent draw from its marginal distribution.

When $X^{(f)}$ lies in the support of $X$, the quantity \eqref{eq:vimp} has the following population interpretation. Since $X_f'$ is independent of $(X, Z)$, we have $\BE[1_{(Z_t = j)} \mid X, X_f'] = p_j(t \mid X)$, so that expanding both risks around $p(t \mid X)$ makes the cross terms vanish and yields
\begin{equation}\label{eq:vimpdecomposition}
    \operatorname{Imp}_f(p) = \frac{1}{\bar{t}}\int_0^{\bar{t}} \BE\Big[\sum_{j \in \CZ} c_j \big(p_j(t \mid X) - p_j(t \mid X^{(f)})\big)^2\Big]\dd t.
\end{equation}
Consequently $\operatorname{Imp}_f(p) \geq 0$, with equality if and only if, for almost every $t \in [0, \bar{t}]$ and every state $j$ with $c_j > 0$, the conditional occupation probabilities satisfy $p_j(t \mid X) = p_j(t \mid X^{(f)})$ almost surely. Thus this importance measures changes in occupation probabilities, jointly across states; it need not detect changes in transition rates that leave those probabilities unchanged. A standard drawback is that when the covariates are dependent, $X^{(f)}$ may fall outside the support of $X$, in which case \eqref{eq:vimpdecomposition} compares predictions at covariate values that do not occur; see \citet{HookerMentchZhouS} for a discussion of this well-known feature of permutation importance.

Two sample versions are proposed, and then used in Supplementary Section~\ref{subsection:Cox}, both obtained by replacing the risks in \eqref{eq:vimp} by their IPCW estimators. The first is computed on held-out data. Suppose the forest is fitted on a training sample, let $\CD^{\text{te}}$ denote an independent test sample of size $n^{\text{te}}$, let $\rho$ be a permutation of $\CD^{\text{te}}$ and write $X^{i,(f)}$ for the covariate vector $X^i$ with its $f$th entry replaced by $X^{\rho(i)}_f$. With the IPCW weights $w_i(t)$ of Supplementary Section~\ref{subsection:error} computed on the test sample, the held-out importance is
\begin{equation}\label{eq:vimphat}
\begin{split}
    \operatorname{Imp}_f^{(n)} := \frac{1}{\bar{t}}\int_0^{\bar{t}} \frac{1}{n^{\text{te}}}\sum_{i \in \CD^{\text{te}}} w_i(t) \sum_{j \in \CZ} c_j \Big\{&\big(1_{(Z_t^i = j)} - \mathbb{p}_{\operatorname{RF},j}^{(n)}(t \mid X^{i,(f)})\big)^2 \\
    &- \big(1_{(Z_t^i = j)} - \mathbb{p}_{\operatorname{RF},j}^{(n)}(t \mid X^i)\big)^2 \Big\}\dd t.
\end{split}
\end{equation}
The second version is internal in the jump forest algorithm and requires no held-out data. For each tree $b$, let $\mathcal{O}_b := \{i : I_{i,b} = 1\}$ denote the observations that are out-of-bag for that tree, let $\rho_b$ be a permutation of $\mathcal{O}_b$ and let $X^{i,(f),b}$ denote $X^i$ with its $f$th entry replaced by $X^{\rho_b(i)}_f$. Writing $\mathbb{p}^{(n),*}_b(\cdot \mid x)$ for the occupation probabilities of tree $b$, obtained from $\mathbb{\Lambda}^{(n),*}_b$ and $\mathbb{p}^{(n),*}_b(0 \mid x)$ through \eqref{eq:occprob}, the out-of-bag score of that tree is
\begin{equation*}
    \ell_b(t) := \frac{1}{|\mathcal{O}_b|}\sum_{i \in \mathcal{O}_b} w_i(t) \sum_{j \in \CZ} c_j \big(1_{(Z_t^i = j)} - \mathbb{p}^{(n),*}_{b,j}(t \mid X^i)\big)^2,
\end{equation*}
and $\ell_b^{(f)}(t)$ denotes the same quantity with $X^{i,(f),b}$ in place of $X^i$. The internal importance is the average over trees of the resulting increases in score,
\begin{equation}\label{eq:vimpoob}
    \operatorname{Imp}_f^{(n),\text{OOB}} := \frac{1}{B}\sum_{b = 1}^B \frac{1}{\bar{t}}\int_0^{\bar{t}} \big\{\ell_b^{(f)}(t) - \ell_b(t)\big\}\dd t.
\end{equation}
Trees that never split on the $f$th covariate contribute zero to \eqref{eq:vimpoob}, since the permutation leaves their predictions unchanged.

Replacing the Brier loss by $\ell_\text{KL}$ yields logarithmic-loss importances; uniform state weights give the usual logarithmic score up to a constant. The Brier version with uniform weights is the default in the implementation. Averaging over several independent permutations can reduce the Monte Carlo variability of either importance estimator. 

\clearpage

\section{Additional asymptotic results}\label{sec:additionalasymptotics}
This supplement section develops the single-tree and bootstrap results supporting the forest guarantees in Section~\ref{sec:theory}, together with additional results for Markov transition probabilities. The standing assumptions, fixed covariate value $x$, time horizon $t$ and matrix norm of Subsection~\ref{subsection:assumptions} apply throughout. All process limits use the uniform norm on the stated index set. Proofs are collected in Supplementary Section~\ref{sec:additionalproofs}.

\subsection{Single-tree and bootstrap consistency}\label{subsection:additionalconsistency}
The following sufficient condition can be used to verify Assumption~\ref{asm:jumpsout}.

\begin{proposition}[A sufficient condition for Assumption~\ref{asm:jumpsout}]\label{prop:jumpsout}
If Assumption~\ref{asm:rrc} holds and for every non-absorbing state $j$ and every $x \in \CX$ there exists $k \neq j$ with
\begin{equation*}
    \int_0^{T_{R,x}} p_{jk}(t \mid x)F_R(\dd t \mid x) > 0,
\end{equation*}
then Assumption~\ref{asm:jumpsout} holds.
\end{proposition}

\begin{theorem}[Consistency of a single tree]\label{thm:constree}
Under the standing assumptions,
\begin{equation*}
    \sup_{s \in [0, t]}|\mathbb{\Lambda}^{(n)}(s \mid x) - \Lambda(s \mid x)| \longrightarrow 0 \quad \BP\text{-a.s.}
\end{equation*}
\end{theorem}

\begin{corollary}[Consistency of occupation probabilities]\label{cor:occupationsingletree}
Under the standing assumptions,
\begin{equation*}
    \sup_{s \in [0, t]}|\mathbb{p}^{(n)}(s \mid x) - p(s \mid x)| \longrightarrow 0 \quad \BP\text{-a.s.},
\end{equation*}
where
\begin{equation*}
    \mathbb{p}^{(n)}(t \mid x) = \mathbb{p}^{(n)}(0 \mid x)\pint\limits_0^t (\operatorname{Id} + \mathbb{\Lambda}^{(n)}(\dd s \mid x)).
\end{equation*}
\end{corollary}

Consistency of a forest follows from consistency of a single \emph{bootstrap tree}, that is, a tree grown on a bootstrap sample. A bootstrap sample is identified with a realisation of a vector $\CW_n := (W_1, \dots, W_n)$ of non-negative exchangeable weights, where $W_i$ is the number of times observation $i$ appears in the sample. Let $\mathbb{\Lambda}^{(n),*}(\cdot \mid x)$ denote the prediction of a generic bootstrap tree for the covariate value $x$. It is obtained from the weighted estimators in the terminal node $h$ containing $x$,
\begin{equation*}
    \BI_j^{h,*}(s) = \frac{1}{n_h^*}\sum_{i \in h} 1_{(Z_s^i = j)}1_{(s < R^i)}W_i, \quad \BN_{jk}^{h,*}(s) = \frac{1}{n_h^*}\sum_{i \in h} N_{jk}^i(s \land R^i) W_i, \quad n_h^* := \sum_{i \in h}W_i,
\end{equation*}
as $\mathbb{\Lambda}_{jk}^{h,*}(t) = \int_0^t \BN_{jk}^{h,*}(\dd s)/\BI_j^{h,*}(s-)$. The argument of Theorem~\ref{thm:constree} requires the bootstrap sample to consist of observations from the original data, so that the bootstrapped covariate values lie in $\CX$. We therefore consider the two subsampling schemes most common in practice: a) multinomial subsampling, which draws $m_n$ observations with replacement; and b) subsampling without replacement of size $m_n$. The bootstrap sample is then a generic subsample
\begin{equation*}
    \CD_{m_n}^* := \{(X^{i, *}, (Z_t^{i,*})_{0 \leq t \leq R^{i,*}}, \tau_{Z^{i,*}} \land R^{i, *})\}_{i = 1}^{m_n}
\end{equation*}
of the original data
\begin{equation}\label{eq:data}
    \CD_n := \{(X^i, (Z_t^i)_{0 \leq t \leq R^i}, \tau_{Z^i} \land R^i)\}_{i = 1}^n,
\end{equation}
determined by the outcome of the bootstrap weights. The background theory in Supplementary Section~\ref{subsection:bootstrap} covers more general weights.

\begin{theorem}[Consistency of a bootstrap tree]\label{thm:consbootstraptree}
Suppose that $\BE[N_{jk}(t)^2] < \infty$ for all $j \neq k$. Let the bootstrap sample be obtained by multinomial subsampling or subsampling without replacement, with subsample size $m_n\to\infty$. Then
\begin{equation*}
    \sup_{s \in [0, t]}|\mathbb{\Lambda}^{(n),*}(s \mid x) - \Lambda(s \mid x)| \overset{\BP}{\longrightarrow} 0.
\end{equation*}
If, in addition, $\BE[N_{jk}(t)^4] < \infty$ for all $j \neq k$ and the subsample size satisfies $\sum_{n = 1}^\infty m_n^{-2} < \infty$, the convergence holds almost surely.
\end{theorem}

Uniform consistency of occupation probabilities follows by the product-integral argument in Corollary~\ref{cor:occupationsingletree}. Averaging a fixed number of bootstrap trees gives Theorem~\ref{cor:forestconsistency}.

\subsection{Single-tree and conditional bootstrap limits}\label{subsection:additionallimits}
Let $h(x)$ denote the terminal node containing $x$, $n_{h(x)}$ its sample size, and $p_x=\BP(X=x)$.

\begin{theorem}[Weak convergence of a single tree]\label{thm:weakconvergence}
Suppose that $\BE[N_{jk}(t)^2] < \infty$ for all $j \neq k$. Then, for any such pair,
\begin{equation*}
    \sqrt{n_{h(x)}}(\mathbb{\Lambda}_{jk}^{(n)}(\cdot \mid x) - \Lambda_{jk}(\cdot \mid x)) \overset{d}{\longrightarrow} \zeta_{jk}^x,
\end{equation*}
where $(\xi_{jk}^x, \xi_j^x)$ is the tight mean-zero Gaussian process of Proposition~\ref{prop:jointweakconvergence} and
\begin{equation*}
    \zeta_{jk}^x(s) = \frac{1}{\sqrt{p_x}}\left(\int_0^s \frac{1}{p_j^c(u- \mid x)}\xi_{jk}^x(\dd u) - \int_0^s \frac{\xi_j^x(u-)}{p_j^c(u- \mid x)}\Lambda_{jk}(\dd u \mid x) \right), \quad s \in [0, t],
\end{equation*}
and the integral with respect to $\xi_{jk}^x$ is defined through integration by parts (see Remark~\ref{rmk:leftlimit}).
\end{theorem}

We now turn to a bootstrap tree and to the whole forest and study the asymptotic distribution of
\begin{equation*}
    \sqrt{n_{h(x)}^*}\big(\mathbb{\Lambda}^{(n),*}(\cdot \mid x) - \mathbb{\Lambda}^{(n)}(\cdot \mid x) \big),
\end{equation*}
where $n_{h(x)}^*$ denotes the number of observations in the terminal node containing $x$ in the bootstrap tree. We keep the data $\CD_n$ fixed and consider randomness from resampling and tree construction. Under the standing assumptions, the trees reduce to stratified estimators with conditional probability tending to one, leaving bootstrap weights as the source of the limiting fluctuation.

\begin{definition}[Conditional weak convergence]\label{def:condweak}
Let $T^{(n)}$ be a statistic of $\CD_n$, let $T^{(n),*}$ denote its bootstrap version based on $(\CD_n, \CW_n)$, and suppose that both take values in a normed space $\BD$. Let $\text{BL}_1$ denote the set of functions $f: \BD \to \BR$ with $\|f\|_\infty \leq 1$ and Lipschitz constant bounded by 1. We say that $\sqrt{k_n}(T^{(n),*} - T^{(n)})$ converges weakly to a tight process $\mathbb{G}$ conditionally on the data, in probability or almost surely, if
\begin{equation*}
    \sup_{f \in \text{BL}_1} \big|\BE_{\CW_n}[f(\sqrt{k_n}(T^{(n),*} - T^{(n)}))] - \BE[f(\mathbb{G})]\big| \longrightarrow 0
\end{equation*}
in probability or almost surely, respectively, where $\BE_{\CW_n}$ denotes expectation over the resampling and, when applicable, tree-construction randomness.
\end{definition}

The supremum in Definition~\ref{def:condweak} is the bounded Lipschitz metric, which metrises weak convergence of separable Borel probability measures on $\BD$ \citep[Theorem 1.12.4]{VaartWellnerS}; in principle, expectations should be outer expectations; as is customary, we suppress these measurability issues and tacitly assume that the maps involved are measurable. Almost sure conditional convergence means weak convergence for almost every fixed data sequence. We work on a product probability space carrying the data and the independent resampling randomness.

\begin{theorem}[Weak convergence of bootstrap trees and forests]\label{thm:weakconvergencebootstrap}
Suppose that $\BE[N_{jk}(t)^2] < \infty$ for all $j \neq k$. Let the bootstrap sample be obtained by multinomial subsampling or subsampling without replacement, with subsample size $m_n\to\infty$. For subsampling without replacement, assume also that $m_n/n\to\alpha\in[0,1)$. Let $\zeta_{jk}^x$ be the limit process of Theorem~\ref{thm:weakconvergence}, and let $\sigma_\alpha := 1$ for multinomial subsampling and $\sigma_\alpha := \sqrt{1 - \alpha}$ for subsampling without replacement.
\begin{enumerate}
    \item[(i)] For a single bootstrap tree,
    \begin{equation*}
        \sqrt{n_{h(x)}^*}\big(\mathbb{\Lambda}_{jk}^{(n),*}(\cdot \mid x) - \mathbb{\Lambda}_{jk}^{(n)}(\cdot \mid x) \big) \overset{d}{\longrightarrow} \sigma_\alpha \zeta_{jk}^x
    \end{equation*}
    conditionally on the data, almost surely.
    \item[(ii)] For a forest of a fixed number $B$ of trees, each grown on an independent bootstrap sample,
    \begin{equation*}
        \sqrt{B m_n} (\mathbb{\Lambda}_{\operatorname{RF}, jk}^{(n)}(\cdot \mid x) - \mathbb{\Lambda}_{jk}^{(n)}(\cdot \mid x)) \overset{d}{\longrightarrow} \frac{\sigma_\alpha}{\sqrt{p_x}}\zeta_{jk}^x
    \end{equation*}
    conditionally on the data, almost surely.
\end{enumerate}
\end{theorem}
The conditional limits describe fluctuation around the full-data estimator; Theorem~\ref{thm:weakconvergencebootstrapunconditional} also accounts for its estimation error around the population target.

\subsection{Variance in the Markov case}\label{subsection:additionalvariance}
For the Gaussian process in Theorem~\ref{thm:weakconvergence}, the Markov assumption yields the following variance formula.

\begin{proposition}[Asymptotic variance and its estimator in the Markov case]\label{prop:markovvariance}
Suppose that $\BE[N_{jk}(t)^2] < \infty$ for all $j \neq k$ and that $Z$ is Markov conditionally on $X$. Then, for $s \in [0, t]$,
\begin{equation*}
    \Var(\zeta_{jk}^x(s)) = \int_0^s \frac{1 - \Delta \Lambda_{jk}(u \mid x)}{p_j^c(u- \mid x)}\Lambda_{jk}(\dd u \mid x) = \int_0^s \frac{1 - \Delta \Lambda_{jk}(u \mid x)}{p_j^c(u- \mid x)^2} p_{jk}^c(\dd u \mid x) =: V_{jk}(s \mid x).
\end{equation*}
If, in addition, $\Lambda_{jk}(\cdot \mid x)$ is continuous on $[0, t]$, the plug-in estimator
\begin{equation*}
    \mathbb{V}_{jk}^{(n)}(s \mid x) := \int_0^s \frac{1}{\BI_j^{h(x)}(u-)^2}\BN_{jk}^{h(x)}(\dd u)
\end{equation*}
satisfies $\sup_{s \in [0, t]}|\mathbb{V}_{jk}^{(n)}(s \mid x) - V_{jk}(s \mid x)| \to 0$, $\BP$-a.s.
\end{proposition}

For a smooth Markov model, $V_{jk}(s\mid x)=\int_0^s\mu_{jk}(u\mid x)/p_j^c(u-\mid x)\dd u$.

\subsection{Transition probabilities in the Markov case}\label{subsection:additionaltransitions}
The special case of a Markov model is worth highlighting. For a jump process which is Markov conditionally on $X$, we may consider the Aalen--Johansen estimator of the transition probabilities,
\begin{equation*}
    \mathbb{p}^{(n)}(s, t \mid x) = \pint\limits_s^t (\text{Id} + \mathbb{\Lambda}^{(n)}(\dd v \mid x)),
\end{equation*}
which is justified by \eqref{eq:Kolmogorov} and the following result. The proof uses Theorem 7 of \citet{GillJohansenS}, which gives continuity of the product integral uniformly in both endpoints $s_1 \leq s_2$.

\begin{corollary}[Consistency of transition probabilities in the Markov case]\label{cor:transitionsingletree}
If $Z$ is Markov conditionally on $X$, then
\begin{equation*}
    \sup_{0 \leq s_1 \leq s_2 \leq t} |\mathbb{p}^{(n)}(s_1, s_2 \mid x) - p(s_1, s_2 \mid x)| \longrightarrow 0 \quad \BP\text{-a.s.}
\end{equation*}
\end{corollary}

The argument in the proof of Corollary~\ref{cor:occupationsingletree}, using continuity of the product integral uniformly in both endpoints, proves this result. The corresponding weak convergence result is as follows.

\begin{theorem}[Weak convergence of transition probabilities]\label{thm:weakconvergencetransition}
Suppose that $\BE[N_{jk}(t)^2] < \infty$ for all $j \neq k$ and that $Z$ is Markov conditionally on $X$. Then
\begin{equation*}
    (\sqrt{n_{h(x)}}(\mathbb{p}^{(n)}(s_1, s_2 \mid x) - p(s_1, s_2 \mid x)))_{0 \leq s_1 \leq s_2 \leq t} \overset{d}{\longrightarrow} (\gamma^x(s_1, s_2))_{0 \leq s_1 \leq s_2 \leq t},
\end{equation*}
with
\begin{equation*}
    \gamma^x(s_1, s_2) = \int_{s_1}^{s_2} p(s_1, u- \mid x)\zeta^x(\dd u) p(u, s_2 \mid x)
\end{equation*}
a tight mean-zero Gaussian process defined through integration by parts, where $\zeta^x=(\zeta_{jk}^x)_{j,k\in\CZ}$ and $\zeta_{jj}^x=-\sum_{k:k\neq j}\zeta_{jk}^x$.
\end{theorem}

\clearpage

\section{Proofs of the main text results}\label{sec:mainproofs}
\subsection{Proof of Theorem~\ref*{cor:forestconsistency}}\label{subsection:forestconsistencyproof}
\begingroup\pushQED{\qed}
Theorem~\ref{thm:consbootstraptree} gives uniform consistency of each bootstrap-tree rate estimator, in probability or almost surely under the stated additional conditions. Averaging over the fixed number $B$ of trees preserves this convergence. The same bootstrap law of large numbers applied to the initial-state indicators gives consistency of the empirical initial distribution in each tree and hence of their average. Continuity of the product integral, as in the proof of Corollary~\ref{cor:occupationsingletree}, then gives uniform consistency of the forest occupation probabilities.
\popQED\endgroup

\subsection{Proof of Theorem~\ref*{thm:weakconvergencebootstrapunconditional}}\label{subsection:unconditional}
\begingroup\pushQED{\qed}
For $b = 1, \dots, B$, let $\CW_n^{(b)}$ collect the bootstrap weights of the $b$th tree together with the random selection of its candidate covariates. The vectors $\CW_n^{(1)}, \dots, \CW_n^{(B)}$ are iid and independent of the data $\CD_n$, and, as in Supplementary Section~\ref{subsection:additionallimits}, the underlying probability space is a product of a space carrying the data and spaces carrying the weights. Let $\sigma_\alpha$ be as in Theorem~\ref{thm:weakconvergencebootstrap}, where for multinomial subsampling $\sigma_\alpha = 1$ for every $\alpha \in [0, 1]$. Denote by $\mathbb{\Lambda}_{jk, b}^{(n),*,x}$ the stratified Nelson--Aalen estimator computed from the observations of the $b$th bootstrap sample with covariate value $x$, as in the proof of Theorem~\ref{thm:weakconvergencebootstrap}(i), and define
\begin{equation*}
    \zeta_{jk}^{(n),x} := \sqrt{n}\big(\mathbb{\Lambda}_{jk}^{(n),x} - \Lambda_{jk}(\cdot \mid x)\big), \qquad \zeta_{jk, b}^{(n),*,x} := \sqrt{m_n}\big(\mathbb{\Lambda}_{jk, b}^{(n),*,x} - \mathbb{\Lambda}_{jk}^{(n),x}\big), \quad b = 1, \dots, B.
\end{equation*}

\emph{Step 1: Reduction to stratified estimators.} By the proof of Theorem~\ref{thm:consbootstraptree}, for each $b$ the terminal node containing $x$ in the $b$th tree consists of observations with covariate value $x$ only, with probability tending to one. On this event, $\mathbb{\Lambda}_{jk, b}^{(n),*}(\cdot \mid x) = \mathbb{\Lambda}_{jk, b}^{(n),*,x}$ on $[0, t]$. Since $B$ is fixed, it follows that, with probability tending to one,
\begin{equation}\label{eq:forestdecomposition}
    \sqrt{m_n}\big(\mathbb{\Lambda}_{\operatorname{RF}, jk}^{(n)}(\cdot \mid x) - \Lambda_{jk}(\cdot \mid x)\big) = \frac{1}{B}\sum_{b = 1}^B \zeta_{jk, b}^{(n),*,x} + \sqrt{\frac{m_n}{n}}\,\zeta_{jk}^{(n),x}.
\end{equation}
It therefore suffices to derive the limit of the right-hand side.

\emph{Step 2: Marginal limits.} By the proof of Theorem~\ref{thm:weakconvergence}, $\zeta_{jk}^{(n),x} \overset{d}{\longrightarrow} p_x^{-1/2}\zeta_{jk}^x$. By the proof of Theorem~\ref{thm:weakconvergencebootstrap}(i), $\zeta_{jk, 1}^{(n),*,x}$ converges weakly to $\sigma_\alpha p_x^{-1/2}\zeta_{jk}^x$ conditionally on the data, almost surely, that is,
\begin{equation*}
    \Delta_n := \sup_{f \in \text{BL}_1}\Big|\BE_{\CW_n}\big[f\big(\zeta_{jk, 1}^{(n),*,x}\big)\big] - \BE\big[f\big(\sigma_\alpha p_x^{-1/2}\zeta_{jk}^x\big)\big]\Big| \longrightarrow 0 \quad \BP\text{-a.s.}
\end{equation*}
For each $n$, the weights take finitely many values, so that $w \mapsto f(\zeta_{jk, 1}^{(n),*,x}(\CD_n, w))$ is measurable for every $f \in \text{BL}_1$ and the minimal measurable majorant with respect to the weights in Lemma~3.1(c) of \citet{BucherKojadinovicS} may be omitted; see the second remark following that lemma. With the measurability convention stated after Definition~\ref{def:condweak}, under which $(\CD_n, w) \mapsto f(\zeta_{jk, 1}^{(n),*,x}(\CD_n, w))$ is jointly measurable for every $f \in \text{BL}_1$, Fubini's theorem gives $\BE[f(\zeta_{jk, 1}^{(n),*,x})] = \BE[\BE_{\CW_n}[f(\zeta_{jk, 1}^{(n),*,x})]]$, and hence
\begin{equation*}
    \sup_{f \in \text{BL}_1}\Big|\BE\big[f\big(\zeta_{jk, 1}^{(n),*,x}\big)\big] - \BE\big[f\big(\sigma_\alpha p_x^{-1/2}\zeta_{jk}^x\big)\big]\Big| \leq \BE[\Delta_n] \longrightarrow 0
\end{equation*}
by dominated convergence, since $0 \leq \Delta_n \leq 2$. Thus $\zeta_{jk, 1}^{(n),*,x}$ converges weakly to a tight Borel measurable limit and is, in particular, asymptotically measurable.

\emph{Step 3: Joint limit.} Set $\mathbb{G}_n := \sigma_\alpha \zeta_{jk}^{(n),x}$ and $\mathbb{G}_n^{(b)} := \zeta_{jk, b}^{(n),*,x}$, $b = 1, \dots, B$. Then $\mathbb{G}_n^{(1)}, \dots, \mathbb{G}_n^{(B)}$ are bootstrap replicates of $\mathbb{G}_n$ in the sense of Section~3 of \citet{BucherKojadinovicS}, obtained by evaluating the same map at $\CD_n$ and at the iid weights $\CW_n^{(b)}$. By Step 2, $\mathbb{G}_n$ converges weakly to the tight Gaussian process $\mathbb{G} := \sigma_\alpha p_x^{-1/2}\zeta_{jk}^x$, and $\mathbb{G}_n^{(1)}$ satisfies assertion (c) of their Lemma~3.1, almost sure conditional convergence implying conditional convergence in outer probability. Owing to the factor $\sigma_\alpha$ in the definition of $\mathbb{G}_n$, the bootstrap replicates and $\mathbb{G}_n$ have the same limit $\mathbb{G}$. The equivalence of assertions (c) and (b) of that lemma, with $M = \max(B, 2)$, yields
\begin{equation*}
    \big(\mathbb{G}_n, \mathbb{G}_n^{(1)}, \dots, \mathbb{G}_n^{(B)}\big) \overset{d}{\longrightarrow} \big(\mathbb{G}, \mathbb{G}^{(1)}, \dots, \mathbb{G}^{(B)}\big) \quad \text{in } \ell^\infty([0, t])^{B + 1},
\end{equation*}
where $\mathbb{G}, \mathbb{G}^{(1)}, \dots, \mathbb{G}^{(B)}$ are iid; for $B = 1$ the statement follows from the case $M = 2$ by projection.

\emph{Step 4: The forest.} Since $\sigma_\alpha = 1$ for multinomial subsampling and $\sigma_\alpha = \sqrt{1 - \alpha} > 0$ for subsampling without replacement, where $\alpha < 1$, the map $(g_0, g_1, \dots, g_B) \mapsto B^{-1}\sum_{b = 1}^B g_b + \sigma_\alpha^{-1}\sqrt{\alpha}\, g_0$ is well defined, linear and continuous on $\ell^\infty([0, t])^{B + 1}$, and $\sqrt{m_n/n} \to \sqrt{\alpha}$. By \eqref{eq:forestdecomposition}, Step 3, the continuous mapping theorem and Slutsky's lemma,
\begin{equation*}
    \sqrt{m_n}\big(\mathbb{\Lambda}_{\operatorname{RF}, jk}^{(n)}(\cdot \mid x) - \Lambda_{jk}(\cdot \mid x)\big) \overset{d}{\longrightarrow} \frac{1}{B}\sum_{b = 1}^B \mathbb{G}^{(b)} + \frac{\sqrt{\alpha}}{\sigma_\alpha}\mathbb{G} \overset{d}{=} \frac{1}{\sqrt{p_x}}\Big(\frac{\sigma_\alpha}{B}\sum_{b = 1}^B \zeta_{jk, b}^x + \sqrt{\alpha}\,\zeta_{jk, B + 1}^x\Big),
\end{equation*}
where $\zeta_{jk, 1}^x, \dots, \zeta_{jk, B + 1}^x$ are iid copies of $\zeta_{jk}^x$. The limit is a tight mean-zero Gaussian process whose covariance function equals $\alpha + \sigma_\alpha^2/B$ times that of $p_x^{-1/2}\zeta_{jk}^x$. Since a tight Gaussian process in $\ell^\infty([0, t])$ is determined in law by its covariance function, the limit has the law of $(\alpha + \sigma_\alpha^2/B)^{1/2} p_x^{-1/2}\zeta_{jk}^x$. Finally, $m_n \kappa_n \to 1/B + \alpha$ for multinomial subsampling and $m_n \kappa_n \to 1/B + (B - 1)\alpha/B$ for subsampling without replacement, and both limits equal $\alpha + \sigma_\alpha^2/B$. Since $\kappa_n^{-1/2} = (m_n \kappa_n)^{-1/2}\sqrt{m_n}$, Slutsky's lemma yields the claim.
\popQED\endgroup

\subsection{Proof of Corollary~\ref*{cor:forestCI}}\label{subsection:forestCIproof}
\begingroup\pushQED{\qed}
\emph{Consistency of the variance estimator.} By the proof of Theorem~\ref{thm:consbootstraptree}, for each $b$ the terminal node $h_b(x)$ consists of observations with covariate value $x$ only, with probability tending to one, and the corresponding weighted estimators $\BI_j^{h_b(x),*}$ and $\BN_{jk}^{h_b(x),*}$ converge uniformly on $[0, t]$ in probability to $p_j^c(\cdot \mid x)$ and $p_{jk}^c(\cdot \mid x)$. Assumption~\ref{asm:nonvanishing} and $t<T_{R,x}$ bound the limiting at-risk probability away from zero. The continuity of the integration functional with $\varphi(u)=u^{-2}$, established in Lemma~\ref{lem:gHD} and Remark~\ref{rmk:leftlimit}, therefore gives
\begin{equation*}
    \sup_{s \in [0, t]}\Big|\mathbb{V}_{jk, b}^{(n),*}(s \mid x) - \int_0^s \frac{1}{p_j^c(u- \mid x)^2}p_{jk}^c(\dd u \mid x)\Big| \overset{\BP}{\longrightarrow} 0,
\end{equation*}
and the integral equals $V_{jk}(s \mid x)$ because $\Lambda_{jk}(\cdot \mid x)$ is continuous. Moreover, $n_{h_b(x)}^*/m_n \to p_x$ in probability by the proof of Theorem~\ref{thm:weakconvergencebootstrap}(i). Since $B$ is fixed, $\tilde{\mathbb{V}}_{\operatorname{RF}, jk}^{(n)}(\cdot \mid x) \to V_{jk}(\cdot \mid x)/p_x$ uniformly on $[0, t]$ in probability.

\emph{Part (i).} Evaluation at $s$ is continuous on $\ell^\infty([0, t])$, and $\zeta_{jk}^x(s)$ is normally distributed with mean zero and variance $V_{jk}(s \mid x)$ by Proposition~\ref{prop:markovvariance}. Theorem~\ref{thm:weakconvergencebootstrapunconditional} and the continuous mapping theorem thus give an asymptotically normal estimation error with limiting variance $V_{jk}(s \mid x)/p_x > 0$ after scaling by $\kappa_n^{-1/2}$. Slutsky's lemma and the consistency of $\tilde{\mathbb{V}}_{\operatorname{RF}, jk}^{(n)}(s \mid x)$ show that the studentised difference is asymptotically standard normal, which yields (i).

\emph{Part (ii).} In the proof of Proposition~\ref{prop:markovvariance} it is shown that $\mathbb{U}_{jk}^x$ is a mean-zero Gaussian process with orthogonal increments and that $\zeta_{jk}^x(s) = p_x^{-1/2}\int_0^s p_j^c(u- \mid x)^{-1}\mathbb{U}_{jk}^x(\dd u)$ is the $L^2$-limit of linear combinations of increments of $\mathbb{U}_{jk}^x$ over $(0, s]$. Hence, for $0 \leq s_1 \leq s_2 \leq t$, the increment $\zeta_{jk}^x(s_2) - \zeta_{jk}^x(s_1)$ is the $L^2$-limit of linear combinations of increments of $\mathbb{U}_{jk}^x$ over $(s_1, s_2]$ and is therefore uncorrelated with, and by joint Gaussianity independent of, $(\zeta_{jk}^x(u))_{u \leq s_1}$. Consequently, $\Cov(\zeta_{jk}^x(s_1), \zeta_{jk}^x(s_2)) = V_{jk}(s_1 \land s_2 \mid x)$, and $\zeta_{jk}^x$ has the same finite-dimensional distributions as $W \circ V_{jk}(\cdot \mid x)$ for a standard Brownian motion $W$. Both processes are tight and Borel measurable in $\ell^\infty([0, t])$, the latter having continuous paths because $V_{jk}(\cdot \mid x)$ is continuous, so that they have the same law, since the law of a tight Borel measurable map into $\ell^\infty([0, t])$ is determined by its finite-dimensional marginals \citep[Section 1.5]{VaartWellnerS}. Since $V_{jk}(\cdot \mid x)$ is continuous and non-decreasing with $V_{jk}(0 \mid x) = 0$, it maps $[0, t]$ onto $[0, V_{jk}(t \mid x)]$, and Brownian scaling gives
\begin{equation*}
    \sup_{s \in [0, t]}|\zeta_{jk}^x(s)| \overset{d}{=} \sup_{0 \leq u \leq V_{jk}(t \mid x)}|W(u)| \overset{d}{=} V_{jk}(t \mid x)^{1/2}\sup_{0 \leq v \leq 1}|W(v)|.
\end{equation*}
The supremum norm is continuous on $\ell^\infty([0, t])$, so Theorem~\ref{thm:weakconvergencebootstrapunconditional}, the continuous mapping theorem and Slutsky's lemma yield
\begin{equation*}
    \frac{\sup_{s \in [0, t]}|\mathbb{\Lambda}_{\operatorname{RF}, jk}^{(n)}(s \mid x) - \Lambda_{jk}(s \mid x)|}{\big(\kappa_n \tilde{\mathbb{V}}_{\operatorname{RF}, jk}^{(n)}(t \mid x)\big)^{1/2}} \overset{d}{\longrightarrow} \sup_{0 \leq v \leq 1}|W(v)|.
\end{equation*}
The distribution function of the limit is continuous, which yields (ii). The value $c_{0.95} = 2.2414$ is the solution of $\BP(\sup_{0 \leq v \leq 1}|W(v)| \leq c) = 0.95$, computed from the series representation
\begin{equation*}
    \BP\Big(\sup_{0 \leq v \leq 1}|W(v)| \leq c\Big) = \frac{4}{\pi}\sum_{k = 0}^\infty \frac{(-1)^k}{2k + 1}\exp\Big(-\frac{(2k + 1)^2\pi^2}{8c^2}\Big). \qedhere
\end{equation*}
\popQED\endgroup

\clearpage

\section{Proofs of the additional asymptotic results}\label{sec:additionalproofs}
\subsection{Consistency of trees and bootstrap trees}\label{sec:proofs}
\begin{proof}[of Proposition~\ref{prop:jumpsout}]
Fix $x \in \CX$ and a non-absorbing state $j$. It suffices to show that $\BE[N_{jk}(R) \mid X = x] > 0$ for some $k \neq j$. We compute
\begin{align*}
    \BE[N_{jk}(R) \mid X = x] &= \int_0^\infty \BE[N_{jk}(r) \mid R = r, X = x] F_R(\dd r \mid x) \\
    &= \int_0^{T_{R,x}} p_{jk}(r \mid x)F_R(\dd r \mid x),
\end{align*}
where we used Assumption~\ref{asm:rrc} in the second equality to remove the conditioning $R = r$. The result now follows.
\end{proof}

\begin{proof}[of Theorem~\ref{thm:constree}]
Let $A \subseteq \CX$ be nonempty, let $j$ be non-absorbing, $k \neq j$, and let $t \in (0, T_{R,x} \land T(x))$. From the decomposition
\begin{equation*}
    \BP(X \in A, N_j^\bullet > 0) = \sum_{x \in A}\BP(N_j^\bullet > 0 \mid X = x)\BP(X = x)
\end{equation*}
and Assumptions~\ref{asm:featurespace} and~\ref{asm:jumpsout}, we see that $\BP(X \in A, N_j^\bullet > 0) > 0$. Letting $N_j^{\bullet, i} = \sum_{k \neq j} N_{jk}^i(R^i)$ denote the total number of observed jumps out of state $j$ for the $i$th process $Z^i$, it follows from the second Borel--Cantelli lemma that
\begin{equation*}
    \BP(X^i \in A, N_j^{\bullet, i} > 0 \text{ i.o.}) = 1,
\end{equation*}
using that the variables $(X^i, N_j^{\bullet, i})$ are iid. To summarise, for any nonempty $A \subseteq \CX$, $\{X^i \in A, N_j^{\bullet, i} > 0\}$ occurs infinitely often with probability one. In particular, taking $A = \{x\}$, every $x \in \CX$ eventually has more than $d_j$ observed jumps out of every non-absorbing state $j$. Consequently, if a node contains two different values from $\CX$, any split separating them is admissible for all large $n$, and by Assumption~\ref{asm:splitting} the node is split. It follows that, almost surely, for all large $n$ every terminal node $h$ can be identified with exactly one $x \in \CX$ (if this is the case, we write $x$ instead of $h$ in the estimators belonging to terminal node $h$). Hence, for any $x$, there exists an almost surely finite $N(x)$ such that for $n \geq N(x)$ we have $\mathbb{\Lambda}_{jk}^{(n)}(s \mid x) = \mathbb{\Lambda}_{jk}^x(s)$ for all $s \leq t$. Thus,
\begin{equation*}
    \lim_{n \to \infty}\sup_{s \in [0, t]}|\mathbb{\Lambda}_{jk}^{(n)}(s \mid x) - \Lambda_{jk}(s \mid x)| = \lim_{n \to \infty} \sup_{s \in [0, t]}|\mathbb{\Lambda}_{jk}^x(s) - \Lambda_{jk}(s \mid x)|.
\end{equation*}

Hence it suffices to show that $\mathbb{\Lambda}^x_{jk}(s)$ converges to $\Lambda_{jk}(s | x)$ uniformly on $[0, t]$ almost surely as $n \to \infty$. Recall that $\mathbb{\Lambda}_{jk}^x(s)$ is grown on the sample
\begin{equation*}
    \CD_{n,x} = \{(X, (Z_t)_{0 \leq t \leq R}, \tau_{Z} \land R) \in \CD_n : X = x\}.
\end{equation*}
Letting $n_x := \#\CD_{n,x} = \sum_{i = 1}^n 1_{(X^i = x)}$, we have
\begin{align*}
    \BN^x_{jk}(s) &= \frac{1}{n_x}\sum_{i:X^i = x} N_{jk}^i(s \land R^i) = \frac{\frac{1}{n}\sum_{i = 1}^n N_{jk}^i(s \land R^i)1_{(X^i = x)}}{\frac{1}{n}\sum_{i = 1}^n 1_{(X^i = x)}} \\
    &\overset{\BP\text{-a.s.}}{\longrightarrow} \frac{\BE[N_{jk}(s \land R) 1_{(X = x)}]}{\BP(X = x)} = \BE[N_{jk}(s \land R) \mid X = x] = p_{jk}^c(s \mid x)
\end{align*}
by the strong law of large numbers and the continuous mapping theorem, and the same argument applied to $N_{jk}((s-) \land R) = \lim_{u \uparrow s} N_{jk}(u \land R)$ shows that also $\BN_{jk}^x(s-) \to p_{jk}^c(s- \mid x)$ almost surely. The same argument applies to show that $\BI_j^x(s) \to p_j^c(s \mid x)$ almost surely. We now apply the key decomposition (see Supplementary Section~\ref{sec:background} for background) and see that by the same argument as for $\BN_{jk}^x$,
\begin{equation*}
    \BC_j^x(s) := \frac{1}{n_x}\sum_{i:X^i = x} 1_{(R^i \leq s)}1_{(Z_{R^i}^i = j)} \to C_j(s \mid x) \quad \BP\text{-a.s.},
\end{equation*}
and similarly $\BC_j^x(s-) \to C_j(s- \mid x)$ almost surely. Lemma~\ref{lem:abstractGC} and the key decomposition therefore give
\begin{align*}
    &\sup_{s \in [0, t]}|\BN_{jk}^x(s) - p_{jk}^c(s \mid x)| \to 0 \quad \BP\text{-a.s.}, \\
    &\sup_{s \in [0, t]}|\BI_j^x(s) - p_j^c(s \mid x)| \to 0 \quad \BP\text{-a.s.}
\end{align*}
Thus $(\BI_j^x,\BN_{jk}^x)$ converges jointly in the product uniform norm, almost surely.
Now consider the function $g:\BD_g \subseteq \BD \to \BF$ as specified further in Supplementary Section~\ref{sec:background} given as follows:
\begin{equation*}
    g(y, z) = \int_0^\cdot \frac{1}{y_-}\dd z.
\end{equation*}
As shown in Lemma~\ref{lem:gHD} and Remark~\ref{rmk:leftlimit}, $g$ is Hadamard differentiable on its domain and in particular continuous there, and $(\BI_j^x, \BN_{jk}^x)$ lies in this domain for all large $n$ almost surely. In our case, $y = p_j^c(\cdot \mid x)$, and $\inf_{s \in [0, t]}p_j^c(s \mid x) > 0$ if $\inf_{s \in [0, t]}p_j(s \mid x) > 0$ and $\inf_{s \in [0, t]}\BP(s < R \mid X = x) > 0$. The first holds by assumption and the second holds by the choice of $t$ since
\begin{equation*}
    \inf_{s \in [0, t]}\BP(s < R \mid X = x) = 1 - \sup_{s \in [0, t]}\BP(R \leq s \mid X = x) > 0.
\end{equation*}
Applying the continuous mapping theorem \citep[see Chapter 1.9 of][]{VaartWellnerS}, we have
\begin{align*}
    \mathbb{\Lambda}_{jk}(\cdot \mid x) &= \int_0^\cdot \frac{1}{\BI_j^x(s-)}\BN_{jk}^x(\dd s) = g(\BI_j^x, \BN_{jk}^x) \overset{\BP\text{-a.s.}}{\longrightarrow} g(p_j^c(\cdot \mid x), p_{jk}^c(\cdot \mid x)) \\
    &= \int_0^\cdot \frac{1}{p_j^c(s- \mid x)}p_{jk}^c(\dd s \mid x) = \Lambda_{jk}(\cdot \mid x)
\end{align*}
in the uniform norm on $[0, t]$. The absorbing rows vanish identically, and the diagonal entries are minus the sums of the off-diagonal entries. Since the state space is finite, coordinatewise uniform convergence implies convergence in the matrix norm.
\end{proof}

\begin{proof}[of Corollary~\ref{cor:occupationsingletree}]
Let $\phi$ denote the mapping from the set of uniformly bounded matrix functions on $[0, t]$ with entries of bounded variation to $\ell^\infty([0, t])$, the space of uniformly bounded functions on $[0, t]$, given by
\begin{equation*}
    \phi(A)(s) = \pint\limits_0^s (\text{Id} + A(\dd v)).
\end{equation*}
From Theorem 7 of~\citet{GillJohansenS}, we have uniform continuity of $\phi$ at $A$ with $A_n \to A$ when
\begin{equation*}
    \limsup_{n \to \infty} \| A_n\|_{\nu_1} < \infty \quad \text{and} \quad \sup_{s \in [0, t]}|A_n(s) - A(s)| \to 0
\end{equation*}
where $\|\cdot \|_{\nu_1}$ denotes the total variation norm over the interval $[0, t]$ given by
\begin{equation*}
    \|A \|_{\nu_1} = \sup_{0 = t_0 < \cdots < t_m = t} \sum_{i = 1}^m |A(t_i) - A(t_{i - 1})|
\end{equation*}
with $|B| = \max_j \sum_k |b_{jk}|$ for a matrix $B = \{b_{jk}\}_{j, k \in \CZ}$ and where the supremum is taken over all possible partitions of $[0, t]$. Both $\Lambda(\cdot \mid x)$ and $\mathbb{\Lambda}^{(n)}(\cdot \mid x)$ are non-decreasing in time in each coordinate $(j, k)$ with $j \neq k$, with diagonal entries equal to minus the row sums. Bounding the maximum over rows by the sum over rows in the definition of $\|\cdot\|_{\nu_1}$, we get
\begin{align*}
    \limsup_{n \to \infty}\|\mathbb{\Lambda}^{(n)}(\cdot \mid x) \|_{\nu_1} &\leq \limsup_{n \to \infty} \sum_{j \in \CZ}\ 2\sum_{\substack{k \in \CZ \\ k \neq j}} \mathbb{\Lambda}^{(n)}_{jk}(t \mid x) \\
    &= 2\sum_{j \in \CZ} \sum_{\substack{k \in \CZ \\ k \neq j}} \Lambda_{jk}(t \mid x) < \infty
\end{align*}
by the assumption $t \in (0, T_{R,x} \land T(x))$. The continuous mapping theorem \citep[see Chapter 1.9 of][]{VaartWellnerS} now yields
\begin{equation*}
    \sup_{s \in [0, t]}\left|\pint\limits_0^s (\text{Id} + \mathbb{\Lambda}^{(n)}(\dd v \mid x)) - \pint\limits_0^s (\text{Id} + \Lambda(\dd v \mid x)) \right| \to 0 \quad \BP\text{-a.s.}
\end{equation*}
as desired. The proof is complete once we verify that $\mathbb{p}^{(n)}(0 \mid x) \to p(0 \mid x)$ almost surely. From the argument of the proof of Theorem~\ref{thm:constree}, we have that $\mathbb{p}^{(n)}(0 \mid x) = \mathbb{p}^x(0)$ for large enough $n$ and so
\begin{equation*}
    \mathbb{p}_j^{(n)}(0 \mid x) = \frac{1}{n_x}\sum_{i:X^i = x} 1_{(Z_0^i = j)} = \frac{\frac{1}{n}\sum_{i = 1}^n 1_{(Z_0^i = j)}1_{(X^i = x)}}{\frac{1}{n}\sum_{i = 1}^n 1_{(X^i = x)}},
\end{equation*}
which converges almost surely to $p_j(0 \mid x)$. 
\end{proof}

\begin{proof}[of Theorem~\ref{thm:consbootstraptree}]
Let $A \subseteq \CX$ be non-empty, $j$ non-absorbing, $k \neq j$ and $t \in (0, T_{R,x} \land T(x))$. We have
\begin{equation*}
    \frac{1}{m_n}\sum_{i = 1}^{m_n} 1_{(X^{i,*} \in A, N_j^{\bullet,i,*} > 0)} = \frac{1}{n}\sum_{i = 1}^n 1_{(X^i \in A, N_j^{\bullet, i} > 0)} \tilde{W}_i
\end{equation*}
which converges to $\BP(X \in A, N_j^\bullet > 0) > 0$ by Lemma~\ref{lem:bootstrapestimatorweak} and Proposition~\ref{prop:subsampling}, in probability if $m_n \to \infty$ and almost surely if $m_n^{-2}$ is summable (Lemma~\ref{lem:bootstrapestimatorstrong}). Multiplying by $m_n \to \infty$, the number of observations in the bootstrap sample with covariate value in $A$ and an observed jump out of $j$ exceeds any fixed level with probability tending to one (eventually, almost surely). Taking $A = \{x\}$ for each $x \in \CX$ and arguing as in the proof of Theorem~\ref{thm:constree}, Assumption~\ref{asm:splitting} implies that every terminal node of the bootstrap tree contains a single covariate value, so that $\mathbb{\Lambda}^{(n), *}_{jk}(s \mid x) = \mathbb{\Lambda}_{jk}^{x, *}(s)$ for all $s \leq t$, with probability tending to one (eventually, almost surely). Thus, uniform weak (strong) consistency of $\mathbb{\Lambda}^{(n),*}(\cdot \mid x)$ follows from uniform weak (strong) consistency of $\mathbb{\Lambda}^{x, *}(\cdot)$. We have
\begin{equation*}
    n_x^* = \sum_{i = 1}^n W_i 1_{(X^i = x)},
\end{equation*}
and thus for a fixed $s \leq t$,
\begin{equation*}
    \BN_{jk}^{x,*}(s) = \frac{1}{n_x^*} \sum_{i = 1}^n N_{jk}^i(s \land R^i)W_i 1_{(X^i = x)} = \frac{\frac{1}{n}\sum_{i = 1}^n N_{jk}^i(s \land R^i) \tilde{W}_i 1_{(X^i = x)}}{\frac{1}{n}\sum_{i = 1}^n \tilde{W}_i 1_{(X^i = x)}}.
\end{equation*}
From the continuous mapping theorem and Lemma~\ref{lem:bootstrapestimatorweak} and Proposition~\ref{prop:subsampling}, if $m_n \to \infty$, this converges in probability to $p_{jk}^c(s \mid x)$. If $m_n^{-2}$ is summable, the convergence is almost sure. The same argument applies to $\BI_j^{x,*}(s)$. The same convergence holds at left limits and for the censoring contribution in Lemma~\ref{lem:keydecomposition}. Lemma~\ref{lem:abstractGC} therefore gives joint uniform convergence of the node estimators in the respective mode. Continuity of the Nelson--Aalen functional, as in the proof of Theorem~\ref{thm:constree}, completes the proof.
\end{proof}

\subsection{Single-tree Gaussian limits}\label{sec:proofs2}
We here present all the necessary background results leading up to the proof of Theorem~\ref{thm:weakconvergence}. To this end, we apply methods from empirical processes, see Supplementary Section~\ref{sec:background}. In the proofs, the Nelson--Aalen estimator grown on the data where the covariates only take the value $x$,
\begin{equation*}
    \CD_{n,x} = \{(X, (Z_t)_{0 \leq t \leq R}, \tau_{Z} \land R) \in \CD_n : X = x\},
\end{equation*}
is used throughout. Recall that $\CD_n$ denotes the full data set, see \eqref{eq:data}. For the stratified Nelson--Aalen estimator at $x$, define, for $j \neq k$,
\begin{equation*}
    \BN_{jk}^{(n),x}(s) := \frac{1}{n}\sum_{i = 1}^n N_{jk}^i(s \land R^i)1_{(X^i = x)}, \quad \BI_j^{(n),x}(s) := \frac{1}{n}\sum_{i = 1}^n 1_{(Z_s^i = j)}1_{(s < R^i)}1_{(X^i = x)}
\end{equation*}
for $s \leq t$ with $t \in (0, T(x) \land T_{R,x})$ fixed throughout. Then
\begin{equation*}
    \mathbb{\Lambda}_{jk}^{(n),x}(s) = \int_0^s \frac{1}{\BI_j^{(n),x}(u-)}\BN_{jk}^{(n),x}(\dd u).
\end{equation*}
Define the following empirical processes:
\begin{align*}
    (\xi_{jk}^{(n),x}(s))_{s \in [0, t]} &:= (\sqrt{n}(\BN_{jk}^{(n),x}(s) - p_xp^c_{jk}(s \mid x)))_{s \in [0, t]}, \quad j, k \in \CZ, \ j \neq k, \\
    (\xi_j^{(n),x}(s))_{s \in [0, t]} &:= (\sqrt{n}(\BI_j^{(n),x}(s) - p_xp_j^c(s \mid x)))_{s \in [0, t]}, \quad j \in \CZ, \\
    (\zeta_{jk}^{(n),x}(s))_{s \in [0, t]} &:= (\sqrt{n}(\mathbb{\Lambda}^{(n),x}_{jk}(s) - \Lambda_{jk}(s \mid x)))_{s \in [0, t]}, \quad j, k \in \CZ, \ j \neq k,
\end{align*}
where $p_x = \BP(X = x)$ as before. The strategy to prove weak convergence of the tree estimator $\mathbb{\Lambda}_{jk}^{(n)}(\cdot \mid x)$ goes as follows. Due to the discrete data assumption, a terminal node eventually contains only observations with a specific covariate combination. Hence it suffices to prove weak convergence of $\zeta_{jk}^{(n),x}$. To do so, we prove weak convergence of $\xi_{jk}^{(n),x}$ and $\xi_j^{(n),x}$. This is the content of the following propositions. The function classes belonging to $\xi_{jk}^{(n),x}$ and $\xi_j^{(n),x}$ are
\begin{align*}
    \CN_{jk}^x &:= \{f_{jk, s}:D[0, \infty) \times [0, \infty) \times \CX \to \{0, 1, 2, \dots\} : \\
    & \quad \quad f_{jk, s}(z, r, y) = 1_{(y = x)} \cdot \#\{u \in (0, s \land r] : z_{u-} = j, z_u = k\}, s \in [0, t]\}, \\
    \CI_j^x &:= \{g_{j,s}:D[0, \infty) \times [0, \infty) \times \CX \to \{0, 1\}: g_{j,s}(z, r, y) = 1_{(z_s = j)}1_{(s < r)}1_{(y = x)}, s \in [0, t] \},
\end{align*}
where $D[0,\infty)$ denotes the space of càdlàg paths on $[0,\infty)$.

\begin{proposition}[Weak convergence of $\xi_{jk}^{(n),x}$]\label{prop:NDonsker}
Fix $j,k\in\CZ$ with $j\neq k$ and suppose that $\BE[N_{jk}(t)^2]<\infty$. Then $\CN_{jk}^x$ is Donsker and
\begin{equation*}
    \xi_{jk}^{(n),x} \overset{d}{\longrightarrow} \xi_{jk}^x
\end{equation*}
where $\xi_{jk}^x$ is a tight mean-zero Gaussian process with covariance structure
\begin{equation*}
    \Cov(\xi_{jk}^x(s_1), \xi_{jk}^x(s_2)) = \Cov(N_{jk}(s_1 \land R)1_{(X = x)}, N_{jk}(s_2 \land R)1_{(X = x)})
\end{equation*}
for $s_1, s_2 \in [0, t]$.
\end{proposition}
\begin{proof}[of Proposition~\ref{prop:NDonsker}]
Define the function $F_{jk}^x(s) := \BE[N_{jk}(s \land R)N_{jk}(t \land R)1_{(X = x)}]$ for $s \in [0, t]$. $F_{jk}^x$ is càdlàg and non-decreasing from $F_{jk}^x(0) = 0$ to $F_{jk}^x(t) = \BE[N_{jk}(t \land R)^2 1_{(X = x)}]$. If $F_{jk}^x(t)=0$, every function in $\CN_{jk}^x$ vanishes almost surely and the conclusion is immediate. Otherwise, let $0 = s_0 < s_1 < \cdots < s_m = t$ denote a partition of $[0, t]$ and fix $\varepsilon \in (0, \sqrt{F_{jk}^x(t)}]$. We choose this sequence by dividing the second axis for $F_{jk}^x$ into intervals of length at most $\varepsilon^2$ and including all points where jumps of size larger than or equal to $\varepsilon^2$ occur. Then
\begin{equation*}
    F_{jk}^x(s_p-) - F_{jk}^x(s_{p - 1}) < \varepsilon^2, \quad p = 1, \dots, m.
\end{equation*}
Define brackets $[l_p, u_p]$ for $p = 1, \dots, m$ by
\begin{align*}
    l_p(z, r, y) &= 1_{(y = x)} \cdot \#\{u \in (0, s_{p - 1} \land r] : z_{u-} = j, z_u = k\}, \\
    u_p(z, r, y) &= 1_{(y = x)} \cdot \#\{u \in (0, s_p) \cap (0, r]: z_{u-} = j, z_u = k\}.
\end{align*}
Then $l_p \leq f_{jk, s} \leq u_p$ for $s \in [s_{p - 1}, s_p)$, so that the brackets $[l_p, u_p]$, $p = 1, \dots, m$, together with the degenerate bracket $[f_{jk, t}, f_{jk, t}]$ cover $\CN_{jk}^x$, and 
\begin{align*}
    \BE[(u_p(Z, R, X) - l_p(Z, R, X))^2]^{1/2} &\leq \BE[1_{(X = x)}N_{jk}(t \land R)(N_{jk}((s_p-) \land R) - N_{jk}(s_{p - 1} \land R))]^{1/2} \\
    &= (F_{jk}^x(s_p-) - F_{jk}^x(s_{p - 1}))^{1/2} < \varepsilon,
\end{align*}
so that the bracketing number satisfies $N_{[]}(\varepsilon, \CN_{jk}^x, L^2(\BP)) \leq \lceil C/\varepsilon^2 \rceil \leq C/\varepsilon^2 + 1$ where $C$ can be chosen as $C = 2F_{jk}^x(t)$. The 2 comes from the fact that we need at most one additional point for the $\varepsilon^2$ or larger jumps for each interval along the second axis of $F_{jk}^x$. Now,
\begin{align*}
    \int_0^\infty \sqrt{\log N_{[]}(\varepsilon, \CN_{jk}^x, L^2(\BP))}\dd \varepsilon &\leq \int_0^{\sqrt{F_{jk}^x(t)}} \sqrt{\log\Big(\frac{C}{\varepsilon^2} + 1 \Big)}\dd \varepsilon \\
    &+ \int_{\sqrt{F_{jk}^x(t)}}^\infty \sqrt{\log N_{[]}(\varepsilon, \CN_{jk}^x, L^2(\BP))}\dd \varepsilon,
\end{align*}
and if we can show that the right hand side is finite, the proof is complete by Theorem 2.5.6 in~\citet{VaartWellnerS}. The second integral is zero, since if $\varepsilon \geq \sqrt{F_{jk}^x(t)}$, we only need one bracket. As for the first,
\begin{align*}
    \int_0^{\sqrt{F_{jk}^x(t)}} \sqrt{\log\Big(\frac{C}{\varepsilon^2} + 1\Big)}\dd \varepsilon & \leq \int_0^{\sqrt{F_{jk}^x(t)}} \log\Big(\frac{C}{\varepsilon^2} + e \Big)\dd \varepsilon \\
    &= \Big[\varepsilon \log\Big(\frac{C}{\varepsilon^2} + e\Big) \Big]_0^{\sqrt{F_{jk}^x(t)}} + \int_0^{\sqrt{F_{jk}^x(t)}} \frac{2C}{C + e\varepsilon^2}\dd \varepsilon \\
    &\leq \sqrt{F_{jk}^x(t)}\log(2 + e) + 2\sqrt{F_{jk}^x(t)} = \sqrt{F_{jk}^x(t)}\big(2 + \log(2 + e)\big) < \infty,
\end{align*}
where we used that $C = 2F_{jk}^x(t)$. The proof is now complete.
\end{proof}

\begin{proposition}[Weak convergence of $\xi_j^{(n),x}$]\label{prop:IDonsker}
Suppose that $\BE[N_{jk}(t)^2] < \infty$ for all $j \neq k$. Then, for each $j\in\CZ$, the class $\CI_j^x$ is Donsker and
\begin{equation*}
    \xi_j^{(n),x} \overset{d}{\longrightarrow} \xi_j^x
\end{equation*}
where $\xi_j^x$ is a tight mean-zero Gaussian process with covariance structure
\begin{equation*}
    \Cov(\xi_j^x(s_1), \xi_j^x(s_2)) = \Cov(1_{(Z_{s_1} = j)}1_{(s_1 < R)}1_{(X = x)}, 1_{(Z_{s_2} = j)}1_{(s_2 < R)}1_{(X = x)})
\end{equation*}
for $s_1, s_2 \in [0, t]$.
\end{proposition}
\begin{proof}[of Proposition~\ref{prop:IDonsker}]
We apply the key decomposition as described in Lemma~\ref{lem:keydecomposition}. Consider the function class
\begin{equation*}
    \CC_j^x := \{h_{j,s} : D[0, \infty) \times [0, \infty) \times \CX \to \{0, 1\}: h_{j, s}(z, r, y) = 1_{(r \leq s)}1_{(z_r = j)}1_{(y = x)}, s \in [0, t]\}
\end{equation*}
which we shall prove is Donsker. Let $0 = s_0 < s_1 < \cdots < s_m = t$ denote a partition of $[0, t]$ and consider the functions
\begin{equation*}
    l_p(z, r, y) = 1_{(r \leq s_{p - 1})}1_{(z_r = j)}1_{(y = x)}, \quad u_p(z, r, y) = 1_{(r < s_p)}1_{(z_r = j)}1_{(y = x)}, \quad p = 1, \dots, m.
\end{equation*}
The brackets $[l_p, u_p]$ for $p = 1, \dots, m$, together with the degenerate bracket $[h_{j, t}, h_{j, t}]$, cover $\CC_j^x$. Let $\varepsilon \in (0, 1]$ be arbitrary. Defining $C_j^x(s) = \BP(R \leq s, Z_R = j, X = x)$, we choose the sequence $s_0, s_1, \dots, s_m$ in such a way that all jumps of $C_j^x$ in $[0, t]$ of size larger than or equal to $\varepsilon^2$ are included and
\begin{equation*}
    C_j^x(s_p-) - C_j^x(s_{p - 1}) < \varepsilon^2.
\end{equation*}
This is possible since $C_j^x$ is càdlàg and non-decreasing. Then
\begin{equation*}
    \BE[(u_p(Z, R, X) - l_p(Z, R, X))^2]^{1/2} \leq (C_j^x(s_p-) - C_j^x(s_{p - 1}))^{1/2} < \varepsilon,
\end{equation*}
and thus $N_{[]}(\varepsilon, \CC_j^x, L^2(\BP)) \leq \lceil 2/\varepsilon^2 \rceil \leq 2/\varepsilon^2 + 1$. This is possible since we may first choose a sequence that partitions the second axis of $C_j^x$ into intervals of length at most $\varepsilon^2$ and then include the possible jumps. This yields at most two points for each of the (at most) $\lceil 1/\varepsilon^2 \rceil$ intervals. Now,
\begin{equation*}
    \int_0^\infty \sqrt{\log N_{[]}(\varepsilon, \CC_j^x, L^2(\BP))}\dd \varepsilon \leq \int_0^1 \sqrt{\log\Big(\frac{2}{\varepsilon^2} + 1 \Big)}\dd \varepsilon + \int_1^\infty \sqrt{\log N_{[]}(\varepsilon, \CC_j^x, L^2(\BP))}\dd \varepsilon.
\end{equation*}
We want to show that the right hand side is finite. Since each $h_{j, s} \in \CC_j^x$ is bounded by one, we have $N_{[]}(\varepsilon, \CC_j^x, L^2(\BP)) = 1$ for $\varepsilon \geq 1$, so the second integral is zero. As for the first, we repeat the calculations from the proof of Proposition~\ref{prop:NDonsker} and get
\begin{equation*}
    \int_0^1 \sqrt{\log\Big(\frac{2}{\varepsilon^2} + 1\Big)}\dd \varepsilon\leq 2 + \log(2 + e) < \infty.
\end{equation*}
By Theorem 2.5.6 in~\citet{VaartWellnerS}, $\CC_j^x$ is Donsker as claimed. It follows that the empirical process $\eta_j^{(n),x}$ belonging to
\begin{equation*}
    \BC_j^{(n), x}(s) := \frac{1}{n}\sum_{i = 1}^n 1_{(R^i \leq s)}1_{(Z^i_{R^i} = j)}1_{(X^i = x)}
\end{equation*}
converges weakly to a tight mean-zero Gaussian process. Proposition~\ref{prop:NDonsker} gives weak convergence of $\xi_{jk}^{(n),x}$, and the same holds for the empirical process corresponding to
\begin{equation*}
    \BI_j^{(n),x}(0) := \frac{1}{n}\sum_{i = 1}^n 1_{(Z_0^i = j)}1_{(X^i = x)}
\end{equation*}
by the classical central limit theorem. For any finite collection of time points, the centred and scaled initial-state indicators, censoring counts and transition counts form a vector of averages of iid random variables with finite second moments. The multivariate central limit theorem therefore gives joint finite-dimensional convergence. Together with marginal asymptotic tightness, this yields joint weak convergence of the centred empirical processes \citep[Section 1.5]{VaartWellnerS}. The continuous mapping theorem applied to the linear identity in Lemma~\ref{lem:keydecomposition} now gives the asserted limit for $\xi_j^{(n),x}$ and its covariance function.
\end{proof}

The following proposition combines the results of the two previous propositions into a joint weak convergence statement.

\begin{proposition}[Joint weak convergence of components]\label{prop:jointweakconvergence}
Suppose that $\BE[N_{jk}(t)^2]<\infty$ for all $j\neq k$. Then
\begin{equation*}
    (\xi_{jk}^{(n), x}, \xi_j^{(n),x})_{j, k \in \CZ, j \neq k} \overset{d}{\longrightarrow} (\xi_{jk}^x, \xi_j^x)_{j, k \in \CZ, j \neq k},
\end{equation*}
where the limit is a tight mean-zero Gaussian process with the following covariance structure:
\begin{align*}
    \Cov(\xi_{jk}^x(s_1), \xi_{lm}^x(s_2)) &= \Cov(N_{jk}(s_1 \land R)1_{(X = x)}, N_{lm}(s_2 \land R)1_{(X = x)}), \\
    \Cov(\xi_{jk}^x(s_1), \xi_l^x(s_2)) &= \Cov(N_{jk}(s_1 \land R)1_{(X = x)}, 1_{(Z_{s_2} = l)} 1_{(s_2 < R)}1_{(X = x)}), \\
    \Cov(\xi_j^x(s_1), \xi_l^x(s_2)) &= \Cov(1_{(Z_{s_1} = j)}1_{(s_1 < R)}1_{(X = x)}, 1_{(Z_{s_2} = l)}1_{(s_2 < R)}1_{(X = x)}),
\end{align*}
where $s_1, s_2 \leq t$. 
\end{proposition}

\begin{proof}[of Proposition~\ref{prop:jointweakconvergence}]
Propositions~\ref{prop:NDonsker} and \ref{prop:IDonsker} give marginal asymptotic tightness. Joint finite-dimensional convergence follows from the multivariate central limit theorem applied to the centred observation vectors at any finite collection of time points. Their covariances are the displayed expressions, which identify the joint Gaussian limit.
\end{proof}

\begin{proof}[of Theorem~\ref{thm:weakconvergence}]
By the same arguments as in Theorem~\ref{thm:constree}, there exists an almost surely finite $N(x)$ such that for $n \geq N(x)$, we have $\mathbb{\Lambda}^{(n)}_{jk}(s \mid x) = \mathbb{\Lambda}_{jk}^{(n),x}(s)$ for all $s \leq t$ and $n_{h(x)} = n_x$. Hence it suffices to study the limit behaviour of 
\begin{equation}\label{eq:process_nx}
    \sqrt{n_x}(\mathbb{\Lambda}^{(n),x}_{jk}(\cdot) - \Lambda_{jk}(\cdot \mid x)).
\end{equation}
By the strong law of large numbers, $n_x/n \to p_x$ almost surely, and hence by Slutsky's lemma, the limit behaviour of \eqref{eq:process_nx} is equivalent to that of
\begin{equation*}
    \sqrt{n p_x}(\mathbb{\Lambda}^{(n),x}_{jk}(\cdot) - \Lambda_{jk}(\cdot \mid x)) = \sqrt{p_x}\zeta_{jk}^{(n),x}(\cdot).
\end{equation*}
We apply the result of Proposition~\ref{prop:jointweakconvergence} together with the functional delta method, see Theorem 3.10.4 in~\citet{VaartWellnerS}. The function $g:\BD_g \to \BF$ given by
\begin{equation*}
    g(y, z) = \int_0^\cdot \frac{1}{y_-} \dd z
\end{equation*}
considered in the proof of Theorem~\ref{thm:constree} is Hadamard differentiable in $(y, z)$ with derivative
\begin{equation*}
    g'_{(y, z)}(h_1, h_2)(\cdot) = \int_0^\cdot \frac{1}{y(s-)}h_2(\dd s) - \int_0^\cdot \frac{h_1(s-)}{y(s-)^2} z(\dd s),
\end{equation*}
where the first integral is defined through integration by parts, by Lemma~\ref{lem:gHD} and Remark~\ref{rmk:leftlimit}. Hence by the functional delta method and Proposition~\ref{prop:jointweakconvergence},
\begin{align*}
    \zeta_{jk}^{(n),x} &= \sqrt{n}(g(\BI_j^{(n),x}, \BN_{jk}^{(n),x}) - g(p_j^c(\cdot \mid x)p_x, p_{jk}^c(\cdot \mid x)p_x)) \overset{d}{\longrightarrow} g'_{(p_j^c(\cdot \mid x)p_x, p_{jk}^c(\cdot \mid x)p_x)}(\xi_j^x, \xi_{jk}^x) \\
    &= \int_0^\cdot \frac{1}{p_j^c(s- \mid x)p_x}\xi_{jk}^x(\dd s) - \int_0^\cdot \frac{1}{p_j^c(s- \mid x)^2p_x^2} \xi_j^x(s-)p_x p_{jk}^c(\dd s \mid x) \\
    &= \frac{1}{p_x}\left(\int_0^\cdot \frac{1}{p_j^c(s- \mid x)}\xi_{jk}^x(\dd s) - \int_0^\cdot \frac{\xi_j^x(s-)}{p_j^c(s- \mid x)}\Lambda_{jk}(\dd s \mid x) \right)
\end{align*}
where the integral against $\xi_{jk}^x$ is defined through integration by parts as in Remark~\ref{rmk:leftlimit}. The result of the theorem now follows by Slutsky's lemma.
\end{proof}

\subsection{Conditional bootstrap limits}\label{subsection:bootstraplimitproof}
\begin{proof}[of Theorem~\ref{thm:weakconvergencebootstrap}(i)]
Throughout, the data are fixed and only the bootstrap weights are random; all statements hold for almost every data sequence. For a function $f \geq 0$ of a single observation with $\BE[f(Y)^2] < \infty$, the bootstrap average $\frac{1}{n}\sum_{i = 1}^n \tilde{W}_i f(Y_i)$ has conditional mean $\frac{1}{n}\sum_{i = 1}^n f(Y_i) \to \BE[f(Y)]$ by the strong law of large numbers and conditional variance
\begin{equation*}
    \frac{\Var(\tilde{W}_1)}{n}\cdot\frac{1}{n}\sum_{i = 1}^n f(Y_i)^2 + \Cov(\tilde{W}_1, \tilde{W}_2)\cdot\frac{1}{n^2}\sum_{i \neq l} f(Y_i)f(Y_l) \longrightarrow 0
\end{equation*}
almost surely, since the first term converges to zero by Assumption~\ref{asm:weakbootstrap}(i) and the strong law of large numbers, while the second is non-positive by Proposition~\ref{prop:subsampling}. Hence the bootstrap average converges to $\BE[f(Y)]$ in conditional probability for almost every data sequence. Applied with $f = 1_{(X = x, N_j^\bullet > 0)}$ and arguing as in the proof of Theorem~\ref{thm:consbootstraptree}, the terminal node $h(x)$ of the bootstrap tree contains only the covariate value $x$ with conditional probability tending to one, and on this event
\begin{equation*}
    n_{h(x)}^* = n_x^* = \sum_{i = 1}^n 1_{(X^i = x)}W_i.
\end{equation*}
Applied with $f = 1_{(X = x)}$, we get $n_{h(x)}^*/(m_n p_x) \to 1$ in conditional probability. Hence it suffices to consider the limit results with $\mathbb{\Lambda}^{(n),*,x}_{jk}$ and $\mathbb{\Lambda}^{(n),x}_{jk}$ instead of $\mathbb{\Lambda}^{(n),*}(\cdot \mid x)$ and $\mathbb{\Lambda}^{(n)}(\cdot \mid x)$ and with $m_n$ instead of $n_{h(x)}^*$. Afterwards, we scale by $\sqrt{p_x}$ as was done in the proof of Theorem~\ref{thm:weakconvergence}. It remains to verify the requirements of Theorem 3.7.13 in~\citet{VaartWellnerS} for both the $\BN$ and $\BI$ components and both bootstrap schemes \citep[see also][for background on the bootstrap results used]{PraestgaardWellnerS}. We know that the function class
\begin{equation*}
    \{(f_{jk, s}, g_{j, s})_{j, k \in \CZ, j \neq k} : f_{jk, s} \in \CN_{jk}^x, g_{j, s} \in \CI_j^x, s \in [0, t]\}
\end{equation*}
is Donsker by Proposition~\ref{prop:jointweakconvergence}. Furthermore, by the moment assumption on $N_{jk}(t)$, this function class has an envelope function with finite $L^2(\BP)$-norm. For this particular proof, let
\begin{align*}
    \BN_{jk}^{(n),*,x}(s) &:= \frac{1}{m_n}\sum_{i = 1}^n W_i N_{jk}^i(s \land R^i)1_{(X^i = x)}, \\ \BI_j^{(n),*,x}(s) &:= \frac{1}{m_n}\sum_{i = 1}^n W_i 1_{(Z_s^i = j)}1_{(s < R^i)}1_{(X^i = x)}.
\end{align*}
Now consider the bootstrap empirical processes
\begin{align*}
    (\xi^{(n),*,x}_{jk}(s))_{s \in [0, t]} &:= (\sqrt{m_n}(\BN_{jk}^{(n),*,x}(s) - \BN_{jk}^{(n),x}(s)))_{s \in [0, t]}, \quad j, k \in \CZ, j \neq k, \\
    (\xi_j^{(n),*,x}(s))_{s \in [0, t]} &:= (\sqrt{m_n}(\BI_j^{(n),*,x}(s) - \BI_j^{(n),x}(s)))_{s \in [0, t]}, \quad j \in \CZ.
\end{align*}
We need to incorporate the multinomial subsampling and subsampling without replacement schemes into the framework of Theorem 3.7.13 in~\citet{VaartWellnerS}. Starting with multinomial subsampling, as is done in Example 3.7.10 in~\citet{VaartWellnerS}, consider the weight vector $(\hat{W}_1, \dots, \hat{W}_n)$ equal to $\sqrt{n/m_n}$ times $(W_1, \dots, W_n)$. Let $\bar{W}_n$ denote the empirical average of the weights $\hat{W}_1, \dots, \hat{W}_n$ and note that $\bar{W}_n = \sqrt{m_n/n}$. With this choice of weights, we get
\begin{equation*}
    \xi_{jk}^{(n),*,x}(s) = \sqrt{n}\left(\frac{1}{n}\sum_{i = 1}^n \hat{W}_i N_{jk}^i(s \land R^i) 1_{(X^i = x)} - \frac{\bar{W}_n}{n}\sum_{i = 1}^n N_{jk}^i(s \land R^i)1_{(X^i = x)} \right)
\end{equation*}
which coincides with the weighted bootstrap empirical process in Section 3.7.2 of~\citet{VaartWellnerS} for the function class $\CN_{jk}^x$. The same calculation applies to the process $\xi_j^{(n),*,x}$. The weights $(\hat{W}_1, \dots, \hat{W}_n)$ satisfy the requirements in (3.7.8) of Theorem 3.7.13, and we may conclude that
\begin{equation*}
    (\xi_{jk}^{(n),*,x}, \xi_j^{(n),*,x})_{j, k \in \CZ, j \neq k} \overset{d}{\longrightarrow} (\xi_{jk}^x, \xi_j^x)
\end{equation*}
for almost every fixed sequence of data as specified in the theorem. To conclude the proof in the case of multinomial subsampling, we need only apply an almost sure conditional delta-method. To this end, we apply Theorem 3.10.13 of \citet{VaartWellnerS}. All the conditions in this theorem are satisfied, including the uniform Hadamard differentiability (3.10.12), which is the second statement of Lemma~\ref{lem:gHD}. 

For subsampling without replacement, define the weights $(\hat{W}_1, \dots, \hat{W}_n)$ as a row of $m_n$ times the number $n/\sqrt{m_n(n - m_n)}$ and $n - m_n$ times the number 0 ordered at random. In this case, $\bar{W}_n = \sqrt{m_n/(n - m_n)}$, and elementary calculations show that
\begin{equation*}
    \xi_{jk}^{(n),*,x}(s) = \sqrt{\frac{n - m_n}{n}} \sqrt{n}\left(\frac{1}{n}\sum_{i = 1}^n \hat{W}_i N_{jk}^i(s \land R^i)1_{(X^i = x)} - \frac{\bar{W}_n}{n}\sum_{i = 1}^n N_{jk}^i(s \land R^i)1_{(X^i = x)} \right)
\end{equation*}
which equals the weighted bootstrap empirical process from~\citet{VaartWellnerS} up to multiplication by the factor $\sqrt{(n - m_n)/n}$, which converges to $\sqrt{1 - \alpha} = \sigma_\alpha$ by assumption. The same computation applies to $\xi_j^{(n),*,x}$. Since $\alpha < 1$, also $n - m_n \to \infty$, so that the conditions in (3.7.8) are satisfied, and the joint limit of $(\xi_{jk}^{(n),*,x}, \xi_j^{(n),*,x})_{j, k \in \CZ, j \neq k}$ is $\sigma_\alpha (\xi_{jk}^x, \xi_j^x)_{j, k \in \CZ, j \neq k}$, conditionally on the data, for almost every sequence of data. Applying Theorem 3.10.13 of \citet{VaartWellnerS} once again, and using that the Hadamard derivative is linear, yields the limit $\sigma_\alpha \zeta_{jk}^x$ and completes the proof. 
\end{proof}

\begin{proof}[of Theorem~\ref{thm:weakconvergencebootstrap}(ii)]
Let $n_{h_b(x)}^*$ denote the number of observations in the terminal node belonging to $x$ for the $b$th tree. Since the sets (triangular arrays) of bootstrap weights for each tree are independent, the vector
\begin{equation*}
    \Big(\sqrt{n_{h_1(x)}^*}\big(\mathbb{\Lambda}_{jk, 1}^{(n),*}(\cdot \mid x) - \mathbb{\Lambda}_{jk}^{(n)}(\cdot \mid x) \big), \dots, \sqrt{n_{h_B(x)}^*}\big(\mathbb{\Lambda}_{jk, B}^{(n),*}(\cdot \mid x) - \mathbb{\Lambda}_{jk}^{(n)}(\cdot \mid x) \big)\Big)
\end{equation*}
converges weakly conditionally on the data to $\sigma_\alpha(\zeta_{jk, 1}^x, \dots, \zeta_{jk, B}^x)$, where $\zeta_{jk, 1}^x, \dots, \zeta_{jk, B}^x$ are iid copies of $\zeta_{jk}^x$. By part (i), $n_{h_b(x)}^*/(m_n p_x)\to1$ in conditional probability for almost every data sequence, for each $b$. Conditional Slutsky convergence therefore gives 
\begin{equation*}
    \sqrt{m_n}\Big(\mathbb{\Lambda}_{jk, 1}^{(n),*}(\cdot \mid x) - \mathbb{\Lambda}_{jk}^{(n)}(\cdot \mid x), \dots, \mathbb{\Lambda}_{jk, B}^{(n),*}(\cdot \mid x) - \mathbb{\Lambda}_{jk}^{(n)}(\cdot \mid x) \Big) \overset{d}{\longrightarrow} \frac{\sigma_\alpha}{\sqrt{p_x}}(\zeta_{jk, 1}^x, \dots, \zeta_{jk, B}^x)
\end{equation*}
conditionally on the data. Apply the continuous mapping theorem to conclude
\begin{equation*}
    \sqrt{m_n}(\mathbb{\Lambda}_{\operatorname{RF}, jk}^{(n)}(\cdot \mid x) - \mathbb{\Lambda}_{jk}^{(n)}(\cdot \mid x)) \overset{d}{\longrightarrow} \frac{\sigma_\alpha}{\sqrt{p_x}B} \sum_{b = 1}^B \zeta_{jk, b}^x
\end{equation*}
conditionally on the data. Now, since the components $\zeta_{jk, b}^x$ are iid mean-zero Gaussian processes, the normalised sum $\frac{1}{\sqrt{B}}\sum_{b = 1}^B \zeta_{jk, b}^x$ has the same distribution as $\zeta_{jk}^x$, and we conclude that, conditionally on the data,
\begin{equation*}
    \sqrt{B m_n} (\mathbb{\Lambda}_{\operatorname{RF}, jk}^{(n)}(\cdot \mid x) - \mathbb{\Lambda}_{jk}^{(n)}(\cdot \mid x)) \overset{d}{\longrightarrow} \frac{\sigma_\alpha}{\sqrt{p_x}}\zeta_{jk}^x
\end{equation*}
as desired.
\end{proof}

\subsection{Markov variance and transition probabilities}\label{subsection:weakMarkov}
\begin{proof}[of Proposition~\ref{prop:markovvariance}]
Throughout the proof, let $s \in [0, t]$. Define
\begin{equation*}
    U_{jk}^x(s) := N_{jk}(s \land R)1_{(X = x)} - \int_0^s 1_{(Z_{u-} = j)}1_{(u \leq R)}1_{(X = x)}\Lambda_{jk}(\dd u \mid x)
\end{equation*}
and let $U_{jk}^{x,1}, \dots, U_{jk}^{x,n}$ denote $n$ iid copies of $U_{jk}^x$. Then
\begin{align*}
    \mathbb{U}_{jk}^{(n),x}(s) &:= \frac{1}{\sqrt{n}}\sum_{i = 1}^n U_{jk}^{x,i}(s) = \sqrt{n}\Big(\BN_{jk}^{(n),x}(s) - \int_0^s \BI_j^{(n),x}(u-) \Lambda_{jk}(\dd u \mid x)\Big) \\
    &= \xi_{jk}^{(n),x}(s) - \int_0^s \xi_j^{(n),x}(u-)\Lambda_{jk}(\dd u \mid x).
\end{align*}
From Proposition~\ref{prop:jointweakconvergence}, it holds that 
\begin{equation*}
    \mathbb{U}_{jk}^{(n),x} \overset{d}{\longrightarrow} \mathbb{U}_{jk}^x := \xi_{jk}^x - \int_0^\cdot \xi_j^x(u-)\Lambda_{jk}(\dd u \mid x)
\end{equation*}
uniformly in $[0, t]$. Moreover,
\begin{equation*}
    \zeta_{jk}^x(s) = \frac{1}{\sqrt{p_x}}\int_0^s \frac{1}{p_j^c(u- \mid x)}\mathbb{U}_{jk}^x(\dd u).
\end{equation*}
We now apply the Markov assumption together with Assumption~\ref{asm:rrc}, which ensures that $Z$ remains Markov with the same transition rates with respect to the enlarged filtration $\CG_t := \sigma(X, Z_s, 1_{(s \leq R)} : s \leq t)$, to conclude that the $\CG$-compensator of $N_{jk}(\cdot \land R)1_{(X = x)}$ is
\begin{equation*}
    s \mapsto \int_0^s 1_{(Z_{u-} = j)}1_{(u \leq R)}1_{(X = x)}\Lambda_{jk}(\dd u \mid x).
\end{equation*}
See~\citet{ABGKS} for details, in particular Chapter III.2. It follows from (2.4.2) of~\citet{ABGKS} that the predictable variation process of $U_{jk}^x$ is
\begin{equation*}
    \langle U_{jk}^x \rangle(s) = \int_0^s 1_{(Z_{u-} = j)}1_{(u \leq R)}1_{(X = x)}(1 - \Delta \Lambda_{jk}(u \mid x))\Lambda_{jk}(\dd u \mid x)
\end{equation*}
and hence that the variance of $U_{jk}^x(s)$ is
\begin{equation*}
    \Var(U_{jk}^x(s)) = \int_0^s p_x p_j^c(u- \mid x)(1 - \Delta\Lambda_{jk}(u \mid x))\Lambda_{jk}(\dd u \mid x)
\end{equation*}
which coincides with the variance of $\mathbb{U}_{jk}^{(n),x}(s)$. Also, since $U_{jk}^x$ is a $\CG$-martingale, we have for $0 \leq s_1 < s_2 \leq s_3 < s_4 \leq t$ that
\begin{equation*}
    \Cov(U_{jk}^x(s_2) - U_{jk}^x(s_1), U_{jk}^x(s_4) - U_{jk}^x(s_3)) = \BE[(U_{jk}^x(s_2) - U_{jk}^x(s_1))\BE[U_{jk}^x(s_4) - U_{jk}^x(s_3) \mid \CG_{s_3}]] = 0. \\
\end{equation*}
The multivariate central limit theorem identifies the covariance of any finite vector of increments of $\mathbb{U}_{jk}^x$ with that of the corresponding increments of $U_{jk}^x$. Thus $\mathbb{U}_{jk}^x$ also has orthogonal increments. Now let $g = \sum_{l = 1}^m a_l 1_{I_l}$ be a step function where $I_1, \dots, I_m$ are disjoint intervals. Then if $v_{jk}^x$ denotes the variance measure of $U_{jk}^x$,
\begin{equation*}
    \Var\left(\int g(u) \mathbb{U}_{jk}^x(\dd u) \right) = \sum_{l = 1}^m a_l^2 v_{jk}^x(I_l) = \int g(u)^2 v_{jk}^x(\dd u).
\end{equation*}
The integrand $u \mapsto 1_{(0, s]}(u)/p_j^c(u- \mid x)$ is càglàd, bounded and of bounded variation, hence the uniform limit of step functions $g_m$ of the above form with uniformly bounded variation. By the isometry just established, $\Var(\int (g_m - g_l)\dd \mathbb{U}_{jk}^x) = \int (g_m - g_l)^2 \dd v_{jk}^x \to 0$, so the integrals $\int g_m \dd \mathbb{U}_{jk}^x$ converge in $L^2$. Their limit coincides with the integral defined through integration by parts, since both definitions agree for step functions and are continuous along such sequences (Lemma~\ref{lem:regulated}), and the variances converge, which yields the result.

For the second claim, note that in the proof of Theorem~\ref{thm:constree}, it was established that
\begin{equation*}
    \lim_{n \to \infty} \sup_{s \in [0, t]}|\BN_{jk}^h(s) - p_{jk}^c(s \mid x)| = 0 \quad \text{and} \quad \lim_{n \to \infty} \sup_{s \in [0, t]}|\BI_{j}^h(s) - p_j^c(s \mid x)| = 0 \quad \BP\text{-a.s.}
\end{equation*}
Moreover, $\mathbb{V}_{jk}^{(n)}(\cdot \mid x) = \int_0^\cdot \varphi(\BI_j^h(u-))\BN_{jk}^h(\dd u)$ with $\varphi(u) = u^{-2}$, which is a map of the form considered in Remark~\ref{rmk:leftlimit}. With the constants $c_1$ and $c_2$ given there, $\varphi$ is twice continuously differentiable on $[c_1, c_2]$, so that Lemma~\ref{lem:gHD} and Remark~\ref{rmk:leftlimit} show that this map is continuous at $(p_j^c(\cdot \mid x), p_{jk}^c(\cdot \mid x))$, and $(\BI_j^h, \BN_{jk}^h)$ lies in its domain for all large $n$, almost surely. The second claim now follows from the continuous mapping theorem, since $V_{jk}(s \mid x) = \int_0^s p_j^c(u- \mid x)^{-2}p_{jk}^c(\dd u \mid x)$ when $\Lambda_{jk}(\cdot \mid x)$ is continuous.
\end{proof}

\begin{proof}[of Theorem~\ref{thm:weakconvergencetransition}]
By Proposition~\ref{prop:jointweakconvergence} and the functional delta method applied jointly to all pairs $(j, k)$, exactly as in the proof of Theorem~\ref{thm:weakconvergence}, the matrix-valued process $\sqrt{n_{h(x)}}(\mathbb{\Lambda}^{(n)}(\cdot \mid x) - \Lambda(\cdot \mid x))$ converges weakly to the matrix $\zeta^x = (\zeta^x_{jk})_{j, k \in \CZ}$ with $\zeta^x_{jj} := -\sum_{k \neq j}\zeta^x_{jk}$. By Theorem 8 of \citet{GillJohansenS}, the map $A \mapsto (\pint_{s_1}^{s_2}(\text{Id} + A(\dd v)))_{0 \leq s_1 \leq s_2 \leq t}$ is Hadamard differentiable at $A = \Lambda(\cdot \mid x)$, tangentially to the càdlàg matrix functions, on the set of matrix functions of total variation bounded by a fixed constant, with derivative $h \mapsto (\int_{s_1}^{s_2} p(s_1, u- \mid x)h(\dd u)p(u, s_2 \mid x))_{s_1 \leq s_2}$, the integral being defined through integration by parts. Since $\mathbb{\Lambda}^{(n)}(\cdot \mid x)$ has total variation bounded by a fixed constant for all large $n$ (see the proof of Corollary~\ref{cor:occupationsingletree}), the functional delta method together with \eqref{eq:Kolmogorov} yields the claim.
\end{proof}

\clearpage

\section{Auxiliary results}\label{sec:background}

\subsection{Bootstrap averages}\label{subsection:bootstrap}

When growing a random forest, each tree is grown on its own bootstrap sample of the original data. In order to establish asymptotic results for the random forest, we thus need to consider a sufficiently general mathematical formulation of the bootstrap. Suppose $Y_1, \dots, Y_n$ denotes a generic iid data sample. Each observation in the data is then given a bootstrap weight $W_i$, and we assume that the full vector of weights $(W_1, \dots, W_n)$ is non-negative, exchangeable and independent of $Y_1, \dots, Y_n$. In our treatment of the Nelson--Aalen estimator, we need only consider (functionals of) estimators of the form
\begin{equation*}
    T^{(n)} = \frac{1}{n}\sum_{i = 1}^n f(Y_i)
\end{equation*}
where $f$ is a known measurable function. In the following, we may without loss of generality assume that $f$ is the identity, so that $T^{(n)} = (\sum_{i = 1}^n Y_i)/n$. We then define the bootstrap version of $T^{(n)}$ by
\begin{equation*}
    T^{(n),*} := \frac{1}{n\BE[W_1]}\sum_{i = 1}^n W_i Y_i.
\end{equation*}
The reason for the $\BE[W_1]$ in the denominator is that $n\BE[W_1]$ is the (expected) number of elements in the sample. This will be exemplified via classical bootstrap schemes below. To simplify our analysis and lighten the notation, we will work with \emph{normalised bootstrap weights}
\begin{equation*}
    \tilde{W}_i := \frac{W_i}{\BE[W_1]}
\end{equation*}
so that $T^{(n),*}$ simplifies to
\begin{equation*}
    T^{(n),*} = \frac{1}{n}\sum_{i = 1}^n Y_i \tilde{W}_i.
\end{equation*}
\begin{example}[Multinomial subsampling]\label{ex:Efron}
If we subsample $m_n \leq n$ observations (note that $m_n$ is allowed to depend on $n$) from the original data with replacement, the weights $(W_1, \dots, W_n)$ are multinomial with $m_n$ trials and probabilities $(1/n, \dots, 1/n)$. Note that $\sum_{i = 1}^n W_i = m_n$ and that $n\BE[W_1] = m_n$, which concurs with the interpretation that $n\BE[W_1]$ is the expected number of observations in the sample (here it is the exact number of observations).
\end{example}

\begin{example}[Subsampling without replacement]\label{ex:woreplacment}
Subsampling $m_n \leq n$ observations without replacement from the original data means that $W_i$ is binary with $W_i = 1$ when observation $i$ is in the bootstrap sample and $W_i = 0$ otherwise. As for multinomial subsampling, the subsample size is exactly $m_n = n\BE[W_1]$.
\end{example}

The rest of this subsection is dedicated to stating requirements on the weights and the original data which ensure that consistency of $T^{(n)}$ is transferred to $T^{(n),*}$. 

\begin{assumption}[Weak consistency requirements on bootstrap weights]\label{asm:weakbootstrap}
The bootstrap weights $W_1, \dots, W_n$ satisfy the following:
\begin{enumerate}
    \item[(i)] $\Var(\tilde{W}_1)/n \to 0$ (in particular, $\BE[W_1^2] < \infty$),
    \item[(ii)] $\Cov(\tilde{W}_1, \tilde{W}_2) \to 0$ or $\Cov(\tilde{W}_1, \tilde{W}_2) \leq 0$. 
\end{enumerate}
\end{assumption}

All the bootstrap schemes presented above satisfy these assumptions and the stronger conditions stated below, as verified in the subsequent proofs. The following lemma establishes $L^2$-consistency under Assumption~\ref{asm:weakbootstrap}.

\begin{lemma}[Weak consistency of bootstrap averages]\label{lem:bootstrapestimatorweak}
Suppose $Y_1, \dots, Y_n$ is an iid sample with $\BE[Y_1^2] < \infty$ independent of the bootstrap weights $W_1, \dots, W_n$. If the weights satisfy Assumption~\ref{asm:weakbootstrap}, then
\begin{equation*}
    T^{(n), *} \overset{L^2}{\longrightarrow} \BE[Y_1],
\end{equation*}
and in particular, $T^{(n), *} \overset{\BP}{\longrightarrow} \BE[Y_1]$.
\end{lemma}
\begin{proof}
The law of large numbers for $L^p$-convergence gives $T^{(n)} \overset{L^2}{\to} \BE[Y_1]$; it therefore suffices to show that $T^{(n), *} - T^{(n)} \overset{L^2}{\to} 0$. We compute
\begin{align*}
    \BE[(T^{(n), *} - T^{(n)})^2] &= \frac{1}{n^2}\BE\Big[\Big( \sum_{i = 1}^n Y_i(\tilde{W}_i - 1)\Big)^2 \Big] \\
    &= \frac{1}{n^2}\BE\Big[\sum_{i = 1}^n Y_i^2(\tilde{W}_i - 1)^2 + \sum_{i \neq j} Y_iY_j (\tilde{W}_i - 1)(\tilde{W}_j - 1) \Big] \\
    &= \BE[Y_1^2] \frac{\BE[(\tilde{W}_1 - 1)^2]}{n} + \frac{n - 1}{n} \BE[Y_1]^2 \BE[(\tilde{W}_1 - 1)(\tilde{W}_2 - 1)] \\
    &= \BE[Y_1^2] \frac{\Var(\tilde{W}_1)}{n} + \frac{n - 1}{n}\BE[Y_1]^2 \Cov(\tilde{W}_1, \tilde{W}_2),
\end{align*}
The first term converges to zero by Assumption~\ref{asm:weakbootstrap}(i). Under part (ii), the second term either converges to zero or is non-positive. In the latter case, the first term bounds the non-negative left-hand side from above. Hence the left-hand side converges to zero.
\end{proof}

We now present stronger bootstrap weight assumptions which ensure strong consistency of $T^{(n),*}$.

\begin{assumption}[Strong consistency requirements on bootstrap weights]\label{asm:strongbootstrap}
The bootstrap weights $W_1, \dots, W_n$ satisfy the following:
\begin{enumerate}
    \item[(i)] $\sum_{n = 1}^\infty \BE[(\tilde{W}_1 - 1)^4]/n^3 < \infty$ (in particular, $\BE[W_1^4] < \infty$),
    \item[(ii)] $\sum_{n = 1}^\infty \BE[(\tilde{W}_1 - 1)^3(\tilde{W}_2 - 1)]^+/n^2 < \infty$,
    \item[(iii)] $\sum_{n = 1}^\infty \frac{\BE[(\tilde{W}_1 - 1)^2(\tilde{W}_2 - 1)^2]}{n^2} < \infty$,
    \item[(iv)] $\sum_{n = 1}^\infty \frac{\BE[(\tilde{W}_1 - 1)^2(\tilde{W}_2 - 1)(\tilde{W}_3 - 1)]^+}{n} < \infty$,
    \item[(v)] $\sum_{n = 1}^\infty \BE[(\tilde{W}_1 - 1)(\tilde{W}_2 - 1)(\tilde{W}_3 - 1)(\tilde{W}_4 - 1)]^+ < \infty$.
\end{enumerate}
Here $a^+ = \max(a, 0)$, so that (ii), (iv) and (v) hold in particular whenever the corresponding expectation is non-positive.
\end{assumption}

Under these assumptions, the result of Lemma~\ref{lem:bootstrapestimatorweak} can be strengthened to almost sure convergence.

\begin{lemma}[Strong consistency of bootstrap averages]\label{lem:bootstrapestimatorstrong}
Suppose $Y_1, \dots, Y_n$ is an iid sample of non-negative random variables with $\BE[Y_1^4] < \infty$ independent of the bootstrap weights $W_1, \dots, W_n$. If the weights satisfy Assumption~\ref{asm:strongbootstrap}, then
\begin{equation*}
    T^{(n), *} \longrightarrow \BE[Y_1] \quad \BP\text{-a.s.}
\end{equation*}
\end{lemma}
\begin{proof}
The strong law of large numbers gives $T^{(n)} \to \BE[Y_1]$ almost surely, so it suffices to show that $T^{(n),*} - T^{(n)} \to 0$ almost surely. By the Markov inequality and the Borel--Cantelli criterion for almost sure convergence, it suffices to show that $\BE[(T^{(n),*} - T^{(n)})^4]$ is summable. We compute
\begin{align*}
    n^4\BE[(T^{(n),*} - T^{(n)})^4] &= \BE\Big[\Big(\sum_{i = 1}^n Y_i(\tilde{W}_i - 1)\Big)^4 \Big] \\
    &=\sum_{i, j, k, l = 1}^n \BE[(\tilde{W}_i - 1)(\tilde{W}_j - 1)(\tilde{W}_k - 1)(\tilde{W}_l - 1)]\BE[Y_iY_jY_kY_l] \\
    &= N_1(n)\BE[(\tilde{W}_1 - 1)^4]\BE[Y_1^4] \\
    &+ N_2(n)\BE[(\tilde{W}_1 - 1)^3(\tilde{W}_2 - 1)]\BE[Y_1^3]\BE[Y_1] \\
    &+ N_3(n)\BE[(\tilde{W}_1 - 1)^2(\tilde{W}_2 - 1)^2]\BE[Y_1^2]^2 \\
    &+ N_4(n)\BE[(\tilde{W}_1 - 1)^2(\tilde{W}_2 - 1)(\tilde{W}_3 - 1)]\BE[Y_1^2]\BE[Y_1]^2 \\
    &+ N_5(n)\BE[(\tilde{W}_1 - 1)(\tilde{W}_2 - 1)(\tilde{W}_3 - 1)(\tilde{W}_4 - 1)]\BE[Y_1]^4
\end{align*}
where counting the index multiplicities gives
\begin{gather*}
    N_1(n)=n,\qquad N_2(n)=4n(n-1),\qquad N_3(n)=3n(n-1),\\
    N_4(n)=6n(n-1)(n-2),\qquad N_5(n)=n(n-1)(n-2)(n-3).
\end{gather*}
Substitution yields
\begin{align*}
    \BE[(T^{(n),*} - T^{(n)})^4] &= \BE[Y_1^4] \frac{\BE[(\tilde{W}_1 - 1)^4]}{n^3} + 4\BE[Y_1]\BE[Y_1^3] \frac{n - 1}{n} \frac{\BE[(\tilde{W}_1 - 1)^3(\tilde{W}_2 - 1)]}{n^2} \\
    &+ 3\BE[Y_1^2]^2 \frac{n - 1}{n} \frac{\BE[(\tilde{W}_1 - 1)^2(\tilde{W}_2 - 1)^2]}{n^2} \\
    &+ 6\BE[Y_1^2] \BE[Y_1]^2 \frac{(n - 1)(n - 2)}{n^2} \frac{\BE[(\tilde{W}_1 - 1)^2(\tilde{W}_2 - 1)(\tilde{W}_3 - 1)]}{n} \\
    &+ \frac{(n - 1)(n - 2)(n - 3)}{n^3}\BE[Y_1]^4 \BE[(\tilde{W}_1 - 1)(\tilde{W}_2 - 1)(\tilde{W}_3 - 1)(\tilde{W}_4 - 1)],
\end{align*}
and since $Y_1 \geq 0$, every factor of the form $\BE[Y_1^a]\BE[Y_1]^b$ is non-negative, while the remaining coefficients in $n$ lie in $[0, 1]$. Replacing each weight moment by its positive part therefore yields an upper bound for the non-negative left hand side, and the summability conditions of Assumption~\ref{asm:strongbootstrap} finish the proof.
\end{proof}

The two subsampling schemes satisfy these requirements under the following conditions on their sample sizes.

\begin{proposition}[Bootstrap weights of the two subsampling schemes]\label{prop:subsampling}
Assume either the multinomial subsampling or subsampling without replacement bootstrap weights with subsample size $m_n$. Then:
\begin{enumerate}
    \item[(i)] If $m_n \to \infty$, the weights satisfy the conditions in Assumption~\ref{asm:weakbootstrap}.
    \item[(ii)] If $\sum_{n = 1}^\infty \frac{1}{m_n^2} < \infty$, the weights satisfy the conditions in Assumption~\ref{asm:strongbootstrap}.
\end{enumerate}
\end{proposition}
\begin{proof}[of Proposition~\ref{prop:subsampling}]
We first consider the weak consistency requirements, so assume $m_n \to \infty$. For the multinomial subsampling scheme, $W_1$ is binomial with parameters $m_n$ and $1/n$, so
\begin{equation*}
    \frac{\Var(\tilde{W}_1)}{n} = \frac{\Var(W_1)}{n\BE[W_1]^2} = \frac{m_n(1 - \frac{1}{n})\frac{1}{n}}{\frac{m_n^2}{n}} = \frac{1}{m_n}\Big(1 - \frac{1}{n}\Big) \to 0
\end{equation*}
and by the results of~\citet{OuimetS},
\begin{equation*}
    \Cov(W_1, W_2) = -\frac{m_n}{n^2} \leq 0,
\end{equation*}
which proves the claim for multinomial subsampling. Now consider subsampling without replacement. As $W_1$ is Bernoulli distributed with success probability $m_n/n$,
\begin{equation*}
    \frac{\Var(\tilde{W}_1)}{n} = \frac{\Var(W_1)}{n\BE[W_1]^2} = \frac{\frac{m_n}{n}(1 - \frac{m_n}{n})}{\frac{m_n^2}{n}} = \frac{1}{m_n}\Big(1 - \frac{m_n}{n}\Big) \to 0.
\end{equation*}
As for the covariance, note that $\BE[W_1W_2] = \BP(W_1 = 1, W_2 = 1) = \frac{m_n}{n} \frac{m_n - 1}{n - 1}$ so
\begin{align*}
    \Cov(W_1, W_2) &= \BE[W_1W_2] - \BE[W_1]^2 = \frac{m_n}{n} \frac{m_n - 1}{n - 1} - \frac{m_n^2}{n^2} \\
    &= \frac{m_n}{n} \frac{(m_n - 1)n - m_n(n - 1)}{n(n - 1)} = \frac{m_n(m_n - n)}{n^2(n - 1)} \leq 0,
\end{align*}
which shows that the weak consistency requirements also hold for subsampling without replacement.

Now consider claim (ii) and let $\sum_{n = 1}^\infty \frac{1}{m_n^2} < \infty$. We first consider multinomial subsampling, where we refer to~\citet{OuimetS} for all central moment computations. We compute:
\begin{enumerate}
    \item[(i)]
    \begin{align*}
        \frac{\BE[(\tilde{W}_1 - 1)^4]}{n^3} &= \frac{3\frac{m_n^2}{n^2}(1 - \frac{1}{n})^2 + \frac{m_n}{n}(1 - \frac{1}{n})(1 + \frac{6}{n^2} - \frac{6}{n})}{\frac{m_n^4}{n^4}n^3} \\
        &= 3\Big(1 - \frac{1}{n}\Big)^2 \frac{1}{m_n^2n} + \Big(1 - \frac{1}{n}\Big)\Big(1 + \frac{6}{n^2} - \frac{6}{n} \Big)\frac{1}{m_n^3} = O\Big(\frac{1}{m_n^3}\Big).
    \end{align*}
    \item[(ii)]
    \begin{align*}
        \frac{\BE[(\tilde{W}_1 - 1)^3 (\tilde{W}_2 - 1)]}{n^2} &= \frac{\frac{m_n}{n^2}(3(m_n - 2)\frac{1}{n}(\frac{1}{n} - 1) - 1)}{\frac{m_n^4}{n^4}n^2} = \frac{3(m_n - 2)\frac{1}{n}(\frac{1}{n} - 1) - 1}{m_n^3} \\
        &= O\Big(\frac{1}{nm_n^2} + \frac{1}{m_n^3} \Big) = O\Big(\frac{1}{m_n^3}\Big).
    \end{align*}
    \item[(iii)]
    \begin{align*}
        \frac{\BE[(\tilde{W}_1 - 1)^2(\tilde{W}_2 - 1)^2]}{n^2} &= \frac{m_n(m_n - 2)\frac{1}{n^2}(\frac{3}{n^2} - \frac{2}{n} + 1) + \frac{m_n}{n^2}}{\frac{m_n^4}{n^4}n^2} \\
        &= \frac{(m_n - 2)(\frac{3}{n^2} - \frac{2}{n} + 1) + 1}{m_n^3} = O\Big(\frac{1}{m_n^2}\Big).
    \end{align*}
    \item[(iv)]
    \begin{align*}
        \frac{\BE[(\tilde{W}_1 - 1)^2(\tilde{W}_2 - 1)(\tilde{W}_3 - 1)]}{n} &= \frac{n^3}{m_n^4}\BE\Big[\Big(W_1 - \frac{m_n}{n}\Big)^2\Big(W_2 - \frac{m_n}{n}\Big)\Big(W_3 - \frac{m_n}{n}\Big)\Big] \\
        &= O\Big(\frac{1}{m_n^2}\Big),
    \end{align*}
    since this mixed central moment of the multinomial distribution is of order $m_n^2/n^3$ by the formulas in \citet{OuimetS}.
    \item[(v)]
    \begin{align*}
        \BE[(\tilde{W}_1 - 1)(\tilde{W}_2 - 1)(\tilde{W}_3 - 1)(\tilde{W}_4 - 1)] &= \frac{3m_n(m_n - 2)\frac{1}{n^4}}{\frac{m_n^4}{n^4}} = \frac{3(m_n - 2)}{m_n^3} \\
        &= O\Big(\frac{1}{m_n^2}\Big).
    \end{align*}
\end{enumerate}
Thus each requirement in Assumption~\ref{asm:strongbootstrap} is satisfied by the multinomial subsampling scheme. Now consider subsampling without replacement. We verify each requirement.
\begin{enumerate}
    \item[(i)] Using that $W_1$ is Bernoulli distributed with success probability $m_n/n$,
    \begin{align*}
        \frac{\BE[(\tilde{W}_1 - 1)^4]}{n^3} &= \frac{\frac{m_n}{n}(1 - \frac{m_n}{n})(1 - \frac{3m_n}{n}(1 - \frac{m_n}{n}))}{\frac{m_n^4}{n^4}n^3} \\
        &= \frac{(1 - \frac{m_n}{n})(1 - \frac{3m_n}{n}(1 - \frac{m_n}{n}))}{m_n^3} = O\Big(\frac{1}{m_n^3}\Big).
    \end{align*}
    \item[(ii)] We have $\sum_{i = 1}^n (\tilde{W}_i - 1) = 0$, so
    \begin{align*}
        0 &= \BE\Big[(\tilde{W}_1 - 1)^3 \sum_{i = 1}^n (\tilde{W}_i - 1) \Big] \\
        &= \BE[(\tilde{W}_1 - 1)^4] + (n - 1)\BE[(\tilde{W}_1 - 1)^3(\tilde{W}_2 - 1)]
    \end{align*}
    and rearranging yields
    \begin{equation*}
        \BE[(\tilde{W}_1 - 1)^3(\tilde{W}_2 - 1)] = -\frac{1}{n - 1}\BE[(\tilde{W}_1 - 1)^4] \leq 0.
    \end{equation*}
    \item[(iii)] The joint distribution of $(W_1, W_2)$ is given by
    \begin{align*}
        \BP(W_1 = 1, W_2 = 1) &= \frac{m_n}{n}\frac{m_n - 1}{n - 1}, \\
        \BP(W_1 = 0, W_2 = 1) &= \BP(W_1 = 1, W_2 = 0) = \frac{m_n}{n} \frac{n - m_n}{n - 1}, \\
        \BP(W_1 = 0, W_2 = 0) &= \frac{n - m_n}{n} \frac{n - 1 - m_n}{n - 1}.
    \end{align*}
    Thus,
    \begin{align*}
        \BE[(W_1 - \BE[W_1])^2(W_2 - \BE[W_1])^2] &= \frac{m_n}{n} \frac{m_n - 1}{n - 1}\Big(1 - \frac{m_n}{n}\Big)^4 \\
        &+ 2\frac{m_n}{n} \frac{n - m_n}{n - 1} \frac{m_n^2}{n^2}\Big(1 - \frac{m_n}{n}\Big)^2 \\
        &+ \frac{(n - m_n)(n - m_n - 1)}{n(n - 1)} \frac{m_n^4}{n^4} \\
        &= O\Big(\frac{m_n^2}{n^2} + \frac{m_n^3}{n^3} + \frac{m_n^4}{n^4}\Big) = O\Big(\frac{m_n^2}{n^2}\Big),
    \end{align*}
    which implies $\BE[(\tilde{W}_1 - 1)^2(\tilde{W}_2 - 1)^2]/n^2 = O(1/m_n^2)$. 
    \item[(iv)] As in (ii), the identity $\sum_{i=1}^n(\tilde W_i-1)=0$ gives
    \begin{align*}
        0 &= \BE\Big[(\tilde{W}_1 - 1)^2 (\tilde{W}_2 - 1)\sum_{i = 1}^n (\tilde{W}_i - 1) \Big] \\
        &= \BE[(\tilde{W}_1 - 1)^3(\tilde{W}_2 - 1)] + \BE[(\tilde{W}_1 - 1)^2(\tilde{W}_2 - 1)^2] \\
        &+ (n - 2)\BE[(\tilde{W}_1 - 1)^2(\tilde{W}_2 - 1)(\tilde{W}_3 - 1)],
    \end{align*}
    hence by the result in (ii),
    \begin{align*}
        \frac{\BE[(\tilde{W}_1 - 1)^2(\tilde{W}_2 - 1)(\tilde{W}_3 - 1)]}{n} &= -\frac{1}{n(n - 2)}\big(\BE[(\tilde{W}_1 - 1)^2(\tilde{W}_2 - 1)^2] \\
        &\qquad + \BE[(\tilde{W}_1 - 1)^3(\tilde{W}_2 - 1)]\big) \\
        &\leq \frac{1}{n(n - 1)(n - 2)}\BE[(\tilde{W}_1 - 1)^4] = O\Big(\frac{1}{m_n^3}\Big).
    \end{align*}
    \item[(v)] We have
    \begin{align*}
        0 &= \BE\Big[(\tilde{W}_1 - 1)(\tilde{W}_2 - 1)(\tilde{W}_3 - 1)\sum_{i = 1}^n (\tilde{W}_i - 1) \Big] \\
        &= (n - 3)\BE[(\tilde{W}_1 - 1)(\tilde{W}_2 - 1)(\tilde{W}_3 - 1)(\tilde{W}_4 - 1)] + 3\BE[(\tilde{W}_1 - 1)^2(\tilde{W}_2 - 1)(\tilde{W}_3 - 1)],
    \end{align*}
    implying
    \begin{align*}
        \BE[(\tilde{W}_1 - 1)(\tilde{W}_2 - 1)(\tilde{W}_3 - 1)(\tilde{W}_4 - 1)] &= -\frac{3}{n - 3}\BE[(\tilde{W}_1 - 1)^2(\tilde{W}_2 - 1)(\tilde{W}_3 - 1)] \\
        &= \frac{3}{(n - 2)(n - 3)}\big(\BE[(\tilde{W}_1 - 1)^2(\tilde{W}_2 - 1)^2] \\
        &\qquad + \BE[(\tilde{W}_1 - 1)^3(\tilde{W}_2 - 1)]\big) = O\Big(\frac{1}{m_n^2}\Big),
    \end{align*}
    where we used the identity from (iv) and the orders established in (i)--(iii).
\end{enumerate}
\end{proof}

\subsection{Uniform convergence}

\begin{lemma}[Abstract Glivenko--Cantelli]\label{lem:abstractGC}
Let $I = [a, b] \subseteq \BR$ be a closed interval and $T:I \to \BR$ a non-decreasing, bounded and càdlàg function. Assume that $(T^{(n)})$ is a sequence of random functions $T^{(n)}:I \to \BR$ with non-decreasing càdlàg paths such that $T^{(n)}(x) \to T(x)$ and $T^{(n)}(x-) \to T(x-)$ $\BP$-a.s. (in probability, respectively) for all $x \in I$. At the left endpoint, set $T(a-)=T(a)$ and $T^{(n)}(a-)=T^{(n)}(a)$. Then
\begin{equation*}
    \sup_{x \in I}|T^{(n)}(x) - T(x)| \longrightarrow 0 \quad \BP\text{-a.s.} \quad \text{(in probability, respectively)}.
\end{equation*}
\end{lemma}
\begin{proof}
Fix $\varepsilon>0$. Since $T$ is bounded, non-decreasing and càdlàg, there is a finite partition $a=x_0<\cdots<x_k=b$ such that $T(x_j-)-T(x_{j-1})<\varepsilon$ for every $j$. Define
\begin{equation*}
    \Delta^{(n)}=\max_{0\leq j\leq k}\big\{|T^{(n)}(x_j)-T(x_j)|,\ |T^{(n)}(x_j-)-T(x_j-)|\big\}.
\end{equation*}
For $x\in(x_{j-1},x_j)$, monotonicity gives
\begin{equation*}
    T^{(n)}(x_{j-1})-T(x_{j-1})-\varepsilon
    \leq T^{(n)}(x)-T(x)
    \leq T^{(n)}(x_j-)-T(x_j-)+\varepsilon.
\end{equation*}
At the partition points, the absolute error is bounded by $\Delta^{(n)}$. Hence
\begin{equation*}
    \sup_{x\in I}|T^{(n)}(x)-T(x)|\leq\Delta^{(n)}+\varepsilon.
\end{equation*}
The finite maximum $\Delta^{(n)}$ converges to zero in the assumed mode. For almost sure convergence, apply this bound along a countable sequence $\varepsilon\downarrow0$. For convergence in probability, fix any $\eta>0$ and choose $\varepsilon<\eta$; the probability that the supremum exceeds $\eta$ is at most $\BP(\Delta^{(n)}>\eta-\varepsilon)$, which tends to zero.
\end{proof}

\subsection{Counting-process decomposition}

The following decomposition is used repeatedly in the proofs of Supplementary Sections~\ref{sec:proofs} and \ref{sec:proofs2}. When establishing uniform convergence, we rely on tools such as the abstract Glivenko--Cantelli theorem presented above in Lemma~\ref{lem:abstractGC}. This result only applies to estimators and estimands that are non-decreasing in time. This is the case for the estimator $\BN_{jk}^{(n)}$ of $p_{jk}^c$, but not for $\BI_j^{(n)}$. We therefore decompose the latter into a sum of monotone estimators, as stated in the following lemma. 

\begin{lemma}[Key decomposition]\label{lem:keydecomposition}
For each $j\in\CZ$ and $t\geq0$,
\begin{equation*}
    1_{(Z_t = j)}1_{(t < R)} = 1_{(Z_0 = j)} - 1_{(R \leq t)}1_{(Z_R = j)} + \sum_{k:k \neq j} (N_{kj}(t \land R) - N_{jk}(t \land R)).
\end{equation*}
\end{lemma}
\begin{proof}
We use the inflow--outflow identity for each $j$ given by
\begin{equation*}
    1_{(Z_t = j)} = 1_{(Z_0 = j)} + \sum_{k:k \neq j}(N_{kj}(t) - N_{jk}(t)).
\end{equation*}
The identity follows by counting entries into and exits from state $j$. Since $s \mapsto 1_{(s < R)}$ is a non-increasing process with a single jump of size $-1$ at $R$, integration by parts yields
\begin{align*}
    1_{(Z_t = j)}1_{(t < R)} &= 1_{(Z_0 = j)}1_{(0 < R)} + \int_0^t 1_{(Z_s = j)}\dd 1_{(s < R)} + \int_0^t 1_{(s \leq R)}\dd 1_{(Z_s = j)} \\
    &= 1_{(Z_0 = j)} - 1_{(R \leq t)}1_{(Z_R = j)} + \sum_{k:k \neq j}\Big(\int_0^t 1_{(s \leq R)}N_{kj}(\dd s) - \int_0^t 1_{(s \leq R)}N_{jk}(\dd s) \Big) \\
    &= 1_{(Z_0 = j)} - 1_{(R \leq t)}1_{(Z_R = j)} + \sum_{k:k \neq j} (N_{kj}(t \land R) - N_{jk}(t \land R)). \qedhere
\end{align*}
\end{proof}

Defining the censoring contribution $C_j(t) = \BP(R \leq t, Z_R = j)$ and its estimator
\begin{equation*}
    \BC_j^{(n)}(t) = \frac{1}{n}\sum_{i = 1}^n 1_{(R^i \leq t)}1_{(Z_{R^i}^i = j)},
\end{equation*}
we see that the key decomposition yields
\begin{align*}
    p_j^c(t) &= p_j^c(0) - C_j(t) + \sum_{k:k \neq j}(p_{kj}^c(t) - p_{jk}^c(t)),\\
    \BI_j^{(n)}(t) &= \BI_j^{(n)}(0) - \BC_j^{(n)}(t) + \sum_{k: k \neq j}(\BN_{kj}^{(n)}(t) - \BN_{jk}^{(n)}(t)),
\end{align*}
and so uniform consistency for $\BI_j^{(n)}$ follows from uniform consistency of both $\BN_{jk}^{(n)}$ and $\BC_j^{(n)}$, and this is immediately established from the abstract Glivenko--Cantelli lemma. 

\subsection{Empirical processes and differentiability}

For iid observations $Y_1,\dots,Y_n$ with law $\BP$ and a class $\CH$ of measurable functions, write
\begin{equation*}
    \BP^{(n)}h=\frac{1}{n}\sum_{i=1}^n h(Y_i),\qquad
    \mathbb{G}^{(n)}h=\sqrt{n}(\BP^{(n)}h-\BP h).
\end{equation*}
The class is $\BP$-Donsker if
\begin{equation}\label{eq:Donsker}
    \mathbb{G}^{(n)}\overset{d}{\longrightarrow}\mathbb{G}\quad\text{in }\ell^\infty(\CH),
\end{equation}
where $\mathbb{G}$ is a tight mean-zero Gaussian process with covariance $\BP(h_1h_2)-\BP h_1\BP h_2$. All function spaces below carry the uniform norm. We use the empirical-process conventions of \citet{VaartWellnerS}.

A bracket $[l,u]$ contains the functions $h\in\CH$ satisfying $l\leq h\leq u$; its endpoints need not belong to $\CH$. Let $N_{[]}(\varepsilon,\CH,L^2(\BP))$ be the minimum number of brackets of $L^2(\BP)$-width less than $\varepsilon$ covering $\CH$. The bracketing criterion of \citet[Theorem 2.5.6]{VaartWellnerS} establishes the Donsker property when
\begin{equation*}
    \int_0^\infty\sqrt{\log N_{[]}(\varepsilon,\CH,L^2(\BP))}\dd\varepsilon<\infty.
\end{equation*}
To transfer weak convergence to the rate estimators, we use the functional delta method. The next two lemmas establish the required uniform Hadamard differentiability on a domain of functions with uniformly bounded variation.

\begin{lemma}[Integrals against uniformly vanishing integrators]\label{lem:regulated}
Let $I=[a,b]$ be a compact interval, let $f:I\to\BR$ be regulated, and let $F_n$ be càdlàg functions of bounded variation on $I$ such that $\|F_n\|_\infty\to0$ and $\sup_n V_{F_n}(I)<\infty$. Here $V_F(I)$ denotes total variation. Then
\begin{equation*}
    \sup_{v\in I}\left|\int_a^v f(s)F_n(\dd s)\right|\longrightarrow0.
\end{equation*}
\end{lemma}
\begin{proof}
For $\varepsilon>0$, approximate $f$ uniformly within $\varepsilon$ by a step function $g$ that is constant on each open partition interval and has separately specified values at the partition points. Such an approximation exists because $f$ is regulated. The integral of $g$ against $F_n$ is a finite linear combination of increments and jumps of $F_n$, each bounded in absolute value by $2\|F_n\|_\infty$, uniformly in the upper endpoint. Hence
\begin{equation*}
    \sup_{v\in I}\left|\int_a^v f\dd F_n\right|
    \leq\varepsilon\sup_n V_{F_n}(I)+C_g\|F_n\|_\infty,
\end{equation*}
where $C_g$ is a finite constant depending only on $g$. Let $n\to\infty$ and then $\varepsilon\downarrow0$.
\end{proof}

\begin{lemma}[Uniform Hadamard differentiability]\label{lem:gHD}
Let $I=[a,b]$ be a compact interval, $K>0$, $0\leq c_1<c_2\leq K$, and $\varphi\in C^2([c_1,c_2])$. Let $\BD_g\subseteq D(I)^2$ consist of pairs $(y,z)$ of càdlàg functions with total variation at most $K$ in each coordinate and $y(I)\subseteq[c_1,c_2]$. The map
\begin{equation*}
    g(y,z)(v)=\int_a^v\varphi(y(s))z(\dd s),\qquad v\in I,
\end{equation*}
is Hadamard differentiable tangentially to $D(I)^2$, with derivative
\begin{equation*}
    g'_{(y,z)}(h_1,h_2)(v)
    =[h_2(s)\varphi(y(s))]_{s=a}^{v}
    -\int_a^v h_2(s-)\dd\varphi(y(s))
    +\int_a^v\varphi'(y(s))h_1(s)z(\dd s).
\end{equation*}
Moreover, if $t_n\to0$, $(y_n,z_n)\to(y,z)$ and $h_n\to h$ uniformly, with $(y_n,z_n)$ and $(y_n+t_nh_{1n},z_n+t_nh_{2n})$ in $\BD_g$, then
\begin{equation*}
    \frac{g(y_n+t_nh_{1n},z_n+t_nh_{2n})-g(y_n,z_n)}{t_n}
    \longrightarrow g'_{(y,z)}(h_1,h_2)
\end{equation*}
uniformly on $I$.
\end{lemma}
\begin{proof}
A Taylor expansion, the uniform convergence of $h_{1n}$, and the variation bound on $z_n+t_nh_{2n}$ give, uniformly in $v$,
\begin{align*}
    &\frac{g(y_n+t_nh_{1n},z_n+t_nh_{2n})(v)-g(y_n,z_n)(v)}{t_n}\\
    &\qquad=\int_a^v\varphi(y_n)\dd h_{2n}
       +\int_a^v\varphi'(y_n)h_1\dd(z_n+t_nh_{2n})+o(1).
\end{align*}
Indeed, the Taylor remainder is bounded by
$K\{\|\varphi'\|_\infty\|h_{1n}-h_1\|_\infty
+|t_n|\|\varphi''\|_\infty\|h_{1n}\|_\infty^2/2\}$.

For the first integral, integration by parts yields
\begin{equation*}
    \int_a^v\varphi(y_n)\dd h_{2n}
    =[\varphi(y_n)h_{2n}]_a^v-\int_a^v h_{2n}(s-)\dd\varphi(y_n(s)).
\end{equation*}
The boundary terms converge uniformly. Replacing $h_{2n}(s-)$ by $h_2(s-)$ costs at most
$\|h_{2n}-h_2\|_\infty\|\varphi'\|_\infty K$.
Lemma~\ref{lem:regulated}, applied to the regulated function $h_2(\cdot-)$ and the uniformly vanishing integrator $\varphi(y_n)-\varphi(y)$, handles the remaining difference, since its variation is at most $2\|\varphi'\|_\infty K$.

For the second integral, replacing $\varphi'(y_n)$ by $\varphi'(y)$ costs at most
$K\|h_1\|_\infty\|\varphi'(y_n)-\varphi'(y)\|_\infty$.
Finally, $z_n+t_nh_{2n}-z\to0$ uniformly with variation at most $2K$, so Lemma~\ref{lem:regulated}, now with integrand $\varphi'(y)h_1$, gives the asserted limit. The displayed derivative is linear and continuous in $(h_1,h_2)$ by the variation bounds.
\end{proof}

\begin{remark}\label{rmk:leftlimit}
The Nelson--Aalen functional integrates against the left limit of the at-risk process, that is, it is of the form $(y, z) \mapsto \int_a^\cdot \varphi(y(s-))z(\dd s)$ with $\varphi(u) = 1/u$. The proof of Lemma~\ref{lem:gHD} also applies to this map, using $\|y_{n,-}-y_-\|_\infty\leq\|y_n-y\|_\infty$ and the integration-by-parts identity below. It gives uniform Hadamard differentiability with derivative
\begin{equation*}
    (h_1, h_2) \mapsto \int_a^\cdot \varphi(y(s-))h_2(\dd s) + \int_a^\cdot \varphi'(y(s-))h_1(s-)z(\dd s),
\end{equation*}
where the first integral is defined through integration by parts,
\begin{equation*}
    \int_a^t \varphi(y(s-))h_2(\dd s) = \varphi(y(t))h_2(t) - \varphi(y(a))h_2(a) - \int_a^t h_2(s)\dd \varphi(y(s)).
\end{equation*}
At the left endpoint, set $y(a-)=y(a)$ and $h_1(a-)=h_1(a)$. In the applications, $\varphi(u) = 1/u$, or $\varphi(u) = u^{-2}$ in the proofs of Proposition~\ref{prop:markovvariance} and Corollary~\ref{cor:forestCI}, on $[c_1, c_2]$ with $c_1 = p_x\inf_{s \in [0, t]}p_j^c(s \mid x)/2$, which is positive by Assumption~\ref{asm:nonvanishing} and the choice $t < T_{R,x}$, and $c_2 = K$, where $K \geq 1$ exceeds $1 + \sum_{k \neq j}(p_{kj}^c(t \mid x) + p_{jk}^c(t \mid x))$. The estimators involved converge uniformly to functions bounded below by $2c_1$, namely to $p_j^c(\cdot \mid x)$ for the estimators normalised by the number of observations in the node, and to $p_x p_j^c(\cdot \mid x)$ for those normalised by $n$ or $m_n$. They take values in $[0, 1]$ and have total variation bounded by $1 + \sum_{k \neq j}(\BN_{kj}^x(t) + \BN_{jk}^x(t))$ or its analogue under the respective normalisation, which converges to a limit smaller than $K$ in the respective mode of convergence; the estimators of $p_{jk}^c(\cdot \mid x)$ are non-decreasing, with total variation converging to at most $p_{jk}^c(t \mid x) < K$. Hence the estimators lie in the domain of the map for all large $n$, almost surely (with probability tending to one for their bootstrap versions). This suffices for the continuous mapping theorem and for the functional delta method.
\end{remark}

\clearpage

\section{Additional numerical results}\label{section:moreplots}

Table~\ref{tab:simdesign} summarises all four simulation studies. This section presents the proportional-hazards study with log-linear effects and variable importance, supporting results for the two main prediction studies, the coverage study, and further biomedical analyses. The common prediction protocol is given in Subsection~\ref{subsection:design}.

In the simulation box plots, boxes show quartiles, whiskers extend to the most extreme replications within $1.5$ interquartile ranges, more extreme replications are not displayed, and symbols and connecting lines mark the means. The biomedical relative-score plots use the same conventions; the absolute-score plots additionally display points outside the whiskers.

\begin{table}[tbp]
    \centering
    \footnotesize
    \setlength{\tabcolsep}{4pt}
    \begin{tabular}{>{\raggedright\arraybackslash}p{2.2cm} >{\raggedright\arraybackslash}p{2.9cm} >{\raggedright\arraybackslash}p{3.3cm} >{\raggedright\arraybackslash}p{2.9cm} >{\raggedright\arraybackslash}p{2.5cm}}
        \toprule
        & Proportional hazards & Threshold interaction & Non-proportional hazards & Coverage \\
        \midrule
        Transitions & $1 \to 2$, $1 \to 3$, $2 \to 3$ & $1 \to 2$, $2 \to 1$, $1 \to 3$, $2 \to 3$ & $1 \to 2$, $2 \to 1$, $1 \to 3$, $2 \to 3$ & $1\to2$, $2\to1$, $1\to3$, $2\to3$ \\
        \addlinespace
        Proportional hazards & yes & yes & no & no \\
        \addlinespace
        Covariate effect & log-linear & threshold interaction & piecewise constant in time, with interactions & time-varying effects and interactions \\
        \addlinespace
        Covariates & $X_1$ binary; $X_2$, $X_3$ continuous & $X_1$ binary; $X_2$, $X_3$ continuous & $X_1$, $X_3$ binary; $X_2 \in \{1, \dots, 4\}$ & $16$ signal combinations; $96$ with two noise covariates \\
        \addlinespace
        Added noise covariates & $0, 5, \dots, 50$, standard normal & $0, 5, \dots, 50$ at $n = 1000$, standard normal & none & none, or $X_4$ and $X_5$ \\
        \addlinespace
        Initial distribution & $(0.40, 0.60, 0)$ & $(1, 0, 0)$ & $(1, 0, 0)$ & $(1/2, 1/2, 0)$ \\
        \addlinespace
        Censoring & Administrative at time 8 & none & independent of $(X, Z)$, uniform on $[5, 25]$ & none on $[0,4]$ \\
        \addlinespace
        Sample size $n$ & $1000$ & $50$ to $5000$ without noise; $1000$ with noise & $50$ to $5000$ & $500$ to $5000$ \\
        \addlinespace
        Competitors & Cox, Cox lasso & sample size: Cox, Cox $X_1{:}X_2$, Cox interactions, RSF; noise: Cox, Cox $X_1{:}X_2$, Cox interactions, Cox lasso & Cox, Cox interactions & stratified Nelson--Aalen estimator \\
        \addlinespace
        Criteria & relative IBS, variable importance & relative IBS, occupation probabilities & relative IBS & coverage, bias, root mean squared error, interval width \\
        \addlinespace
        Location & Supplementary Section~\ref{subsection:Cox} & Subsection~\ref{subsection:phinteraction} & Subsection~\ref{subsection:nonph} & Supplementary Section~\ref{subsection:app-coverage} \\
        \bottomrule
    \end{tabular}
    \caption{Overview of the simulation studies. All processes are Markov conditionally on the covariates, and every study uses $100$ replications.}
    \label{tab:simdesign}
\end{table}

\subsection{Proportional hazards: noise covariates and variable importance}\label{subsection:Cox}
\begin{figure}[tbp]
    \centering
    \includegraphics[width=\linewidth]{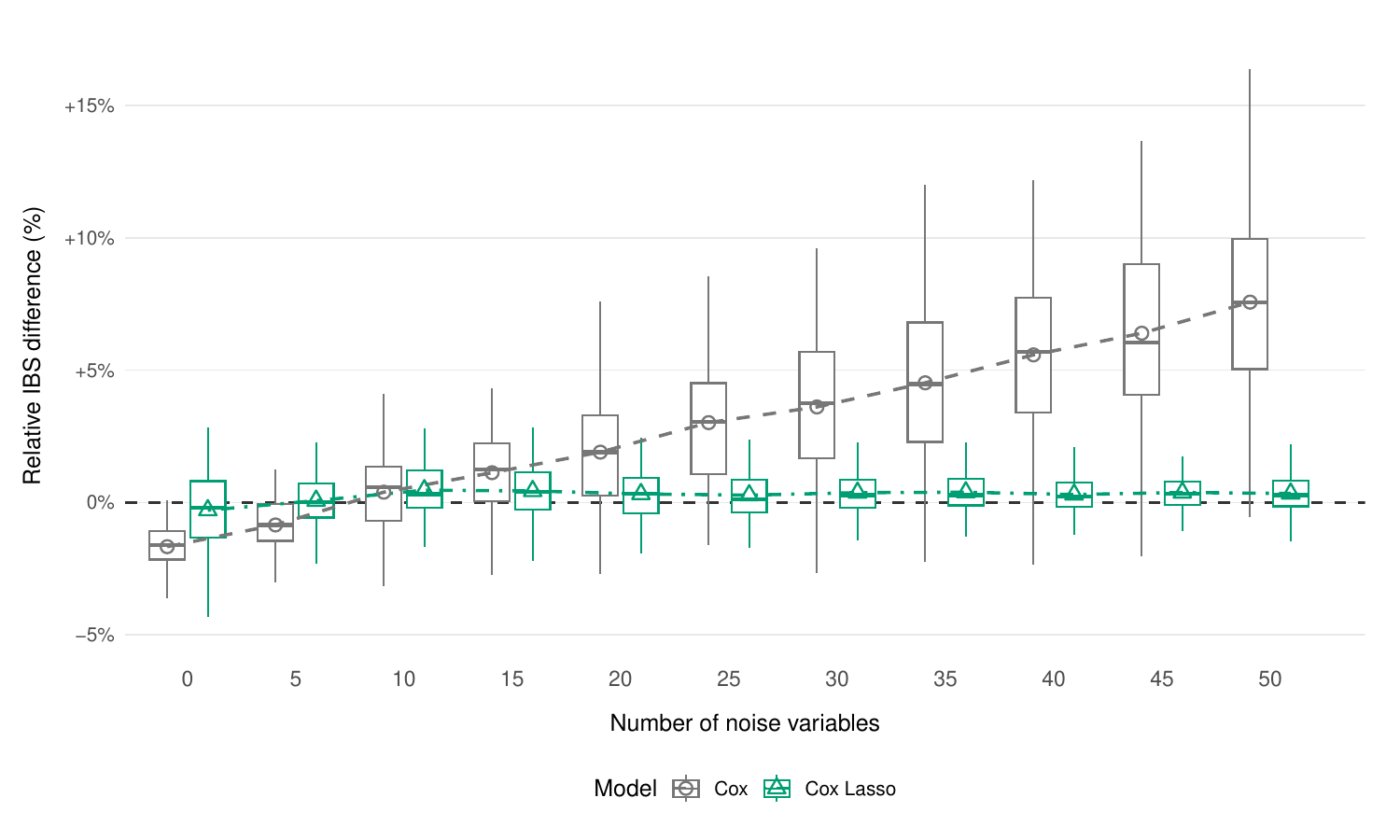}
    \caption{Proportional hazards study with log-linear effects, $n=1000$, $100$ replications. Relative difference (percent) in test-set IBS of the transition-wise Cox model (Cox) and its lasso-penalised version (Cox Lasso) with respect to the jump forest (JF), as a function of the number of added standard normal noise covariates. Positive values favour JF.}
    \label{fig:noiseph}
\end{figure}

We first consider the illness--death model without recovery with transition intensities
\begin{align*}
    \mu_{12}(t \mid X_1, X_2, X_3) &= 0.024(1 + t)e^{0.80(X_1 - 1/2) + 0.60X_2 + 1.10(X_3 - 1/2)}, \\
    \mu_{13}(t \mid X_1) &= \frac{0.16}{\sqrt{1 + t}} e^{0.45(X_1 - 1/2)}, \\
    \mu_{23}(t \mid X_2) &= \frac{27}{325}\sqrt{1 + t}e^{0.35X_2},
\end{align*}
where $X_1 \sim \text{Bern}(1/2)$, $X_2 \sim U(-1, 1)$ and $X_3 \sim U(0, 1)$ are independent and the initial distribution is $(0.40, 0.60, 0)$ independently of the covariates. We use fixed administrative censoring at $R = 8$. All transitions follow proportional hazards models with time-varying baselines, so that the transition-wise Cox model is correctly specified. We simulate $n = 1000$ observations, add $0, 5, \dots, 50$ independent standard normal noise covariates, and compare the jump forest with Cox and Cox lasso; Figure~\ref{fig:noiseph} shows the results.

The comparison with unpenalised Cox changes as noise covariates are added, whereas lasso and the forest remain comparable. The median paired relative IBS difference for Cox is $-1.6$ percent without noise, with interquartile range $[-2.2,-1.1]$, and $7.6$ percent with $50$ noise covariates: a small initial advantage for Cox becomes an advantage for the forest. For lasso, median relative differences remain between $-0.2$ and $0.4$ percent, with interquartile ranges containing zero throughout.

We next rank the covariates by the permutation importance of Supplementary Section~\ref{subsection:vimp}. The held-out importance \eqref{eq:vimphat} is computed for JF, Cox and Cox lasso from the $80/20$ splits, and for the forest we also report the internal importance \eqref{eq:vimpoob}, which requires no test set; each covariate is permuted once in every replication except for the internal importance estimate, which performs its own permutation. The experiment uses $10$, $25$ and $50$ noise covariates and $100$ replications. The effect of $X_3$ acts only through transition $1 \to 2$, which is never experienced by the $60$ percent of individuals who start in state $2$. Across the methods and noise levels considered, $X_3$ has the lowest recovery frequency of the three signal covariates. Table~\ref{tab:vimp} reports how often each signal covariate is ranked among the three most important covariates; summing the three individual percentages and dividing by $100$ gives the mean number of recovered signals.

\begin{table}[tbp]
    \centering
    \footnotesize
    \setlength{\tabcolsep}{4.2pt}
    \begin{tabular}{l rrr rrr rrr rrr}
        \toprule
        & \multicolumn{3}{c}{$X_1$} & \multicolumn{3}{c}{$X_2$} & \multicolumn{3}{c}{$X_3$} & \multicolumn{3}{c}{All three} \\
        \cmidrule(lr){2-4} \cmidrule(lr){5-7} \cmidrule(lr){8-10} \cmidrule(lr){11-13}
        Noise covariates & 10 & 25 & 50 & 10 & 25 & 50 & 10 & 25 & 50 & 10 & 25 & 50 \\
        \midrule
        JF, internal & 93 & 90 & 91 & 96 & 94 & 94 & 44 & 36 & 27 & 35 & 26 & 22 \\
        JF, held-out & 85 & 85 & 83 & 82 & 83 & 81 & 42 & 33 & 27 & 26 & 19 & 13 \\
        Cox, held-out & 88 & 82 & 69 & 86 & 81 & 70 & 41 & 24 & 18 & 24 & 11 & 5 \\
        Cox lasso, held-out & 88 & 80 & 74 & 83 & 76 & 71 & 51 & 46 & 40 & 42 & 38 & 31 \\
        \bottomrule
    \end{tabular}
    \caption{Variable importance in the proportional hazards study of Supplementary Section~\ref{subsection:Cox}, $n = 1000$, $100$ replications. Percentage of replications in which each signal covariate, and all three jointly, are ranked among the three most important covariates, for $10$, $25$ and $50$ added noise covariates. The internal importance \eqref{eq:vimpoob} of the jump forest uses out-of-bag observations; the held-out importances \eqref{eq:vimphat} use the test data. Monte Carlo standard errors of the percentages are at most $5$ percentage points.}
    \label{tab:vimp}
\end{table}

The internal importance of the forest has the largest mean number of recovered signals at all three noise levels, but lasso more often recovers $X_3$ and all three signals jointly. Internal importance recovers $X_1$ and $X_2$ in $90$--$96$ percent of replications, compared with $81$--$85$ percent for its held-out version. At $50$ noise covariates, lasso recovers $X_3$ and all three signals in $40$ and $31$ percent of replications, respectively, compared with $27$ and $22$ percent for internal importance. Thus the comparison does not establish uniformly better variable selection.

\subsection{Supporting results for the threshold interaction}\label{subsection:app-phinteraction}
\begin{figure}[tbp]
    \centering
    \includegraphics[width=\linewidth]{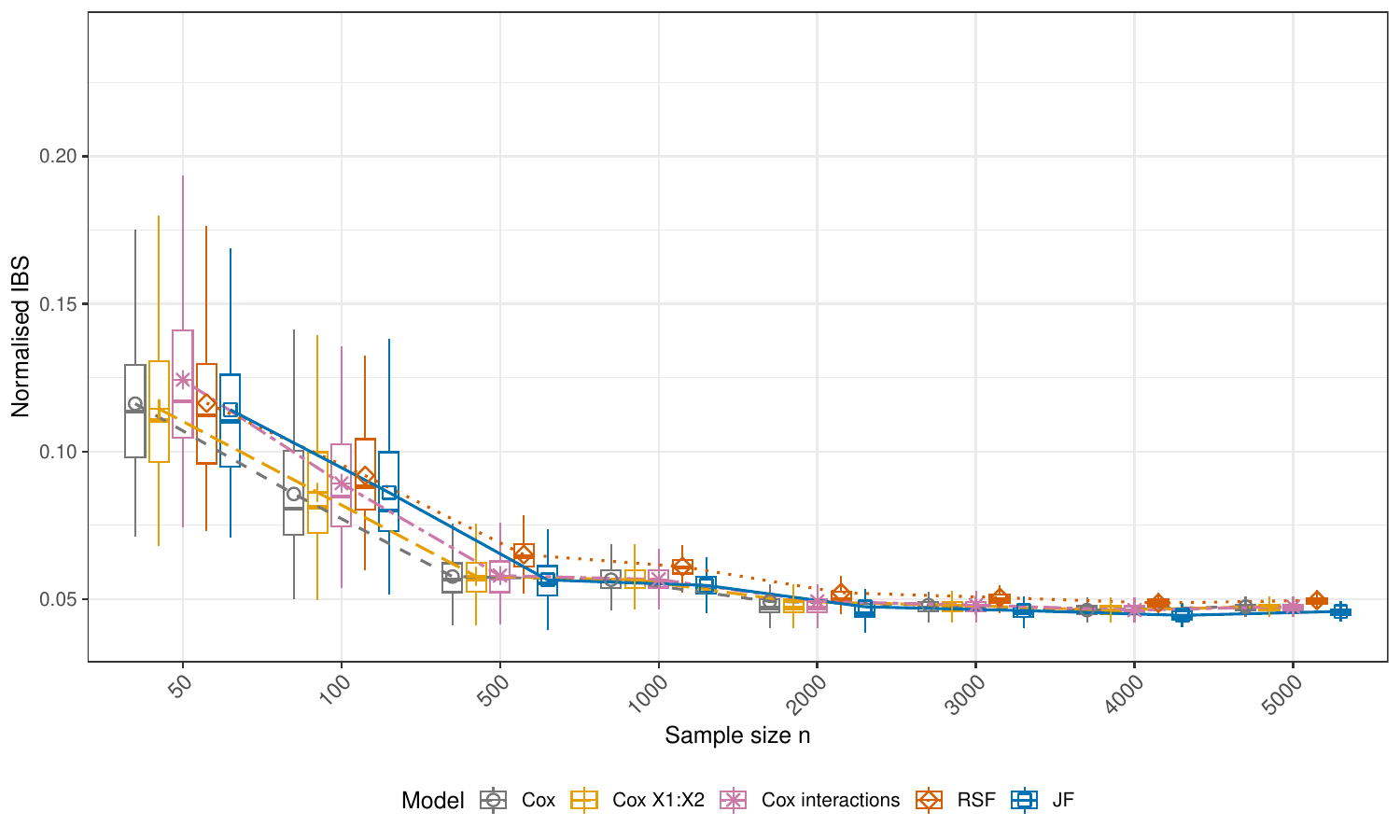}
    \caption{Threshold-interaction study, $100$ replications. Normalised test-set IBS of the jump forest (JF), the transition-wise Cox models (Cox, Cox X1:X2 and Cox interactions) and random survival forest (RSF), as a function of sample size $n$.}
    \label{fig:phint-absolute}
\end{figure}

Figure~\ref{fig:phint-absolute} gives the absolute IBS values underlying Figure~\ref{fig:phint-relative} in Subsection~\ref{subsection:phinteraction}. For this model, $p(s,t\mid x)=\exp(c(x)Q_0(t-s))$, so the covariates act through a change of time scale.

\begin{figure}[tbp]
    \centering
    \includegraphics[width=\linewidth]{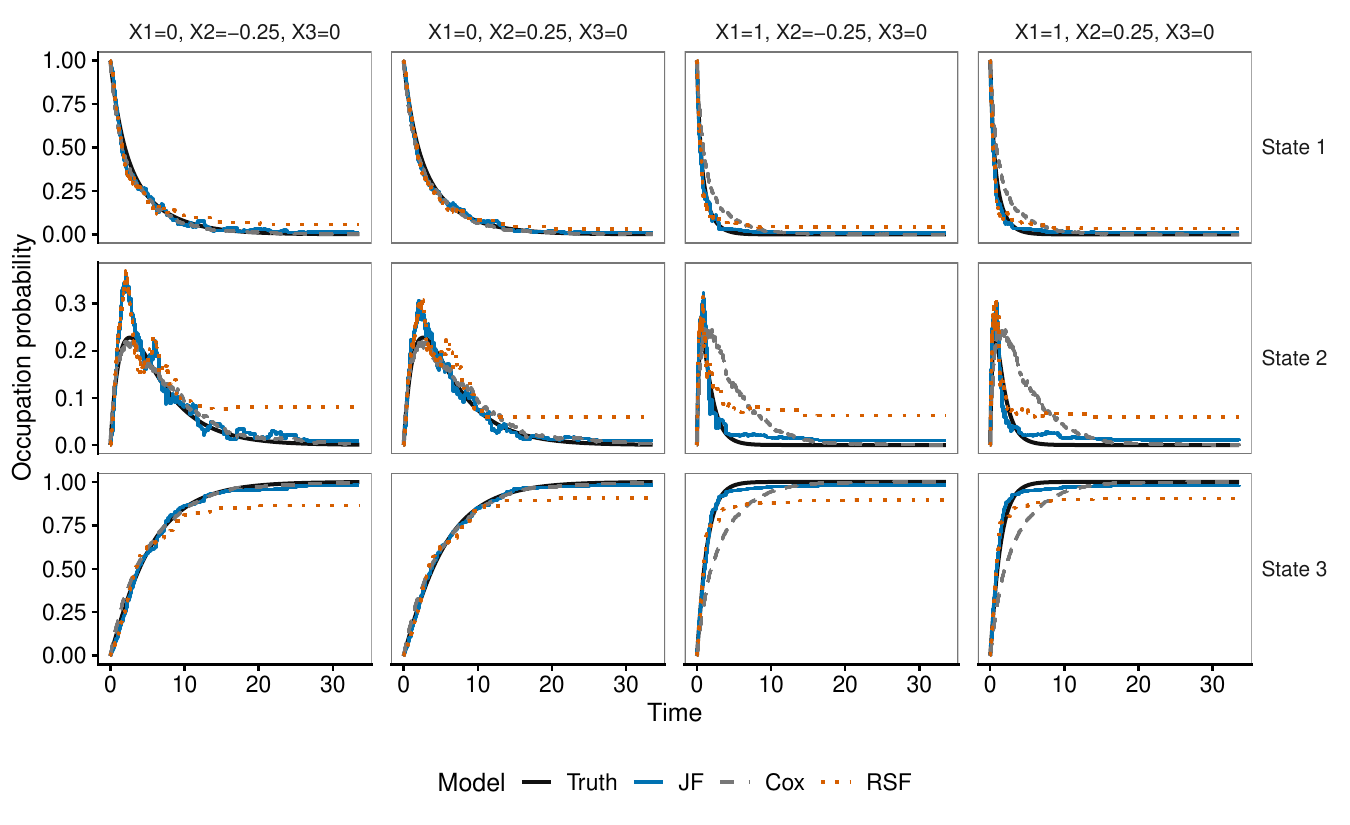}
    \caption{Threshold-interaction study, $n = 1000$. True occupation probabilities (Truth) and estimates from the jump forest (JF), transition-wise Cox model (Cox) and random survival forest (RSF). Rows correspond to the three states; columns to four covariate values with $X_3=0$. The two right-hand columns have $c(x)=4$.}
    \label{fig:phint-occupation}
\end{figure}

The occupation curves in Figure~\ref{fig:phint-occupation} illustrate the functional-form mismatch. Cox approximates the low-intensity group well but is biased in the high-intensity group. The forest captures this separation more closely, although it overestimates the early occupation of state $2$. RSF underestimates absorption more markedly. These curves illustrate the fitted predictions at four covariate values and complement the replicated score comparisons.

\clearpage
\subsection{Non-proportional hazards}\label{subsection:app-nonph}
\begin{figure}[H]
    \centering
    \includegraphics[width=\linewidth]{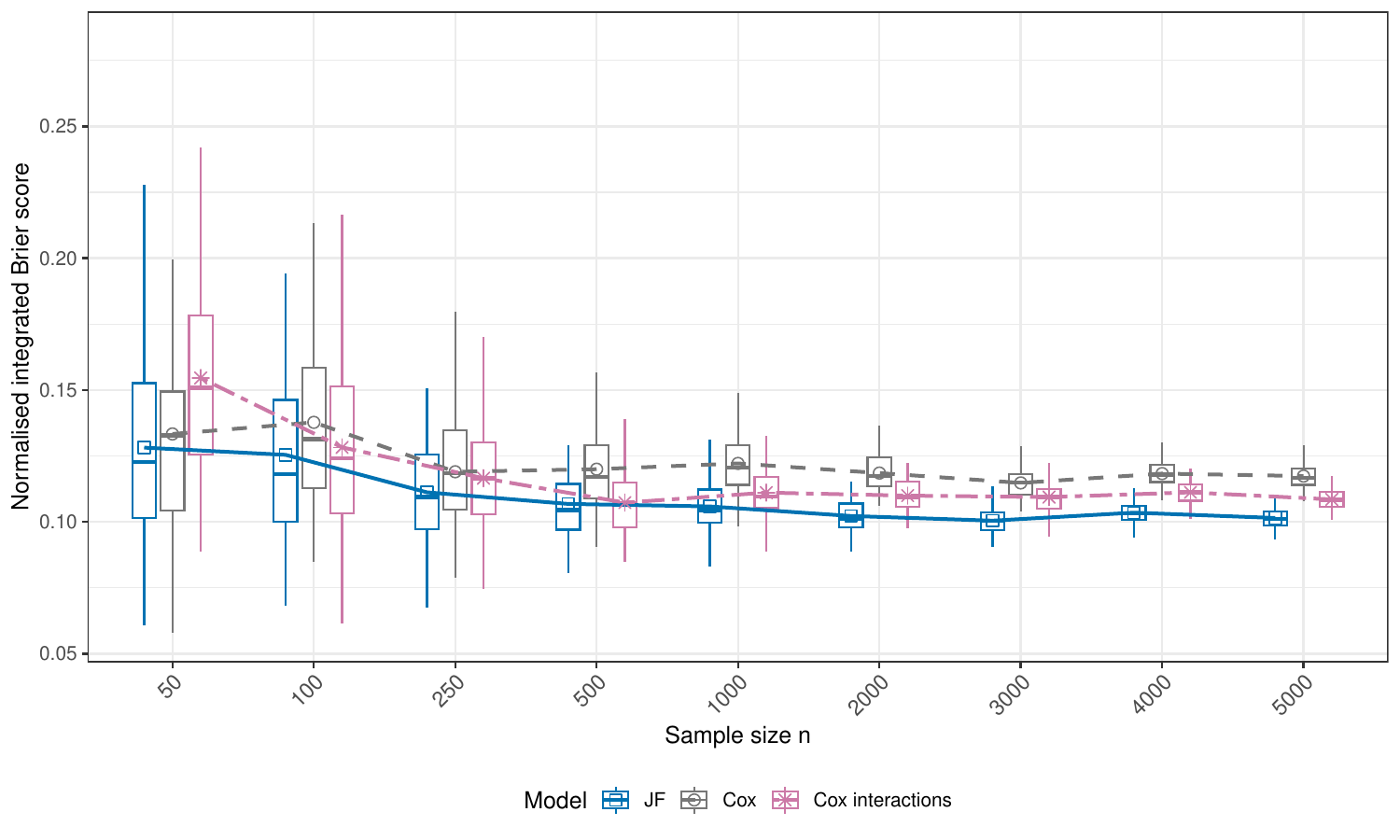}
    \caption{Non-proportional hazards study, $100$ replications. Normalised test-set IBS of the jump forest (JF) and transition-wise Cox models (Cox and Cox interactions), as a function of sample size $n$.}
    \label{fig:nonph-absolute}
\end{figure}

Figure~\ref{fig:nonph-absolute} shows the normalised IBS values underlying Figure~\ref{fig:nonph}. On the absolute scale, the median JF IBS decreases from $0.104$ at $n=500$ to $0.100$--$0.103$ for $n\geq2000$, whereas Cox interactions remains at $0.107$--$0.111$ for $n\geq500$.

\subsection{Coverage study}\label{subsection:app-coverage}\label{subsection:coverage}
The coverage study uses the model of Subsection~\ref{subsection:nonph}, with two modifications. The initial distribution is $(1/2, 1/2, 0)$, so that both non-absorbing states are occupied with positive probability throughout and Assumption~\ref{asm:nonvanishing} holds, and the analysis is restricted to the time interval $[0, 4]$. Occupation probabilities are computed using an estimated initial distribution. Since all censoring times exceed $5$, no observation is censored on $[0, 4]$, and the study isolates the sampling variability of the estimators from the effect of censoring. We consider two covariate settings. Without noise, the covariates are $(X_1, X_2, X_3)$, which take $16$ distinct values; with noise, the covariates $X_4 \sim \text{Bern}(1/2)$ and $X_5$, uniformly distributed on $\{1, 2, 3\}$, which do not affect the process, are added, giving $96$ distinct values. Both settings satisfy Assumption~\ref{asm:featurespace}, and the cumulative transition rates are continuous. For each of the sample sizes $n = 500$, $1000$, $2000$, $3000$, $4000$ and $5000$, the splitting rule of the forest, chosen among the five rules described in Subsection~\ref{subsection:splittingrules} and Supplementary Section~\ref{subsection:alternativesplitting}, and its minimum node size are selected by the OOB integrated Brier score on an independent pilot data set, which is shared by the two covariate settings (Table~\ref{tab:coverage-tuning}); the selected settings are held fixed across evaluation replications. Without noise, the selected minimum node size increases from $50$ at $n = 500$ to $250$ at $n = 5000$. Both estimators use all $n$ observations. The forest uses the number of trees, resampling scheme and candidate split settings of Subsection~\ref{subsection:design}. For $B=500$ and $m_n=0.7n$ with sampling without replacement, \eqref{eq:kappa} gives $\kappa_n\approx1.0009/n$, so the Monte Carlo contribution to the asymptotic variance is negligible. The Nelson--Aalen estimators in the terminal nodes are evaluated at all event times, and each setting is replicated $100$ times.

We compute the nominal $95$ percent pointwise intervals and simultaneous bands of Corollary~\ref{cor:forestCI}, using $\kappa_n$ for subsampling without replacement, $z_{0.975}=1.96$ and $c_{0.95}$ rounded to $2.24$, and truncate negative lower limits at zero. The benchmark is the fully stratified Nelson--Aalen estimator (NA) for each $X=x$, with the variance estimator of Proposition~\ref{prop:markovvariance} computed from the same observations, and intervals and bands of the same form. Coverage is averaged equally over covariate values within the groups $X_1=0$ and $X_1=1$; a band covers only if it contains the true cumulative rate at all event times on $[0,4]$ and their left limits, and unavailable intervals count as non-coverage.

Monte Carlo standard errors are at most $3.6$ percentage points, computed from the $100$ replication-level group averages; forest variances use the node estimator and scaling of Corollary~\ref{cor:forestCI}; band coverage is checked at $t=4$, all event times across trees and their left limits.

Without noise, the selected minimum node size exceeds the expected subsampled count $0.7n/16$ per covariate combination at every sample size: at $n=5000$, the two are $250$ and approximately $219$. This makes separation of all covariate combinations difficult.

Table~\ref{tab:coverage-uniform-full} shows three behaviours. JF coverage is close to nominal for the covariate-independent transition $2\to1$, ranging from $93.4$ to $98.4$ percent across both groups and covariate settings at $n=500$, $2000$ and $5000$. Pooling can also mitigate sparsity. With noise covariates and $n=500$, coverage for $1\to3$ in the group $X_1=1$ is $88.9$ percent for JF but $8.3$ percent for NA, which has no observed transition, and hence zero estimate and variance, in $91.7$ percent of the replication--covariate pairs. Pooling improves coverage substantially here, although JF remains below nominal.

\begin{figure}[p]
    \centering
    \includegraphics[width=\linewidth]{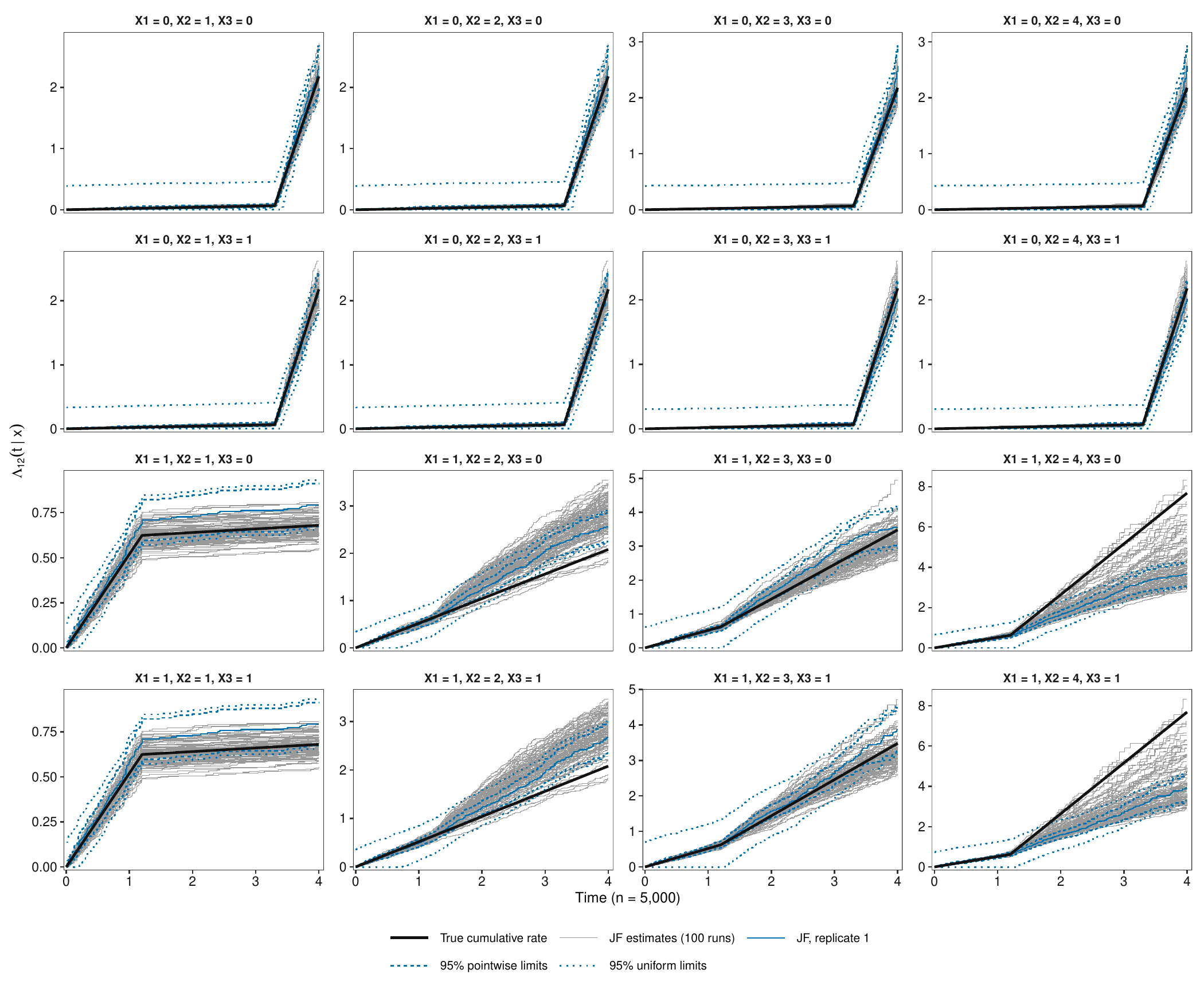}
    \caption{Coverage study without noise covariates, $n=5000$, transition $1\to2$. Cumulative transition rates for all $16$ covariate values. Grey curves show jump forest estimates from $100$ replications; black curves show the true rates. Blue curves show the estimate, nominal $95$ percent pointwise intervals and simultaneous band from the first replication.}
    \label{fig:curves-12}
\end{figure}

Pooling heterogeneous rates can instead create bias that invalidates the bands. Without noise, JF coverage for $1\to2$ in the group $X_1=1$ is only $41.1$--$65.0$ percent. At $n=5000$ and $t=4$, its group-averaged bias is $-0.71$, against $-0.09$ for NA, while its mean interval is less than half as wide ($0.84$ against $1.76$; Table~\ref{tab:coverage-bias}). Figure~\ref{fig:curves-12} reveals opposing errors across $X_2$: JF underestimates the highest cumulative rate and overestimates some lower rates. Pointwise coverage deteriorates after $t=1.2$, when $X_2$ begins to affect the transition (Table~\ref{tab:coverage-times-without}).

Tables~\ref{tab:coverage-uniform-full}--\ref{tab:coverage-times-with} report simultaneous coverage and pointwise coverage over reporting times. Early pointwise intervals often undercover because few transitions have occurred. The constant-width band instead uses the variance at $t=4$. Conversely, a band can fail because of bias at one time even when intervals cover at other times. These are different coverage criteria and need not have the same ordering. Log-scale intervals may improve the normal approximation when the estimated cumulative rate is positive, but do not resolve zero-event estimates or pooling bias.

\begin{figure}[p]
    \centering
    \includegraphics[width=\linewidth]{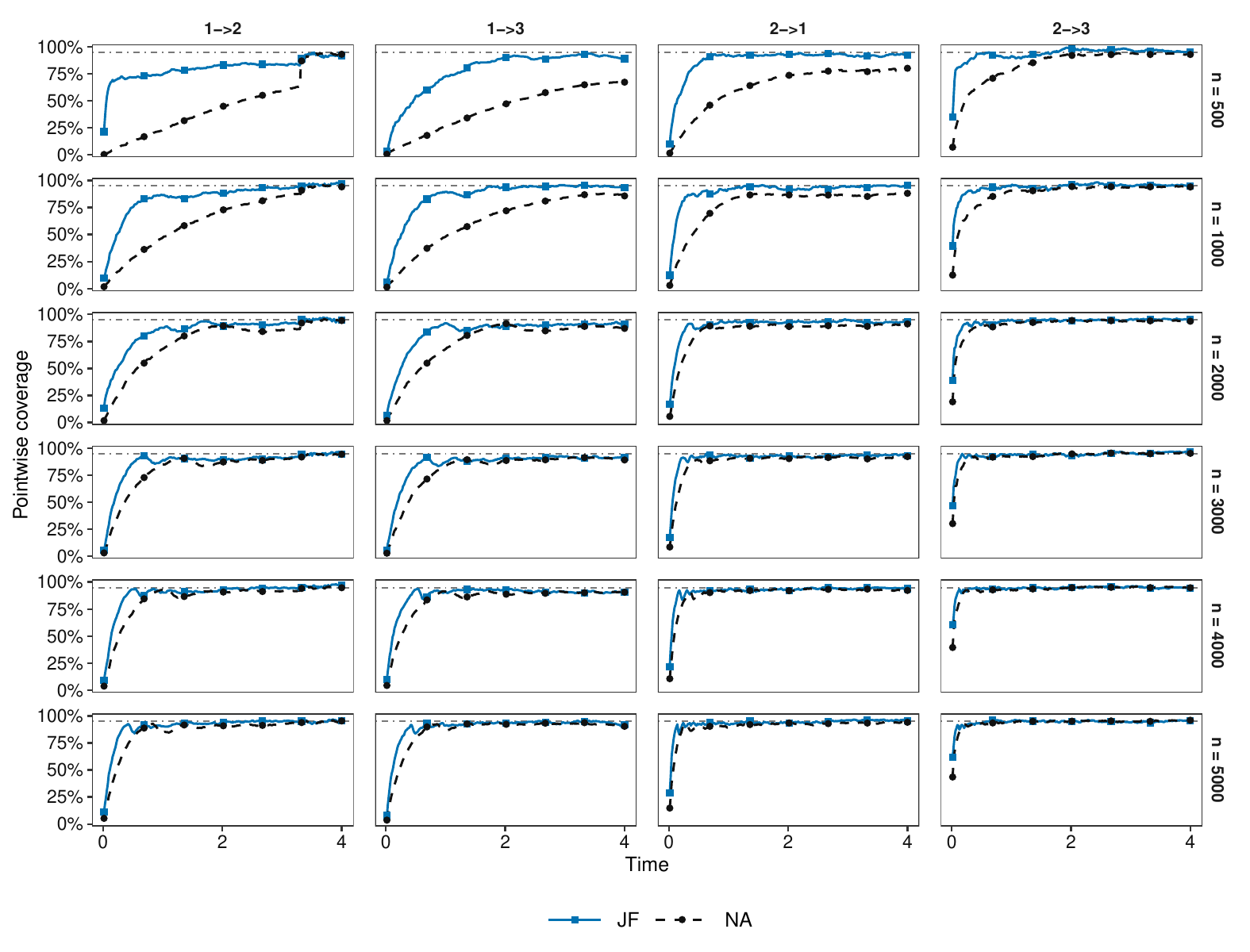}
    \caption{Coverage study without noise covariates, $X_1=0$, $100$ replications per sample size. Pointwise coverage of nominal $95$ percent intervals for the jump forest (JF) and stratified Nelson--Aalen estimator (NA), averaged equally over covariate values within the group. Columns correspond to the four transitions; rows to $n=500$, $1000$, $2000$, $3000$, $4000$ and $5000$.}
    \label{fig:coverage-pointwise-without}
\end{figure}

\begin{figure}[p]
    \centering
    \includegraphics[width=\linewidth]{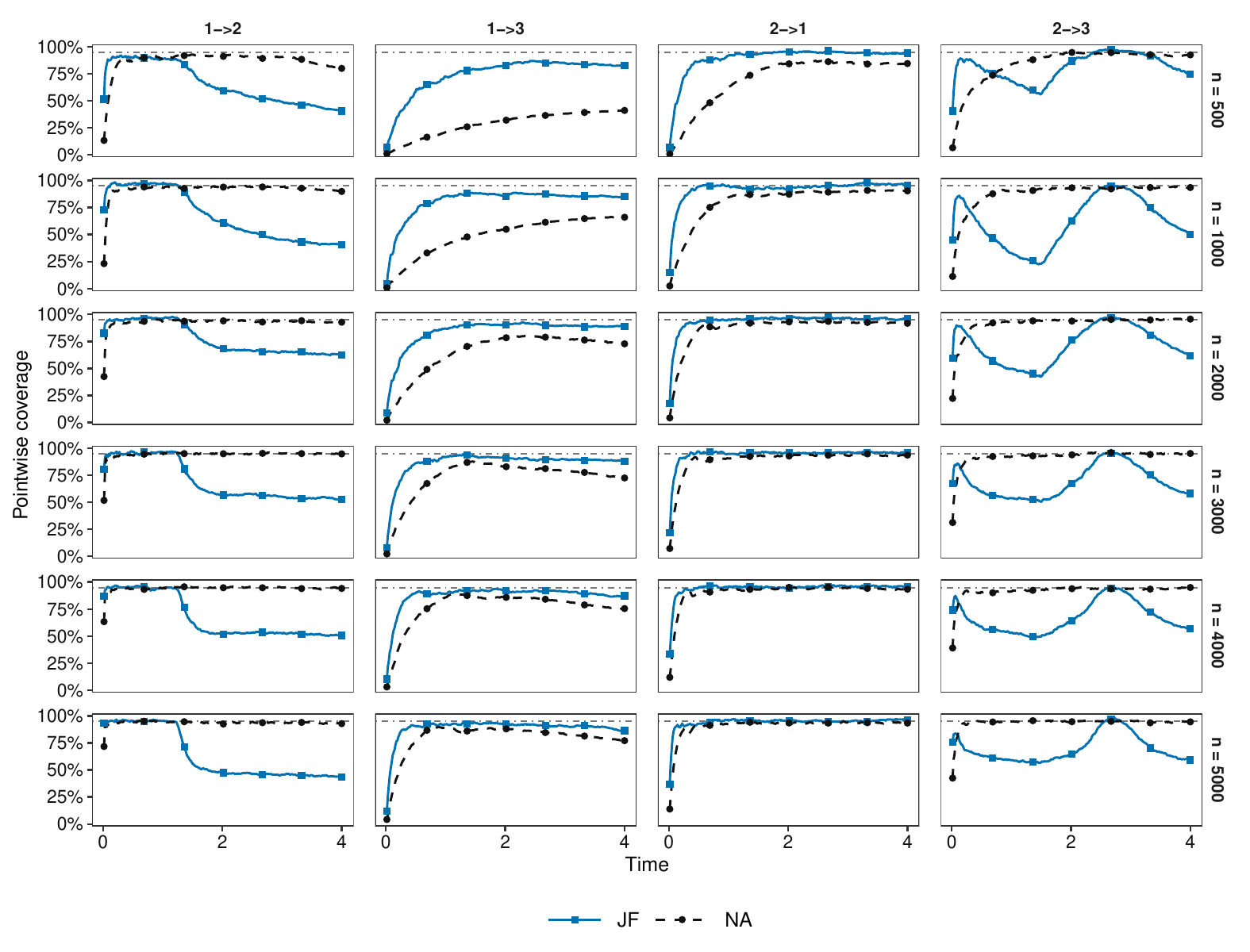}
    \caption{Coverage study without noise covariates, $X_1=1$, $100$ replications per sample size. Pointwise coverage of nominal $95$ percent intervals for the jump forest (JF) and stratified Nelson--Aalen estimator (NA), averaged equally over covariate values within the group. Columns correspond to the four transitions; rows to $n=500$, $1000$, $2000$, $3000$, $4000$ and $5000$.}
    \label{fig:coverage-pointwise-without-1}
\end{figure}

\begin{figure}[p]
    \centering
    \includegraphics[width=\linewidth]{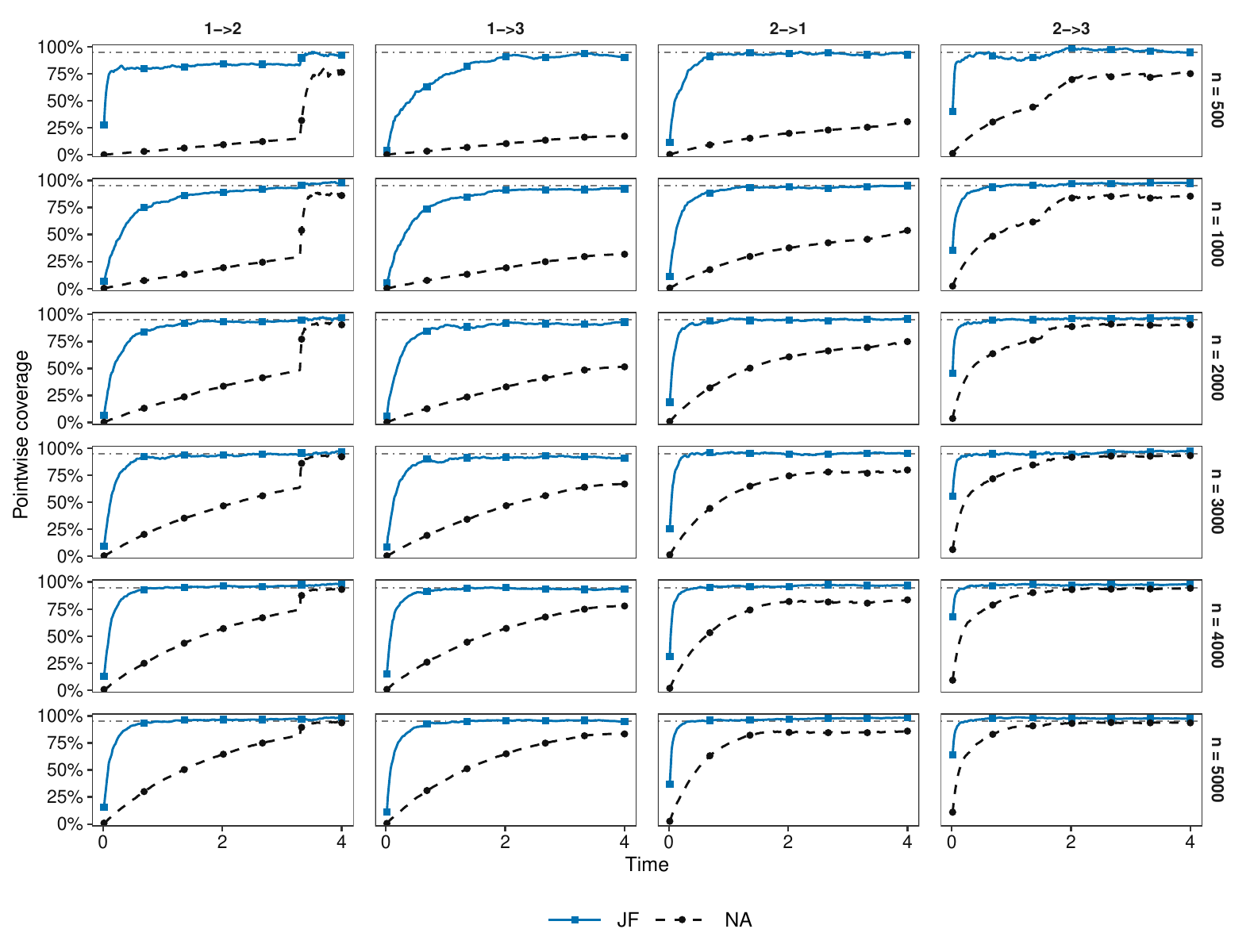}
    \caption{Coverage study with noise covariates, $X_1=0$, $100$ replications per sample size. Pointwise coverage of nominal $95$ percent intervals for the jump forest (JF) and stratified Nelson--Aalen estimator (NA), averaged equally over covariate values within the group. Columns correspond to the four transitions; rows to $n=500$, $1000$, $2000$, $3000$, $4000$ and $5000$.}
    \label{fig:coverage-pointwise-with}
\end{figure}

\begin{figure}[p]
    \centering
    \includegraphics[width=\linewidth]{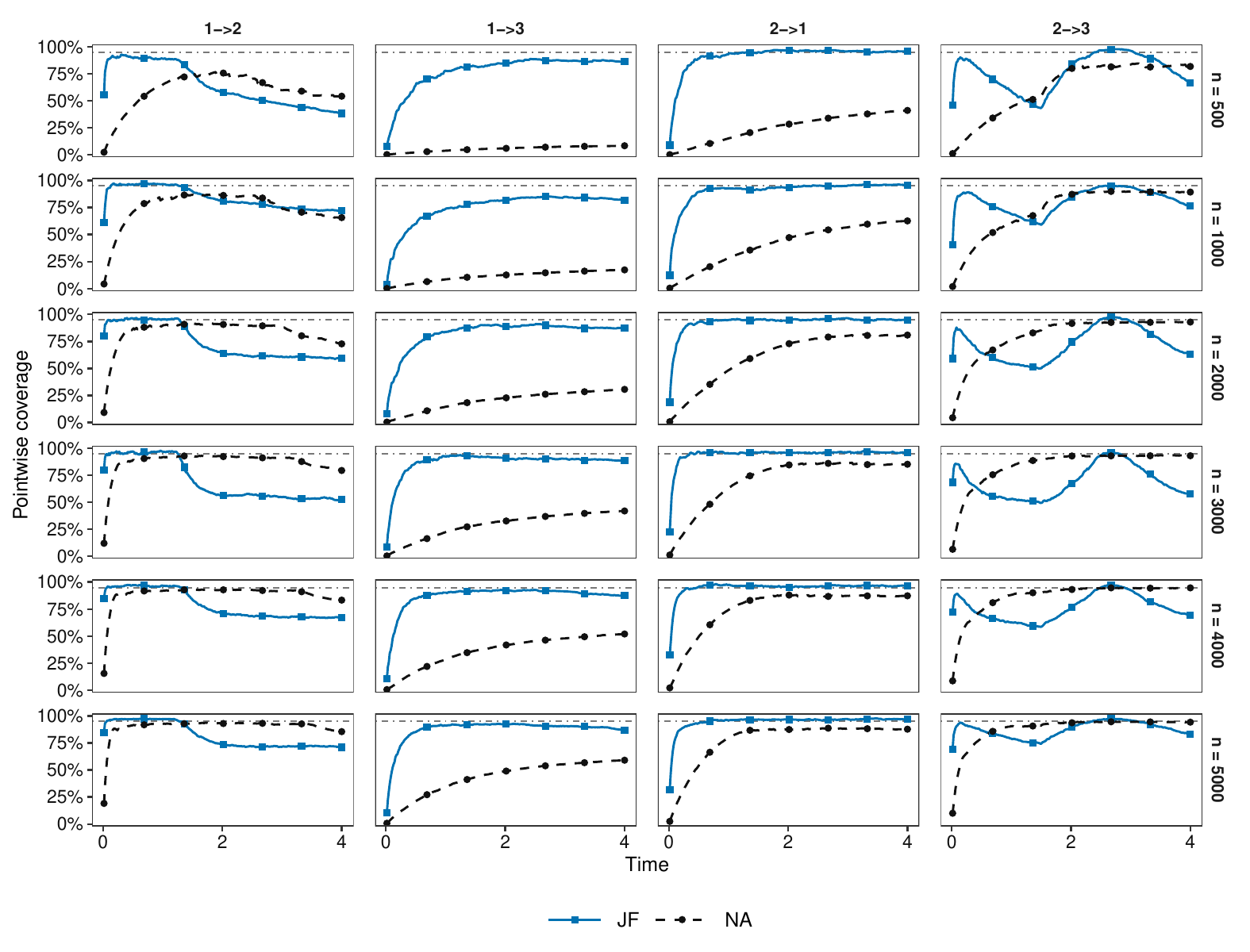}
    \caption{Coverage study with noise covariates, $X_1=1$, $100$ replications per sample size. Pointwise coverage of nominal $95$ percent intervals for the jump forest (JF) and stratified Nelson--Aalen estimator (NA), averaged equally over covariate values within the group. Columns correspond to the four transitions; rows to $n=500$, $1000$, $2000$, $3000$, $4000$ and $5000$.}
    \label{fig:coverage-pointwise-with-1}
\end{figure}

\begin{figure}[p]
    \centering
    \includegraphics[width=\linewidth]{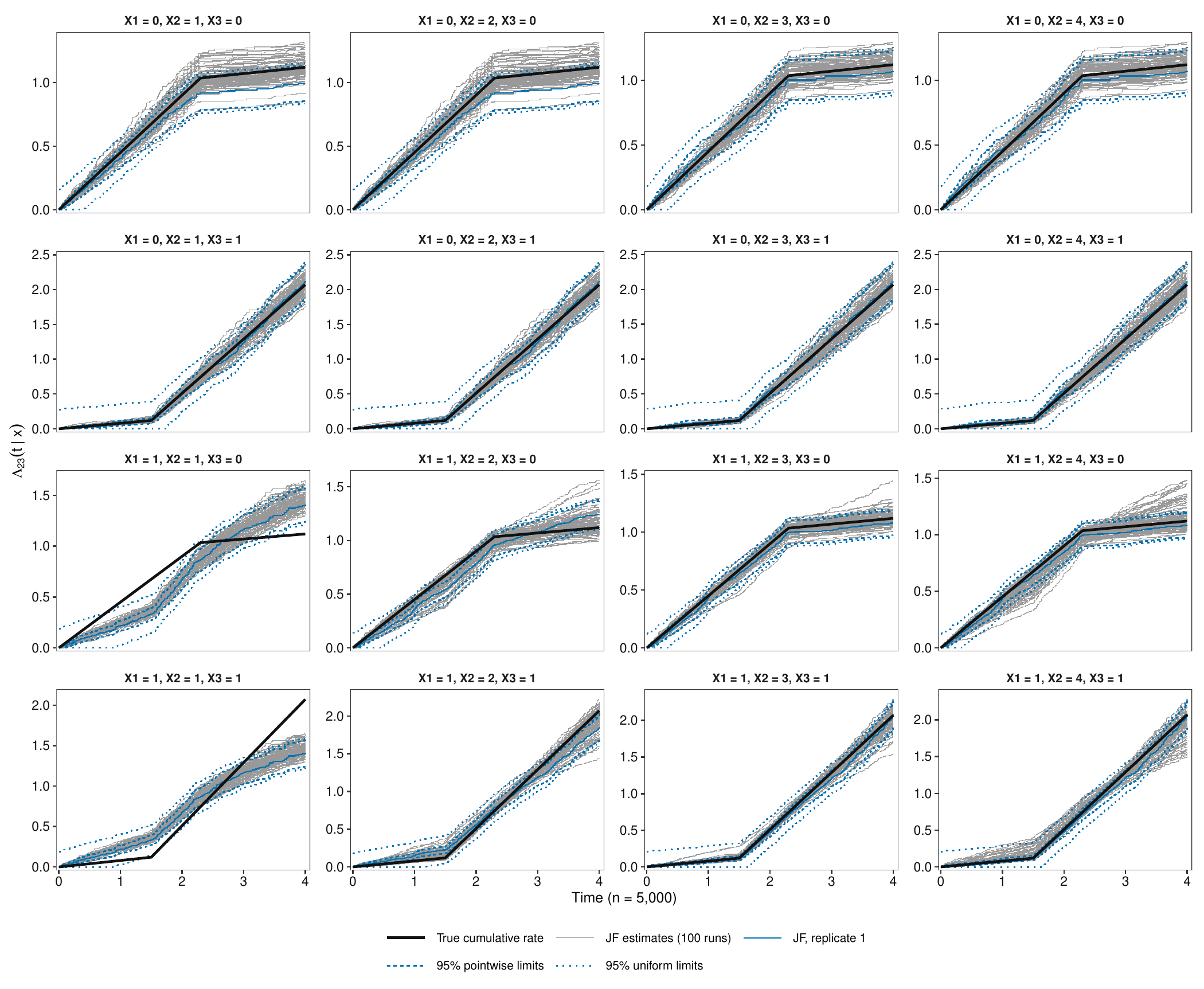}
    \caption{Coverage study without noise covariates, $n=5000$, transition $2\to3$. Cumulative transition rates for all $16$ covariate values. Grey curves show jump forest estimates from $100$ replications; black curves show the true rates. Blue curves show the estimate, nominal $95$ percent pointwise intervals and simultaneous band from the first replication.}
    \label{fig:curves-23}
\end{figure}

Table~\ref{tab:coverage-bias} separates bias, RMSE and interval width. Small group-averaged bias does not imply accurate estimation at every covariate value: for $2\to3$ in the group $X_1=1$, biases of opposite signs largely cancel. Its RMSE at $t=4$ is nevertheless $0.28$ for the forest and $0.14$ for NA. Figures~\ref{fig:curves-12} and~\ref{fig:curves-23} show this heterogeneity. For transition $2\to3$ with $X_1=X_2=1$, the estimates do not separate the two values of $X_3$, whose rates differ. Figures~\ref{fig:coverage-pointwise-without}--\ref{fig:coverage-pointwise-with-1} locate coverage failures in time.

\begin{table}[H]
    \centering
    \footnotesize
    \begin{tabular}{r lrr lrr}
        \toprule
        & \multicolumn{3}{c}{Without noise covariates} & \multicolumn{3}{c}{With noise covariates} \\
        \cmidrule(lr){2-4} \cmidrule(lr){5-7}
        $n$ & Splitting rule & Min.\ node size & OOB IBS & Splitting rule & Min.\ node size & OOB IBS \\
        \midrule
        500 & Peto--Prentice & 50 & 0.1191 & Peto--Prentice & 50 & 0.1190 \\
        1000 & Gehan & 90 & 0.1133 & Gehan & 50 & 0.1129 \\
        2000 & Peto--Prentice & 150 & 0.1085 & Gehan & 125 & 0.1084 \\
        3000 & Gehan & 175 & 0.1081 & Gehan & 200 & 0.1079 \\
        4000 & Gehan & 200 & 0.1074 & Gehan & 175 & 0.1073 \\
        5000 & Gehan & 250 & 0.1069 & Peto--Prentice & 175 & 0.1063 \\
        \bottomrule
    \end{tabular}
    \caption{Tuning parameters of the jump forest in the coverage study. For each sample size, the splitting rule, chosen among the five rules described in Subsection~\ref{subsection:splittingrules} and Supplementary Section~\ref{subsection:alternativesplitting}, and the minimum node size, chosen from a candidate grid depending on the number of observations, minimise the normalised out-of-bag integrated Brier score (OOB IBS) on an independent pilot data set, which is shared by the two covariate settings; they are held fixed over the $100$ evaluation data sets.}
    \label{tab:coverage-tuning}
\end{table}

\begin{table}[tbp]
    \centering
    \footnotesize
    \begin{tabular}{llll rrrrrr}
        \toprule
        Covariates & Group & Transition & Model & $n = 500$ & $1000$ & $2000$ & $3000$ & $4000$ & $5000$ \\
        \midrule
        Without noise & $X_1 = 0$ & $1 \to 2$ & JF & 92.1 & 97.8 & 95.9 & 96.5 & 97.4 & 96.2 \\
         &  &  & NA & 93.9 & 95.0 & 95.5 & 95.1 & 96.2 & 95.2 \\
         &  & $1 \to 3$ & JF & 91.2 & 94.8 & 93.8 & 92.0 & 91.4 & 93.5 \\
         &  &  & NA & 67.1 & 88.5 & 89.6 & 91.4 & 91.6 & 92.5 \\
         &  & $2 \to 1$ & JF & 93.4 & 95.4 & 93.6 & 94.0 & 94.8 & 96.0 \\
         &  &  & NA & 81.0 & 89.0 & 90.8 & 91.6 & 93.4 & 94.9 \\
         &  & $2 \to 3$ & JF & 96.6 & 95.1 & 96.2 & 96.0 & 95.8 & 94.8 \\
         &  &  & NA & 93.6 & 94.2 & 95.6 & 95.9 & 95.4 & 95.9 \\
        \addlinespace
         & $X_1 = 1$ & $1 \to 2$ & JF & 41.1 & 42.9 & 65.0 & 54.0 & 50.6 & 44.0 \\
         &  &  & NA & 82.2 & 92.0 & 93.6 & 94.9 & 95.8 & 93.1 \\
         &  & $1 \to 3$ & JF & 85.1 & 87.2 & 90.4 & 89.2 & 89.0 & 87.8 \\
         &  &  & NA & 41.0 & 66.6 & 75.9 & 75.2 & 78.0 & 79.5 \\
         &  & $2 \to 1$ & JF & 94.4 & 96.9 & 97.4 & 96.5 & 96.6 & 97.2 \\
         &  &  & NA & 85.9 & 91.2 & 93.9 & 94.6 & 93.9 & 94.0 \\
         &  & $2 \to 3$ & JF & 75.4 & 45.8 & 61.9 & 55.8 & 56.5 & 61.1 \\
         &  &  & NA & 93.1 & 93.5 & 95.6 & 95.5 & 94.6 & 94.8 \\
        \midrule
        With noise & $X_1 = 0$ & $1 \to 2$ & JF & 92.9 & 98.5 & 97.8 & 96.7 & 99.0 & 98.6 \\
         &  &  & NA & 80.6 & 89.2 & 92.8 & 93.8 & 94.2 & 94.5 \\
         &  & $1 \to 3$ & JF & 92.6 & 93.5 & 94.2 & 93.0 & 94.8 & 95.7 \\
         &  &  & NA & 17.1 & 31.8 & 51.5 & 66.9 & 77.9 & 83.5 \\
         &  & $2 \to 1$ & JF & 94.4 & 95.7 & 96.8 & 96.5 & 97.6 & 98.4 \\
         &  &  & NA & 30.6 & 53.7 & 74.3 & 80.1 & 84.0 & 86.7 \\
         &  & $2 \to 3$ & JF & 95.7 & 98.1 & 97.4 & 97.4 & 98.2 & 98.3 \\
         &  &  & NA & 76.5 & 87.3 & 91.2 & 93.5 & 94.4 & 94.0 \\
        \addlinespace
         & $X_1 = 1$ & $1 \to 2$ & JF & 39.1 & 73.8 & 60.7 & 54.0 & 68.5 & 73.6 \\
         &  &  & NA & 55.3 & 67.5 & 76.3 & 82.5 & 85.4 & 87.8 \\
         &  & $1 \to 3$ & JF & 88.9 & 84.7 & 88.9 & 89.8 & 89.8 & 89.9 \\
         &  &  & NA & 8.3 & 17.3 & 30.6 & 42.0 & 52.0 & 59.1 \\
         &  & $2 \to 1$ & JF & 96.2 & 96.5 & 96.4 & 96.4 & 97.2 & 97.5 \\
         &  &  & NA & 41.0 & 62.7 & 82.2 & 86.8 & 88.8 & 89.1 \\
         &  & $2 \to 3$ & JF & 70.7 & 78.5 & 61.5 & 55.6 & 68.8 & 85.0 \\
         &  &  & NA & 84.1 & 90.5 & 94.0 & 93.8 & 95.2 & 95.1 \\
        \bottomrule
    \end{tabular}
    \caption{Simultaneous coverage (\%) of the nominal $95$ percent confidence bands on $[0, 4]$ for the jump forest (JF) and the fully stratified Nelson--Aalen estimator (NA), for all sample sizes, without noise covariates ($X_1, X_2, X_3$; $16$ covariate values) and with noise covariates ($X_1, \dots, X_5$; $96$ covariate values). Coverage is averaged with equal weights over the covariate values within each group; intervals that cannot be computed count as non-coverage; $100$ replications per sample size.}
    \label{tab:coverage-uniform-full}
\end{table}

\begin{table}[tbp]
    \centering
    \footnotesize
    \begin{tabular}{llll rrrrrr}
        \toprule
        Covariates & Group & Transition & Model & $n = 500$ & $1000$ & $2000$ & $3000$ & $4000$ & $5000$ \\
        \midrule
        Without noise & $X_1 = 0$ & $1 \to 2$ & JF & 80.4 & 84.9 & 85.0 & 86.8 & 89.9 & 90.6 \\
         &  &  & NA & 45.6 & 65.0 & 75.4 & 81.3 & 85.1 & 86.1 \\
         &  & $1 \to 3$ & JF & 78.2 & 86.3 & 83.8 & 85.8 & 88.3 & 90.2 \\
         &  &  & NA & 42.8 & 63.8 & 74.9 & 80.7 & 83.9 & 86.6 \\
         &  & $2 \to 1$ & JF & 88.2 & 90.6 & 90.3 & 91.2 & 92.1 & 93.3 \\
         &  &  & NA & 63.7 & 78.2 & 85.0 & 87.8 & 89.8 & 90.5 \\
         &  & $2 \to 3$ & JF & 93.6 & 93.6 & 93.3 & 93.8 & 94.2 & 94.3 \\
         &  &  & NA & 83.5 & 88.4 & 91.0 & 92.6 & 93.2 & 93.8 \\
        \addlinespace
         & $X_1 = 1$ & $1 \to 2$ & JF & 65.7 & 67.5 & 77.2 & 69.9 & 67.1 & 63.0 \\
         &  &  & NA & 87.3 & 92.2 & 93.1 & 94.4 & 94.3 & 93.8 \\
         &  & $1 \to 3$ & JF & 74.8 & 81.2 & 84.5 & 87.0 & 87.9 & 89.3 \\
         &  &  & NA & 28.7 & 49.7 & 65.8 & 73.6 & 77.2 & 80.3 \\
         &  & $2 \to 1$ & JF & 89.6 & 92.5 & 94.0 & 94.4 & 94.5 & 94.5 \\
         &  &  & NA & 70.7 & 80.8 & 87.5 & 89.5 & 91.2 & 91.1 \\
         &  & $2 \to 3$ & JF & 81.4 & 61.8 & 71.7 & 68.9 & 67.1 & 69.6 \\
         &  &  & NA & 84.6 & 88.5 & 92.1 & 92.6 & 92.2 & 94.0 \\
        \midrule
        With noise & $X_1 = 0$ & $1 \to 2$ & JF & 83.3 & 83.3 & 87.7 & 89.9 & 92.7 & 93.2 \\
         &  &  & NA & 18.4 & 27.6 & 38.0 & 47.3 & 54.1 & 59.5 \\
         &  & $1 \to 3$ & JF & 79.6 & 82.3 & 85.7 & 87.5 & 91.1 & 92.2 \\
         &  &  & NA & 9.9 & 18.6 & 30.9 & 42.6 & 51.9 & 57.9 \\
         &  & $2 \to 1$ & JF & 89.5 & 89.9 & 92.8 & 93.8 & 95.2 & 96.0 \\
         &  &  & NA & 18.0 & 33.6 & 53.2 & 63.9 & 70.2 & 75.1 \\
         &  & $2 \to 3$ & JF & 93.6 & 94.7 & 94.8 & 95.0 & 96.9 & 97.2 \\
         &  &  & NA & 55.3 & 69.3 & 78.4 & 83.2 & 86.4 & 87.6 \\
        \addlinespace
         & $X_1 = 1$ & $1 \to 2$ & JF & 64.5 & 84.1 & 74.3 & 69.8 & 79.2 & 81.6 \\
         &  &  & NA & 59.9 & 75.2 & 84.0 & 87.8 & 89.5 & 90.7 \\
         &  & $1 \to 3$ & JF & 78.1 & 75.4 & 82.7 & 87.3 & 87.3 & 87.7 \\
         &  &  & NA & 5.5 & 11.6 & 20.4 & 29.3 & 37.0 & 43.3 \\
         &  & $2 \to 1$ & JF & 91.5 & 90.9 & 92.9 & 94.6 & 95.0 & 94.9 \\
         &  &  & NA & 25.3 & 41.6 & 61.8 & 71.1 & 76.1 & 78.2 \\
         &  & $2 \to 3$ & JF & 77.4 & 81.6 & 72.7 & 68.8 & 76.4 & 87.6 \\
         &  &  & NA & 62.5 & 73.5 & 81.5 & 84.7 & 87.1 & 88.6 \\
        \bottomrule
    \end{tabular}
    \caption{Pointwise coverage (\%) of the nominal $95$ percent confidence intervals of Corollary~\ref{cor:forestCI}(i), averaged over a grid of reporting times consisting of $250$ common event times on $(0, 4]$ and the change points $1.2$, $2.3$ and $3.3$, for the jump forest (JF) and the fully stratified Nelson--Aalen estimator (NA). Coverage is averaged with equal weights over the covariate values within each group; $100$ replications per sample size.}
    \label{tab:coverage-gridaverage}
\end{table}

\begin{table}[tbp]
    \centering
    \footnotesize
    \setlength{\tabcolsep}{2.5pt}
    \begin{tabular}{lll rrr rrr rrr rrr}
        \toprule
        & & & \multicolumn{3}{c}{$t = 1.2$} & \multicolumn{3}{c}{$t = 2$} & \multicolumn{3}{c}{$t = 3.3$} & \multicolumn{3}{c}{$t = 4$} \\
        \cmidrule(lr){4-6} \cmidrule(lr){7-9} \cmidrule(lr){10-12} \cmidrule(lr){13-15}
        Group & Transition & Model & $n = 500$ & $2000$ & $5000$ & $n = 500$ & $2000$ & $5000$ & $n = 500$ & $2000$ & $5000$ & $n = 500$ & $2000$ & $5000$ \\
        \midrule
        $X_1 = 0$ & $1 \to 2$ & JF & 78.8 & 85.9 & 92.2 & 83.1 & 89.1 & 93.1 & 82.8 & 91.2 & 94.8 & 91.8 & 94.6 & 95.5 \\
         &  & NA & 27.8 & 74.9 & 88.8 & 44.8 & 88.9 & 91.2 & 63.1 & 86.6 & 92.2 & 93.0 & 94.2 & 95.4 \\
         & $1 \to 3$ & JF & 75.9 & 86.1 & 91.2 & 89.5 & 89.2 & 93.9 & 92.6 & 90.6 & 94.4 & 88.9 & 91.0 & 91.8 \\
         &  & NA & 30.5 & 75.8 & 90.2 & 46.8 & 91.2 & 92.8 & 64.5 & 88.9 & 93.4 & 67.1 & 87.0 & 90.4 \\
         & $2 \to 1$ & JF & 91.5 & 93.0 & 93.2 & 92.4 & 91.9 & 93.8 & 91.4 & 92.8 & 96.2 & 91.9 & 93.2 & 95.8 \\
         &  & NA & 60.4 & 89.4 & 92.0 & 73.1 & 88.6 & 93.2 & 76.5 & 88.9 & 93.6 & 80.0 & 91.1 & 94.1 \\
         & $2 \to 3$ & JF & 90.6 & 93.0 & 95.5 & 98.8 & 93.8 & 94.8 & 96.4 & 95.1 & 94.5 & 95.0 & 95.4 & 95.4 \\
         &  & NA & 82.8 & 92.9 & 94.8 & 92.4 & 93.6 & 95.0 & 93.1 & 94.1 & 95.1 & 92.8 & 93.6 & 96.0 \\
        \addlinespace
        $X_1 = 1$ & $1 \to 2$ & JF & 88.1 & 97.0 & 94.6 & 59.2 & 68.6 & 47.6 & 46.5 & 65.0 & 45.0 & 40.4 & 62.5 & 43.6 \\
         &  & NA & 90.9 & 94.5 & 94.0 & 91.2 & 94.2 & 93.4 & 88.1 & 94.1 & 93.9 & 79.9 & 92.8 & 92.9 \\
         & $1 \to 3$ & JF & 76.1 & 88.1 & 92.0 & 82.8 & 90.4 & 92.2 & 83.6 & 89.0 & 91.0 & 82.9 & 88.8 & 86.4 \\
         &  & NA & 23.9 & 66.4 & 86.1 & 31.9 & 78.4 & 87.5 & 39.1 & 76.2 & 81.4 & 41.0 & 72.8 & 77.1 \\
         & $2 \to 1$ & JF & 93.4 & 95.2 & 95.9 & 94.6 & 96.4 & 95.9 & 94.1 & 95.6 & 94.6 & 94.0 & 96.4 & 96.2 \\
         &  & NA & 68.9 & 91.1 & 93.2 & 84.1 & 93.1 & 93.8 & 83.9 & 92.8 & 93.8 & 84.4 & 91.8 & 93.4 \\
         & $2 \to 3$ & JF & 63.2 & 48.0 & 58.6 & 86.0 & 75.1 & 64.6 & 91.6 & 82.2 & 72.2 & 74.8 & 61.4 & 59.2 \\
         &  & NA & 86.2 & 93.1 & 94.2 & 95.0 & 93.8 & 94.8 & 93.1 & 95.2 & 94.1 & 92.4 & 95.6 & 94.5 \\
        \bottomrule
    \end{tabular}
    \caption{Pointwise coverage (\%) of the nominal $95$ percent confidence intervals at the times $t \in \{1.2, 2, 3.3, 4\}$ for $n \in \{500, 2000, 5000\}$, without noise covariates, for the jump forest (JF) and the fully stratified Nelson--Aalen estimator (NA). Coverage is averaged with equal weights over the covariate values within each group; $100$ replications per sample size.}
    \label{tab:coverage-times-without}
\end{table}

\begin{table}[tbp]
    \centering
    \footnotesize
    \setlength{\tabcolsep}{2.5pt}
    \begin{tabular}{lll rrr rrr rrr rrr}
        \toprule
        & & & \multicolumn{3}{c}{$t = 1.2$} & \multicolumn{3}{c}{$t = 2$} & \multicolumn{3}{c}{$t = 3.3$} & \multicolumn{3}{c}{$t = 4$} \\
        \cmidrule(lr){4-6} \cmidrule(lr){7-9} \cmidrule(lr){10-12} \cmidrule(lr){13-15}
        Group & Transition & Model & $n = 500$ & $2000$ & $5000$ & $n = 500$ & $2000$ & $5000$ & $n = 500$ & $2000$ & $5000$ & $n = 500$ & $2000$ & $5000$ \\
        \midrule
        $X_1 = 0$ & $1 \to 2$ & JF & 82.1 & 89.8 & 95.6 & 84.2 & 93.5 & 96.7 & 82.9 & 94.3 & 96.9 & 92.1 & 97.0 & 98.2 \\
         &  & NA & 5.3 & 21.2 & 46.5 & 9.3 & 33.4 & 64.3 & 15.2 & 48.4 & 81.8 & 76.3 & 90.4 & 93.6 \\
         & $1 \to 3$ & JF & 77.4 & 88.8 & 94.2 & 90.3 & 92.0 & 95.5 & 93.1 & 90.6 & 95.9 & 89.9 & 92.8 & 94.5 \\
         &  & NA & 6.1 & 21.2 & 46.5 & 10.2 & 32.7 & 64.7 & 16.2 & 48.2 & 81.5 & 17.1 & 51.5 & 83.3 \\
         & $2 \to 1$ & JF & 92.8 & 95.7 & 95.9 & 93.4 & 94.9 & 97.0 & 92.5 & 95.4 & 98.0 & 92.5 & 95.8 & 98.3 \\
         &  & NA & 13.9 & 46.9 & 79.0 & 19.7 & 60.4 & 84.6 & 25.4 & 69.2 & 84.4 & 30.6 & 74.8 & 86.0 \\
         & $2 \to 3$ & JF & 87.6 & 95.0 & 98.6 & 98.7 & 95.8 & 98.0 & 96.3 & 96.6 & 97.7 & 94.7 & 96.4 & 97.8 \\
         &  & NA & 41.9 & 73.8 & 90.4 & 69.0 & 88.2 & 93.1 & 71.4 & 89.9 & 93.6 & 75.0 & 90.4 & 93.6 \\
        \addlinespace
        $X_1 = 1$ & $1 \to 2$ & JF & 88.5 & 96.1 & 96.9 & 57.8 & 64.2 & 73.8 & 44.0 & 60.5 & 72.3 & 38.2 & 59.2 & 71.0 \\
         &  & NA & 68.8 & 90.4 & 93.0 & 75.2 & 90.7 & 93.0 & 58.7 & 80.7 & 92.5 & 54.2 & 72.7 & 85.5 \\
         & $1 \to 3$ & JF & 80.3 & 85.8 & 91.7 & 84.8 & 88.6 & 92.4 & 86.5 & 87.6 & 90.1 & 86.6 & 87.2 & 86.9 \\
         &  & NA & 4.3 & 16.8 & 38.5 & 5.9 & 22.7 & 48.7 & 7.8 & 28.4 & 56.6 & 8.3 & 30.6 & 59.0 \\
         & $2 \to 1$ & JF & 94.3 & 95.1 & 96.7 & 96.1 & 94.7 & 96.8 & 95.5 & 94.6 & 97.2 & 95.6 & 94.8 & 96.9 \\
         &  & NA & 18.1 & 55.0 & 84.7 & 28.2 & 72.5 & 87.4 & 37.6 & 80.4 & 88.4 & 41.0 & 80.9 & 87.7 \\
         & $2 \to 3$ & JF & 50.9 & 52.9 & 76.1 & 82.9 & 72.2 & 89.8 & 89.7 & 83.4 & 92.8 & 66.5 & 63.1 & 83.2 \\
         &  & NA & 49.0 & 80.1 & 90.1 & 79.3 & 91.6 & 93.5 & 81.0 & 92.7 & 94.5 & 81.7 & 92.9 & 94.2 \\
        \bottomrule
    \end{tabular}
    \caption{Pointwise coverage (\%) of the nominal $95$ percent confidence intervals at the times $t \in \{1.2, 2, 3.3, 4\}$ for $n \in \{500, 2000, 5000\}$, with noise covariates, for the jump forest (JF) and the fully stratified Nelson--Aalen estimator (NA). Coverage is averaged with equal weights over the covariate values within each group; $100$ replications per sample size.}
    \label{tab:coverage-times-with}
\end{table}

\begin{table}[tbp]
    \centering
    \scriptsize
    \setlength{\tabcolsep}{2.6pt}
    \begin{tabular}{lll rrrr rrrr rrrr}
        \toprule
        & & & \multicolumn{4}{c}{Bias} & \multicolumn{4}{c}{RMSE} & \multicolumn{4}{c}{Mean interval width} \\
        \cmidrule(lr){4-7} \cmidrule(lr){8-11} \cmidrule(lr){12-15}
        Group & Transition & Model & $t = 1.2$ & $2$ & $3.3$ & $4$ & $t = 1.2$ & $2$ & $3.3$ & $4$ & $t = 1.2$ & $2$ & $3.3$ & $4$ \\
        \midrule
        $X_1 = 0$ & $1 \to 2$ & JF & 0.000 & 0.000 & 0.000 & $+$0.001 & 0.009 & 0.011 & 0.014 & 0.154 & 0.033 & 0.043 & 0.056 & 0.595 \\
         &  & NA & 0.000 & 0.000 & 0.000 & $+$0.002 & 0.013 & 0.016 & 0.021 & 0.227 & 0.044 & 0.060 & 0.079 & 0.843 \\
         & $1 \to 3$ & JF & 0.000 & 0.000 & 0.000 & $-$0.001 & 0.009 & 0.011 & 0.014 & 0.019 & 0.033 & 0.044 & 0.056 & 0.071 \\
         &  & NA & 0.000 & 0.000 & 0.000 & $-$0.001 & 0.012 & 0.015 & 0.021 & 0.027 & 0.045 & 0.061 & 0.079 & 0.098 \\
         & $2 \to 1$ & JF & $+$0.001 & 0.000 & $+$0.003 & $+$0.003 & 0.017 & 0.023 & 0.036 & 0.038 & 0.065 & 0.091 & 0.143 & 0.154 \\
         &  & NA & $+$0.001 & 0.000 & $+$0.003 & $+$0.003 & 0.024 & 0.033 & 0.053 & 0.056 & 0.092 & 0.128 & 0.201 & 0.217 \\
         & $2 \to 3$ & JF & $-$0.001 & $-$0.002 & $-$0.003 & $-$0.002 & 0.036 & 0.061 & 0.095 & 0.104 & 0.130 & 0.230 & 0.377 & 0.405 \\
         &  & NA & $-$0.001 & $-$0.002 & $-$0.002 & $-$0.001 & 0.051 & 0.081 & 0.135 & 0.146 & 0.184 & 0.325 & 0.534 & 0.573 \\
        \addlinespace
        $X_1 = 1$ & $1 \to 2$ & JF & $-$0.001 & $-$0.106 & $-$0.475 & $-$0.710 & 0.047 & 0.487 & 1.360 & 1.858 & 0.186 & 0.335 & 0.645 & 0.835 \\
         &  & NA & $-$0.001 & $-$0.014 & $-$0.028 & $-$0.089 & 0.072 & 0.168 & 0.418 & 0.586 & 0.286 & 0.561 & 1.317 & 1.756 \\
         & $1 \to 3$ & JF & 0.000 & 0.000 & $-$0.001 & $-$0.004 & 0.009 & 0.013 & 0.022 & 0.028 & 0.035 & 0.053 & 0.086 & 0.102 \\
         &  & NA & 0.000 & 0.000 & $-$0.002 & $-$0.004 & 0.014 & 0.022 & 0.042 & 0.063 & 0.049 & 0.076 & 0.119 & 0.140 \\
         & $2 \to 1$ & JF & $-$0.001 & $-$0.002 & $-$0.003 & $-$0.005 & 0.012 & 0.016 & 0.024 & 0.028 & 0.053 & 0.068 & 0.097 & 0.114 \\
         &  & NA & $-$0.001 & $-$0.002 & $-$0.003 & $-$0.005 & 0.020 & 0.027 & 0.038 & 0.045 & 0.081 & 0.106 & 0.150 & 0.177 \\
         & $2 \to 3$ & JF & $-$0.009 & $-$0.006 & $-$0.017 & $-$0.047 & 0.124 & 0.113 & 0.136 & 0.278 & 0.107 & 0.168 & 0.256 & 0.301 \\
         &  & NA & $-$0.001 & 0.000 & $-$0.002 & $+$0.002 & 0.045 & 0.066 & 0.101 & 0.135 & 0.162 & 0.260 & 0.401 & 0.486 \\
        \bottomrule
    \end{tabular}
    \caption{Bias and root mean squared error (RMSE) of the point estimators and mean width of the nominal $95$ percent pointwise confidence intervals at the times $t \in \{1.2, 2, 3.3, 4\}$, for $n = 5000$ without noise covariates, for the jump forest (JF) and the fully stratified Nelson--Aalen estimator (NA); $100$ replications. The quantities are averaged with equal weights over the covariate values within each group, so that biases of opposite signs for different covariate values cancel in the average, whereas the RMSE (computed over all replications and covariate values of the group) does not. Entries $0.000$ denote values smaller than $5 \times 10^{-4}$ in absolute value.}
    \label{tab:coverage-bias}
\end{table}

\clearpage

\clearpage
\subsection{Biomedical data}\label{subsection:app-data}
The four data sets of Subsection~\ref{sec:data} arise from the following clinical settings. The \texttt{ebmt3} data are distributed with \texttt{mstate} \citep{mstateS}; the companion website of \citet{AndersenRavnS} provides the remaining data and covariate descriptions.

\begin{description}
    \item[\normalfont\texttt{ebmt3}] Bone marrow transplantation data collected by the European Society for Blood and Marrow Transplantation. Platelet recovery is the intermediate event, and the question is whether reaching it changes the subsequent risk of relapse or death.
    \item[\normalfont\texttt{bmt}] Bone marrow transplantation data for acute leukaemia. Graft-versus-host disease (GvHD) is the intermediate event, and its effect on relapse and death is the clinical question.
    \item[\normalfont\texttt{cphholter}] The Copenhagen Holter Study, whose participants were surveyed in 1998--2000 and followed until 2013. Excessive supraventricular ectopic activity (ESVEA) is a baseline covariate, and the question is how much of the excess stroke risk that it carries can be explained by atrial fibrillation (AF).
    \item[\normalfont\texttt{prova}] The PROVA trial in liver cirrhosis, which randomised patients to propranolol, sclerotherapy, both or neither. Bleeding is the intermediate event, and the question is the effect of treatment on bleeding and death.
\end{description}

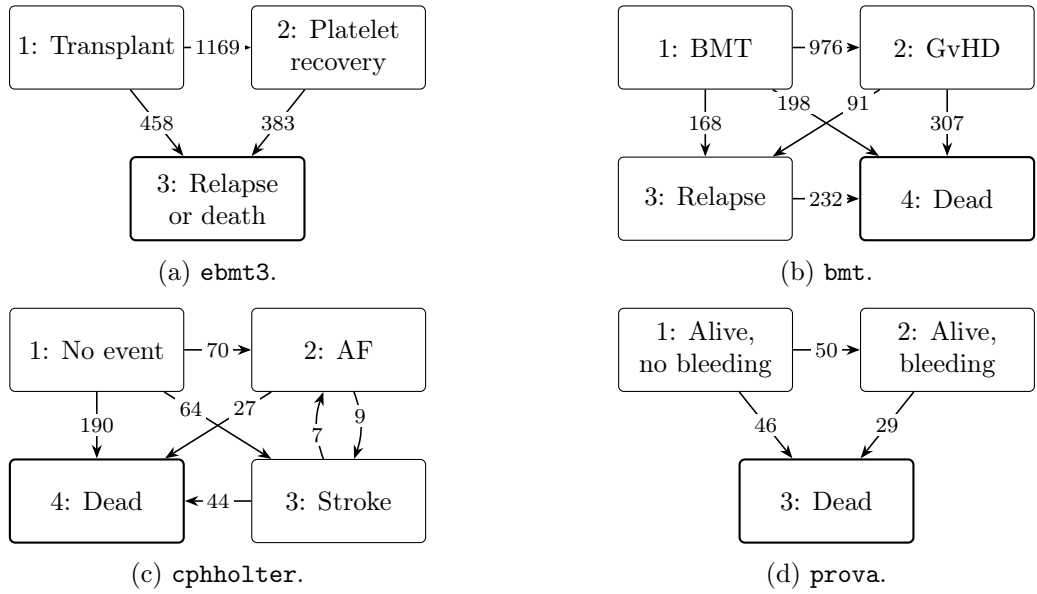
\begin{figure}[tbp]
    \centering
    \tikzset{
      state/.style={rectangle, draw, rounded corners=2pt, align=center,
        minimum width=2.3cm, minimum height=1.1cm, inner sep=3pt, font=\small},
      absorbing/.style={state, thick},
      arr/.style={-{Stealth[length=1.8mm]}, semithick},
      cnt/.style={font=\scriptsize, inner sep=1.5pt, fill=white}
    }
    \begin{subfigure}[b]{0.48\textwidth}
        \centering
        \begin{tikzpicture}
          \node[state]     (s1) at (0,0)      {1: Transplant};
          \node[state]     (s2) at (3.2,0)    {2: Platelet\\recovery};
          \node[absorbing] (s3) at (1.6,-2.0) {3: Relapse\\or death};
          \draw[arr] (s1) -- node[cnt] {1169} (s2);
          \draw[arr] (s1) -- node[cnt] {458} (s3);
          \draw[arr] (s2) -- node[cnt] {383} (s3);
        \end{tikzpicture}
        \caption{\texttt{ebmt3}.}
    \end{subfigure}
    \hfill
    \begin{subfigure}[b]{0.48\textwidth}
        \centering
        \begin{tikzpicture}
          \node[state]     (s1) at (0,0)      {1: BMT};
          \node[state]     (s2) at (3.2,0)    {2: GvHD};
          \node[state]     (s3) at (0,-2.0)   {3: Relapse};
          \node[absorbing] (s4) at (3.2,-2.0) {4: Dead};
          \draw[arr] (s1) -- node[cnt] {976} (s2);
          \draw[arr] (s1) -- node[cnt] {168} (s3);
          \draw[arr] (s2) -- node[cnt] {307} (s4);
          \draw[arr] (s3) -- node[cnt] {232} (s4);
          \draw[arr] (s1) -- node[cnt, pos=0.2] {198} (s4);
          \draw[arr] (s2) -- node[cnt, pos=0.2] {91} (s3);
        \end{tikzpicture}
        \caption{\texttt{bmt}.}
    \end{subfigure}

    \medskip

    \begin{subfigure}[b]{0.48\textwidth}
        \centering
        \begin{tikzpicture}
          \node[state]     (s1) at (0,0)      {1: No event};
          \node[state]     (s2) at (3.2,0)    {2: AF};
          \node[state]     (s3) at (3.2,-2.0) {3: Stroke};
          \node[absorbing] (s4) at (0,-2.0)   {4: Dead};
          \draw[arr] (s1) -- node[cnt] {70} (s2);
          \draw[arr] (s1) -- node[cnt] {190} (s4);
          \draw[arr] (s3) -- node[cnt] {44} (s4);
          \draw[arr] (s1) -- node[cnt, pos=0.25] {64} (s3);
          \draw[arr] (s2) -- node[cnt, pos=0.25] {27} (s4);
          \draw[arr] (s2) to[bend left=20] node[cnt, pos=0.35] {9} (s3);
          \draw[arr] (s3) to[bend left=20] node[cnt, pos=0.35] {7} (s2);
        \end{tikzpicture}
        \caption{\texttt{cphholter}.}
    \end{subfigure}
    \hfill
    \begin{subfigure}[b]{0.48\textwidth}
        \centering
        \begin{tikzpicture}
          \node[state]     (s1) at (0,0)      {1: Alive,\\no bleeding};
          \node[state]     (s2) at (3.2,0)    {2: Alive,\\bleeding};
          \node[absorbing] (s3) at (1.6,-2.0) {3: Dead};
          \draw[arr] (s1) -- node[cnt] {50} (s2);
          \draw[arr] (s1) -- node[cnt] {46} (s3);
          \draw[arr] (s2) -- node[cnt] {29} (s3);
        \end{tikzpicture}
        \caption{\texttt{prova}.}
    \end{subfigure}
    \caption{Multi-state models for the four biomedical data sets. Arrows indicate possible transitions and are labelled with the observed transition counts. Absorbing states have a heavier border.}
    \label{fig:datasets}
\end{figure}

\begin{figure}[H]
    \centering
    \includegraphics[width=\linewidth]{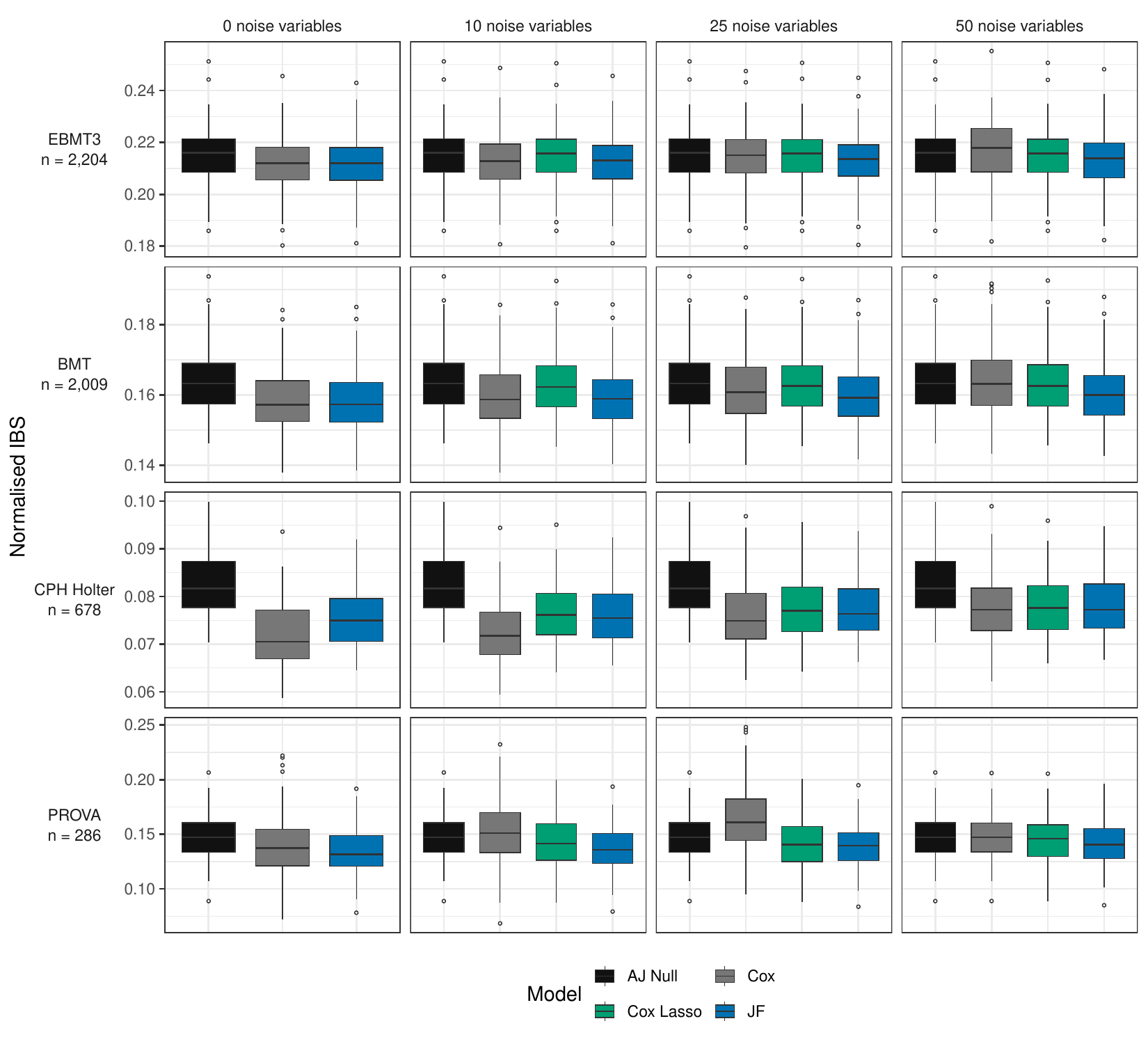}
    \caption{Biomedical data, $100$ random $80/20$ training--test splits. Normalised test-set IBS of the Aalen--Johansen estimator without covariates (AJ Null), the transition-wise Cox model (Cox), its lasso-penalised version (Cox Lasso) and the jump forest (JF). Rows correspond to the four data sets; columns to $0$, $10$, $25$ and $50$ added noise covariates. Points outside the whiskers mark individual replications.}
    \label{fig:ibsnoise}
\end{figure}

Figure~\ref{fig:datasets} shows the state structures and observed transition counts. Figure~\ref{fig:ibsnoise} gives the absolute IBS values underlying Figure~\ref{fig:ibsnoiserelative}.

\clearpage

\clearpage
\subsection{Covariate interactions}\label{subsection:interactions}
\begin{figure}[!htbp]
    \centering
    \begin{subfigure}[t]{0.48\textwidth}
        \centering
        \includegraphics[width=\linewidth]{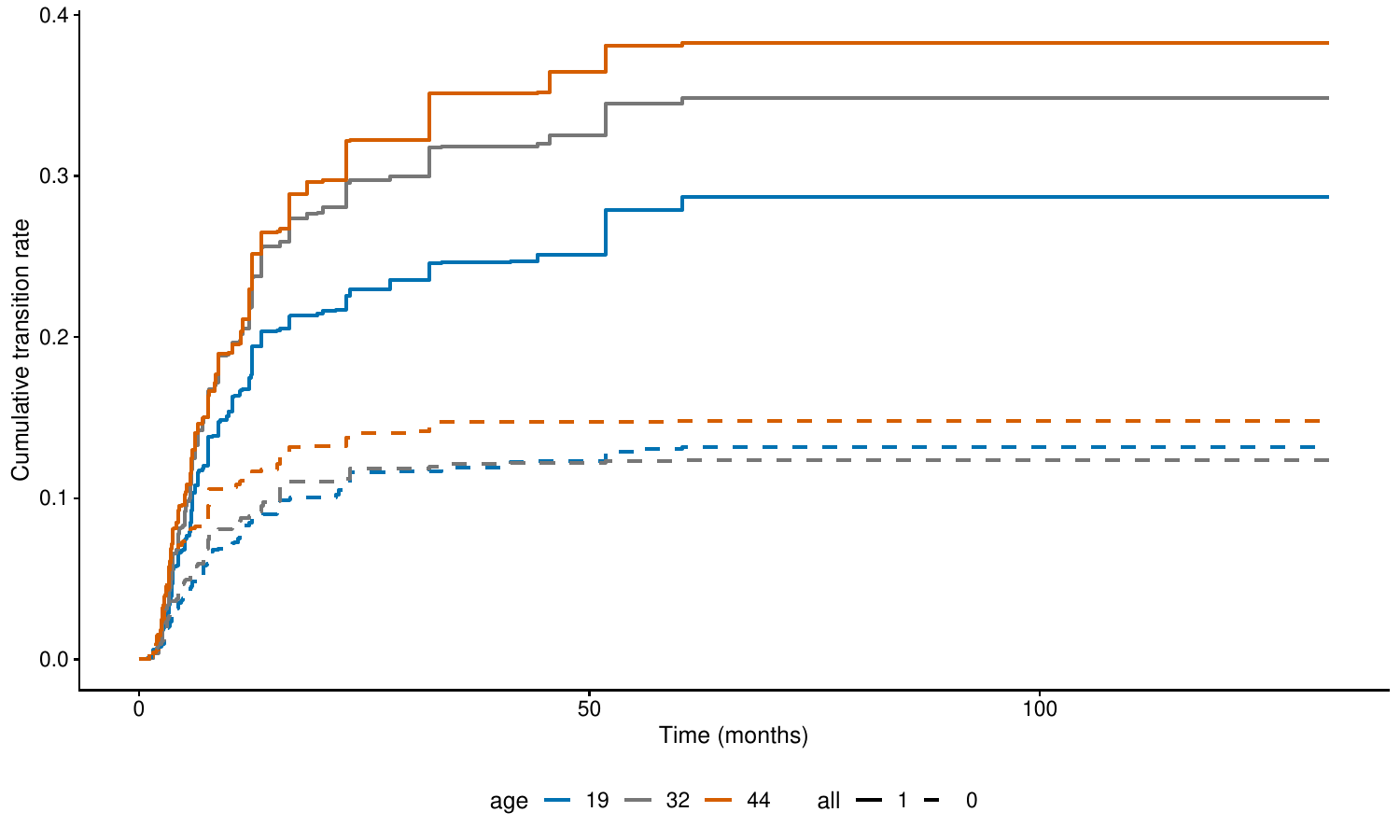}
        \caption{\texttt{age} by \texttt{all}, transition $1 \to 3$.}
        \label{fig:bmt-age-all-13}
    \end{subfigure}
    \hfill
    \begin{subfigure}[t]{0.48\textwidth}
        \centering
        \includegraphics[width=\linewidth]{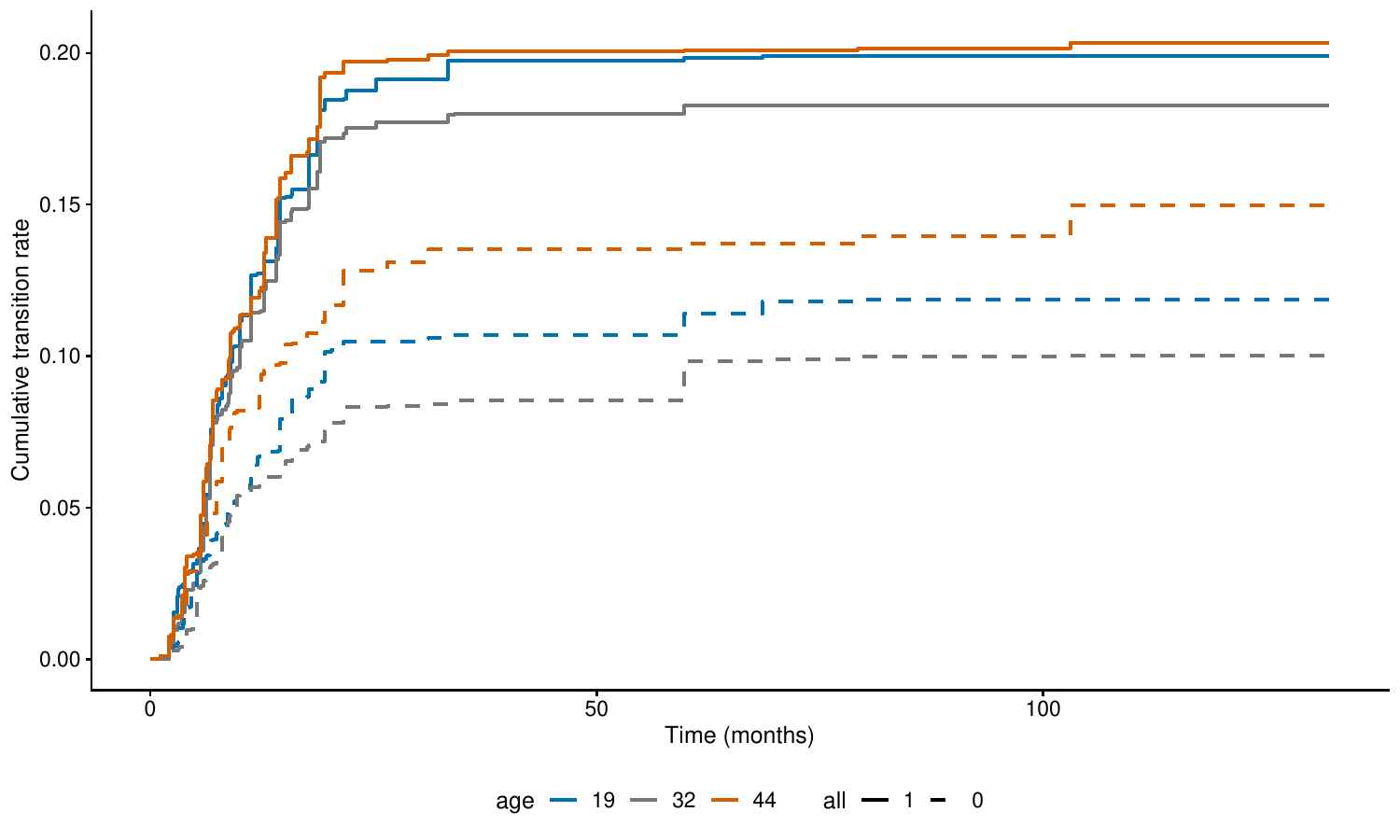}
        \caption{\texttt{age} by \texttt{all}, transition $2 \to 3$.}
        \label{fig:bmt-age-all-23}
    \end{subfigure}

    \medskip

    \begin{subfigure}[t]{0.48\textwidth}
        \centering
        \includegraphics[width=\linewidth]{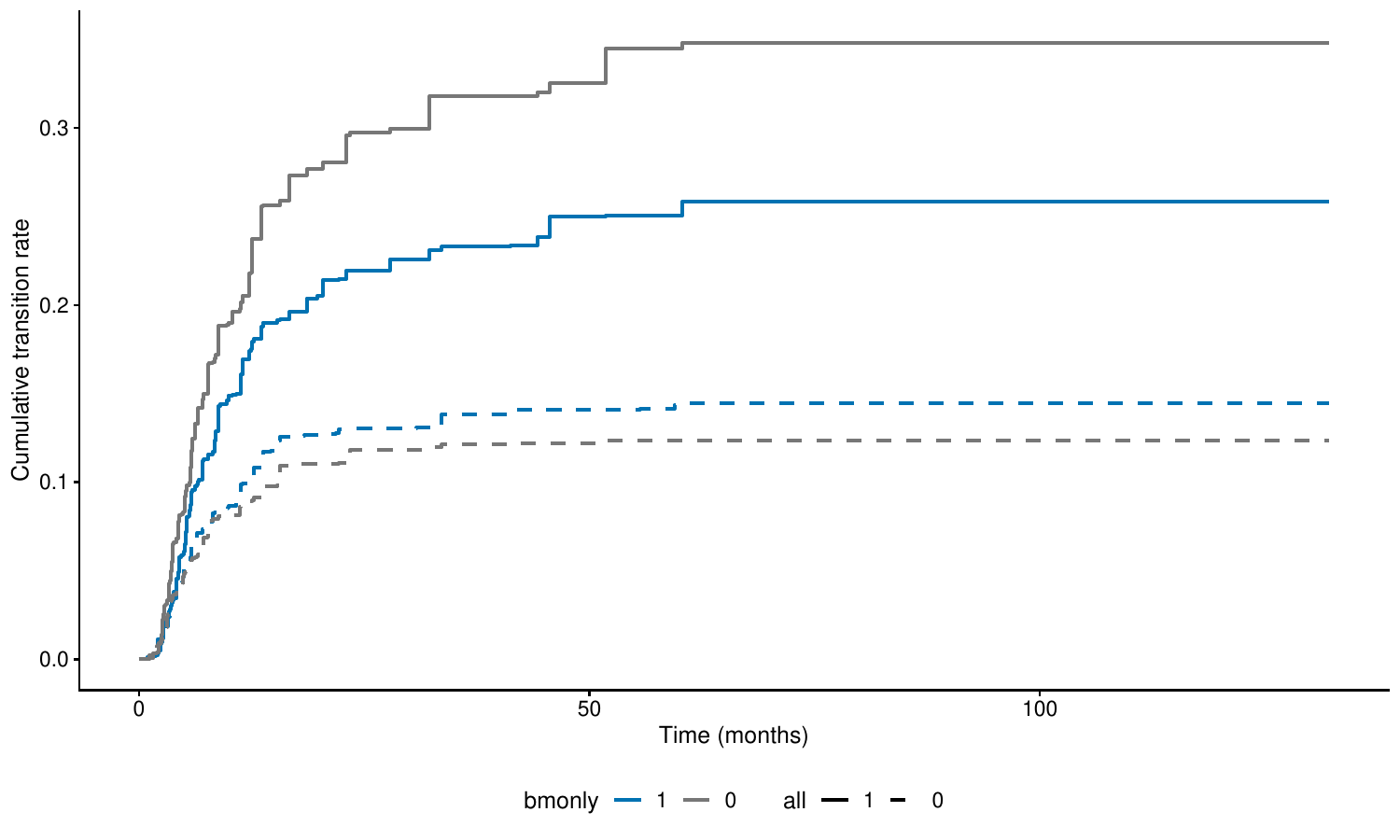}
        \caption{\texttt{bmonly} by \texttt{all}, transition $1 \to 3$.}
        \label{fig:bmt-bmonly-all-13}
    \end{subfigure}
    \hfill
    \begin{subfigure}[t]{0.48\textwidth}
        \centering
        \includegraphics[width=\linewidth]{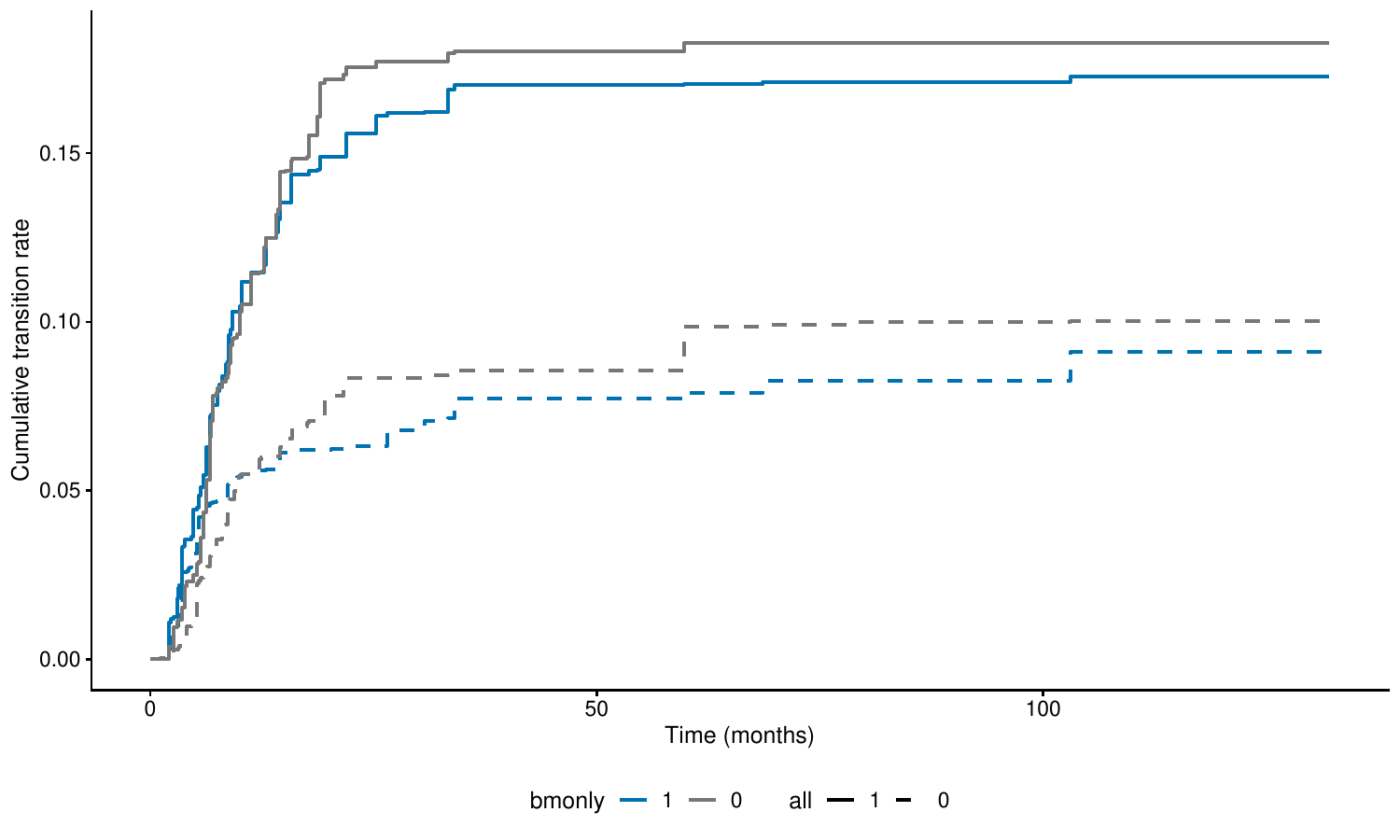}
        \caption{\texttt{bmonly} by \texttt{all}, transition $2 \to 3$.}
        \label{fig:bmt-bmonly-all-23}
    \end{subfigure}
    \caption{Covariate combinations in the \texttt{bmt} data. Cumulative transition rates predicted by the jump forest for transplant to relapse ($1\to3$, left) and GvHD to relapse ($2\to3$, right). Top row: ages $19$ (blue), $32$ (grey) and $44$ years (orange), with \texttt{sex} and \texttt{bmonly} held fixed. Bottom row: \texttt{bmonly} $=1$ (blue) and $0$ (grey), with \texttt{sex} held fixed and \texttt{age} set to $32$ years. Solid and dashed curves denote \texttt{all} $=1$ and $0$, respectively.}
    \label{fig:bmt-transition-interactions}
\end{figure}

Beyond prediction, the fitted forest can be used to explore how the covariates act on the transition rates. We plot the predicted cumulative transition rates for selected covariate combinations; a reversal in their ordering between levels of a second covariate is incompatible with a main-effects Cox model. Differences in absolute spacing alone do not establish interaction on the multiplicative hazard scale. For the \texttt{bmt} data, Figure~\ref{fig:bmt-transition-interactions} displays the predicted rates for combinations of \texttt{age} (the age of the patient at transplantation) and \texttt{all} (an indicator of acute lymphoblastic leukaemia), and of \texttt{all} and \texttt{bmonly} (an indicator of bone marrow being the only graft source). In each panel, the covariates that are not displayed are held fixed at their most frequent levels; in panels (c) and (d), \texttt{age} is set to $32$ years, the median age in the data.

\begin{figure}[!htbp]
    \centering
    \begin{subfigure}[t]{0.48\textwidth}
        \centering
        \includegraphics[width=\linewidth]{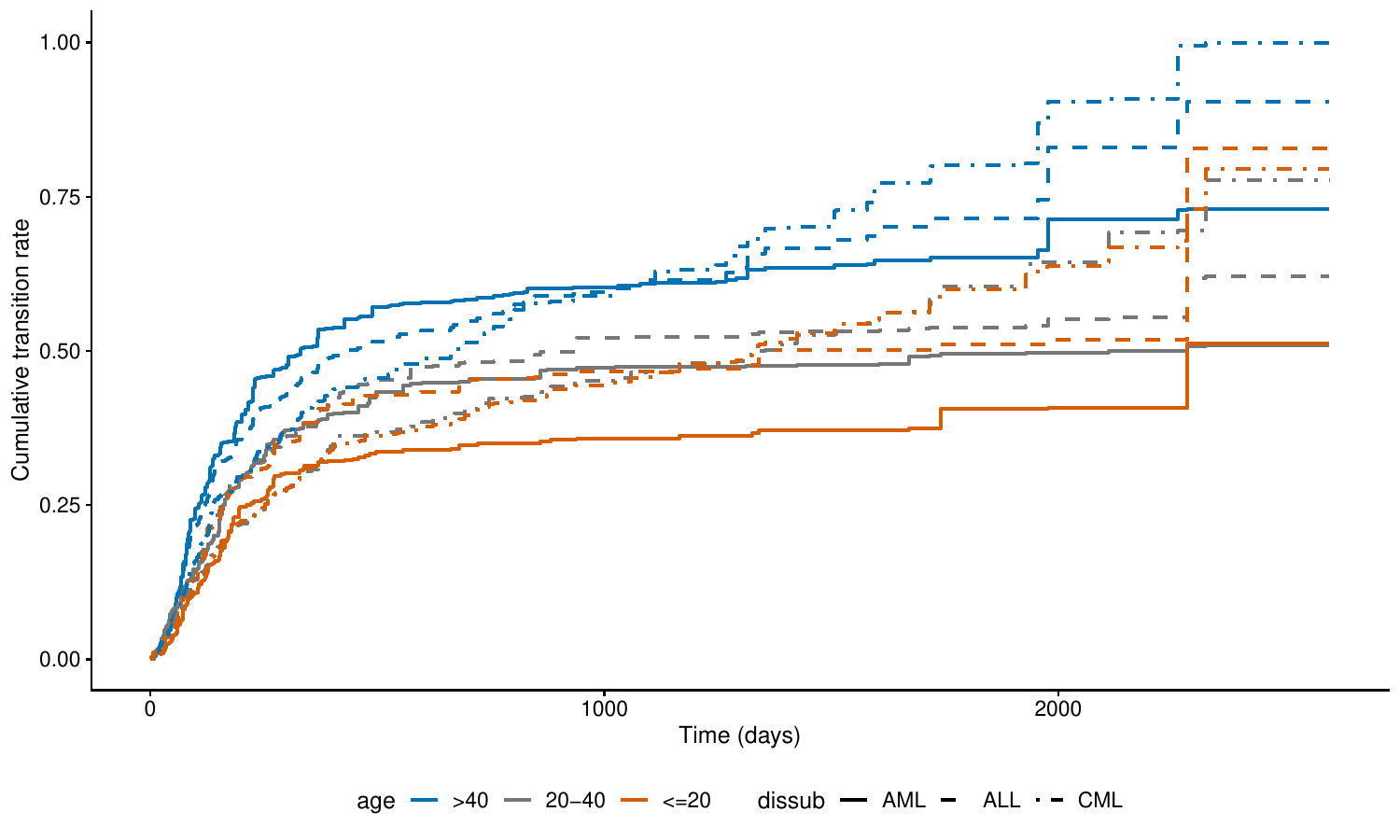}
        \caption{\texttt{age} by \texttt{dissub}, transition $1 \to 3$.}
        \label{fig:ebmt-age-dissub-13}
    \end{subfigure}
    \hfill
    \begin{subfigure}[t]{0.48\textwidth}
        \centering
        \includegraphics[width=\linewidth]{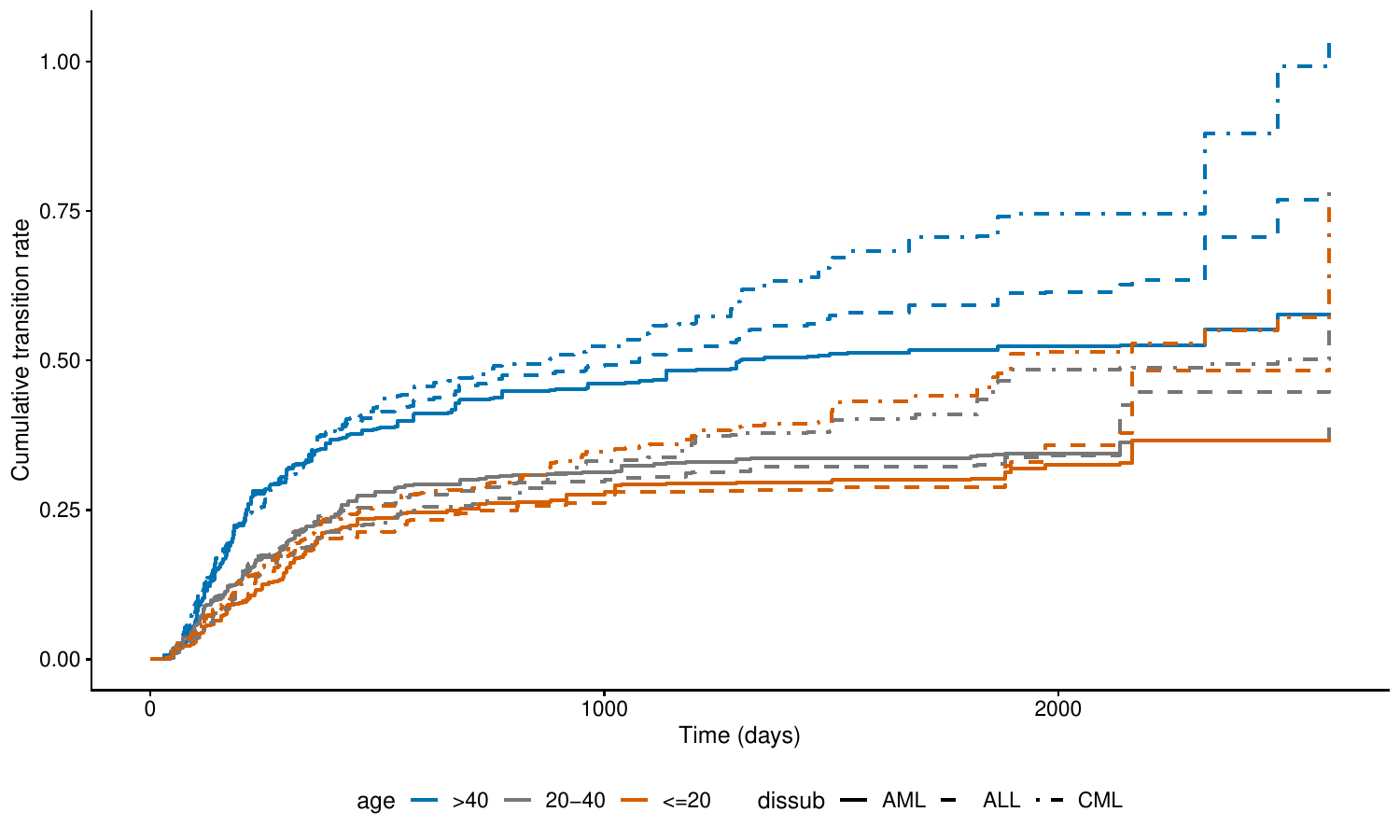}
        \caption{\texttt{age} by \texttt{dissub}, transition $2 \to 3$.}
        \label{fig:ebmt-age-dissub-23}
    \end{subfigure}

    \medskip

    \begin{subfigure}[t]{0.48\textwidth}
        \centering
        \includegraphics[width=\linewidth]{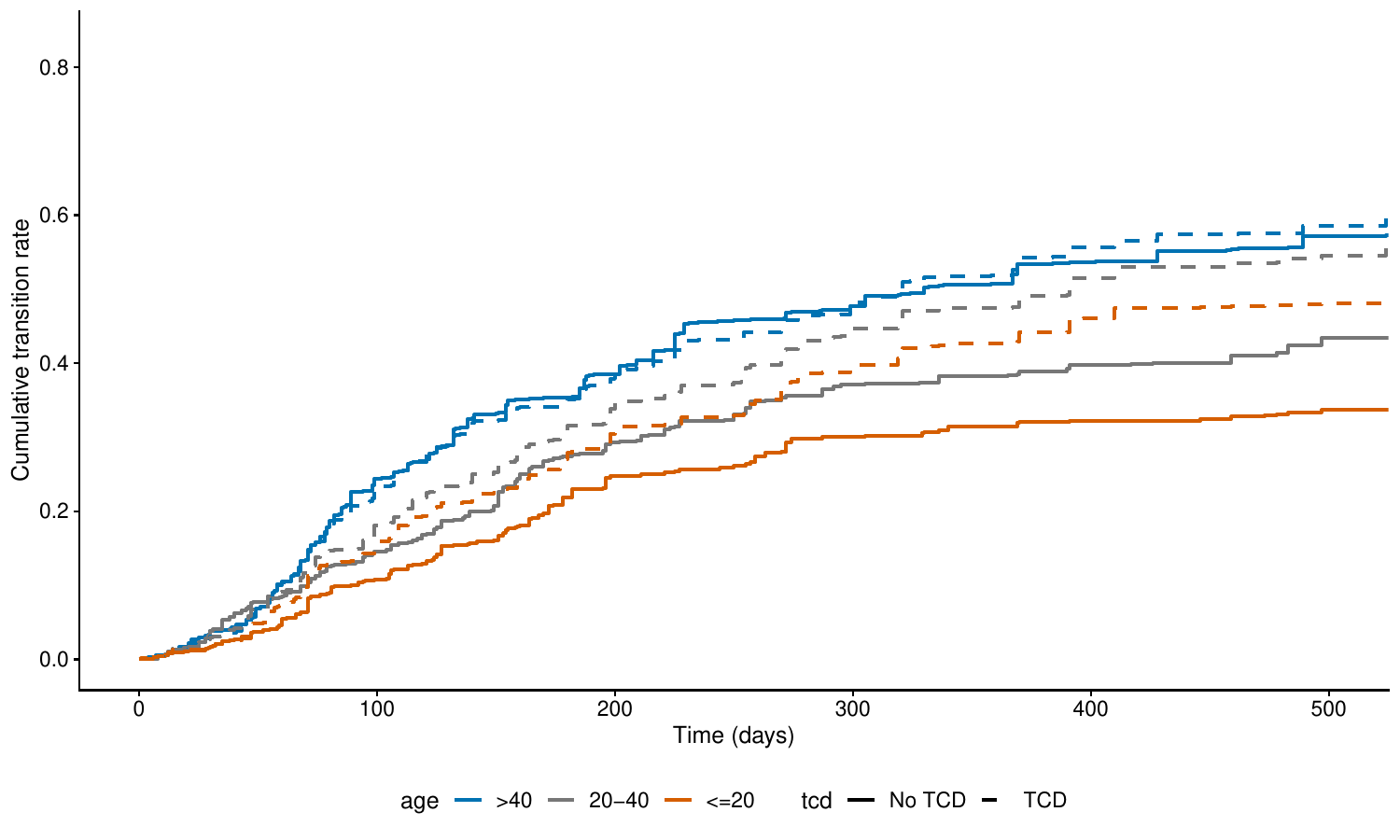}
        \caption{\texttt{age} by \texttt{tcd}, transition $1 \to 3$.}
        \label{fig:ebmt-age-tcd-13}
    \end{subfigure}
    \hfill
    \begin{subfigure}[t]{0.48\textwidth}
        \centering
        \includegraphics[width=\linewidth]{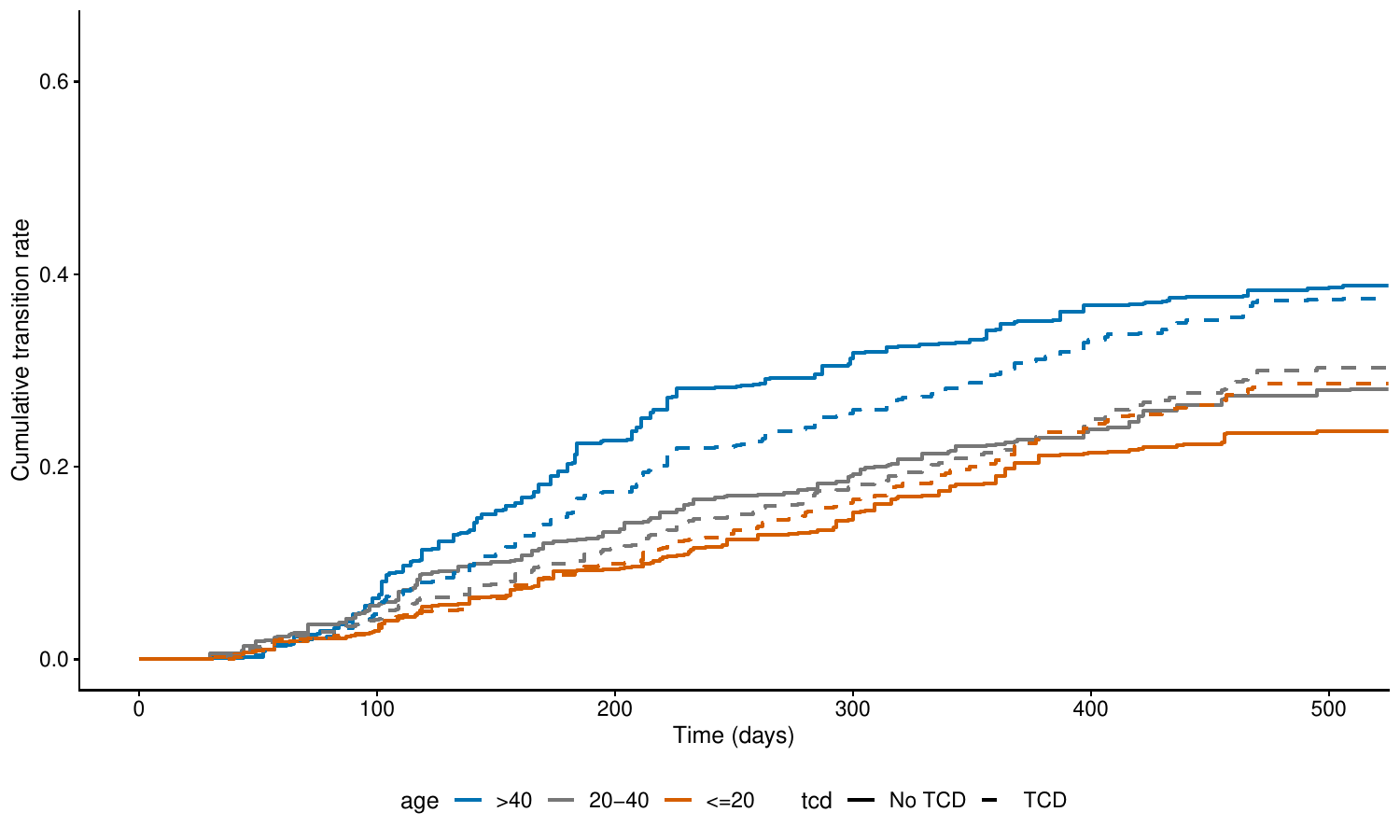}
        \caption{\texttt{age} by \texttt{tcd}, transition $2 \to 3$.}
        \label{fig:ebmt-age-tcd-23}
    \end{subfigure}
    \caption{Covariate combinations in the \texttt{ebmt3} data. Cumulative transition rates predicted by the jump forest for transplant to relapse or death ($1\to3$, left) and platelet recovery to relapse or death ($2\to3$, right). Colours denote age above $40$ years (blue), $20$--$40$ years (grey) and at most $20$ years (orange). Top row, \texttt{dissub}: AML (solid), ALL (dashed) and CML (dash-dotted). Bottom row, \texttt{tcd}: no T-cell depletion (solid) and T-cell depletion (dashed). The remaining covariates are held fixed at their most frequent levels.}
    \label{fig:ebmt-transition-interactions}
\end{figure}

For the transition from transplant to relapse, the ordering of the ages $19$ and $32$ reverses between the two levels of \texttt{all}; the ordering of \texttt{bmonly} also reverses. These fitted patterns suggest interactions that a main-effects Cox model cannot represent. However, the two methods have comparable overall IBS on \texttt{bmt}, so the patterns do not establish a predictive gain and should be treated as exploratory. Figure~\ref{fig:ebmt-transition-interactions} gives corresponding plots for \texttt{ebmt3}.

The \texttt{ebmt3} plots in Figure~\ref{fig:ebmt-transition-interactions} show predicted cumulative rates by \texttt{age}, \texttt{dissub} (AML, ALL or CML) and \texttt{tcd} (T-cell depletion). The absolute differences associated with T-cell depletion vary across age groups, but the ordering of those groups is preserved. Such differences alone do not demonstrate interaction on the multiplicative hazard scale; the plots describe the fitted covariate patterns.

\clearpage
\phantomsection

\end{document}